\documentclass[12pt]{article}
\usepackage{a41}
\usepackage{float}
\usepackage{amsmath}
\usepackage{amssymb}
\usepackage{eufrak}
\usepackage{epsfig}
\usepackage{rotating}
\usepackage{cite}
\usepackage{portland}
\newcommand\SH{\,\mbox{$\sqcup \! \sqcup$}\,}

\newcommand{\NN}{{\nonumber}}

\newcommand{\Li}{{\rm Li}}
\newcommand{\Ti}{{\rm Ti}}

\newcommand{\ds}{\displaystyle}

\newcommand{\bq}{\begin{equation}}
\newcommand{\eq}{\end{equation}}
\newcommand\beq{\begin{equation}}
\newcommand\eeq{\end{equation}}
\newcommand\bea{\begin{eqnarray}}
\newcommand\eea{\end{eqnarray}}

\newcommand\Mvec{\,\mbox{\bf M}}

\newcommand\todo[1][x]{\fbox{\textbf{TODO} \ifx#1x\fi}}

\usepackage{amsthm}

\newcommand{\X}[2]{{S}_{#1}\left(#2\right)}
\newcommand{\sign}[1]{\textnormal{sign}\left(#1\right)}
\newcommand{\abs}[1]{\left|{#1}\right|}
\newcommand{\Z}{\mathbb Z}
\newcommand{\N}{\mathbb N}
\newcommand{\ve}[1]{\textit{\textbf{#1}}}

\begin{document}
\setlength{\baselineskip}{0.515cm}
\sloppy
\thispagestyle{empty}
\begin{flushleft}
DESY 11--033
%\hfill {\tt arXiv:1102.yyyy [math-ph]}
\\
DO--TH 11--12\\
SFB/CPP-11-24\\
LPN 11/24\\
May 2011\\
\end{flushleft}

\mbox{}
\vspace*{\fill}
\begin{center}

{\LARGE\bf Harmonic Sums and Polylogarithms}

\vspace*{2mm}
{\LARGE\bf Generated by Cyclotomic Polynomials}

\vspace{4cm}
\large
Jakob Ablinger$^a$,  Johannes Bl\"umlein$^b$, and Carsten Schneider$^a$

\vspace{1.5cm}
\normalsize
{\it $^a$~Research Institute for Symbolic Computation (RISC),\\
                          Johannes Kepler University, Altenbergerstra\ss{}e 69,
                          A--4040, Linz, Austria}\\

\vspace*{3mm}
{\it  $^b$ Deutsches Elektronen--Synchrotron, DESY,}\\
{\it  Platanenallee 6, D-15738 Zeuthen, Germany}
\\
%%\today

\end{center}
\normalsize
\vspace{\fill}
\begin{abstract}
\noindent
The computation of Feynman integrals in massive higher order perturbative calculations
in renormalizable Quantum Field Theories requires extensions of multiply nested harmonic
sums, which can be generated as real representations by Mellin transforms of Poincar\'e--iterated
integrals including denominators of higher cyclotomic polynomials. We derive the cyclotomic
harmonic polylogarithms and harmonic sums and study their algebraic and structural relations.
The analytic continuation of cyclotomic harmonic sums to complex values of $N$ is performed
using analytic representations. We also consider special values of the cyclotomic harmonic
polylogarithms at argument $x=1$, resp., for the cyclotomic harmonic sums at $N \rightarrow \infty$,
which are related to colored multiple zeta values, deriving various of their relations, based on
the stuffle and shuffle algebras and three multiple argument relations. We also consider infinite
generalized nested harmonic sums at roots of unity which are related to the infinite cyclotomic
harmonic sums. Basis representations are derived for weight {\sf w = 1,2} sums up to cyclotomy
{\sf l = 20}.
\end{abstract}

\vspace*{5mm}
\begin{center}
{\sf Dedicated to Martinus Veltman on the occasion of his 80th birthday.}
\end{center}

\vspace*{\fill}
\noindent
\numberwithin{equation}{section}
%%%%%%%%%%%%%%%%%%%%%%%%%%%%%%%%%%%%%%%%%%%%%%%%%%%%%%%%%%%%%%%%%%%%%%%
\newpage
%%%%%%%%%%%%%%%%%%%%%%%%%%%%%%%%%%%%%%%%%%%%%%%%%%%%%%%%%%%%%%%%%%%%%%%%
\section{Introduction}
\label{sec:1}
%%%%%%%%%%%%%%%%%%%%%%%%%%%%%%%%%%%%%%%%%%%%%%%%%%%%%%%%%%%%%%%%%%%%%%%%

\vspace{1mm}
\noindent
The analytic calculation of Feynman integrals requires the complete understanding 
of the associated mathematical structures at a given loop level. By the pioneering 
work in Refs.~\cite{VELTMAN1,VELTMAN2} this has been thoroughly achieved for one loop integrals 
occurring in renormalizable Quantum Field Theories. The corresponding complete framework 
in case of higher order calculations is, however, not yet available. Which mathematical
functions are of relevance there is revealed stepwise in  specific higher order calculations.
Here higher transcendental functions, like the generalized hypergeometric
functions and their generalizations \cite{HYP}, play a central role. Their series expansion in the dimensional 
parameter \cite{VELTMAN1,DIM} leads to nested infinite sums over products of digamma functions
\cite{DIGAMMA}, cf.~\cite{STRUCT5}. The nested harmonic sums~\cite{HS1,HS2} are defined by
%----------------------------------------------------------------------------------------------
\begin{eqnarray}
S_{b,\vec{a}}(N) &=& \sum_{k=1}^N \frac{({\sf sign}(b))^k}{k^{|b|}}
S_{\vec{a}}(k)~,~~~~S_{\emptyset}(N) = 1~,~~~b, a_i \in {\mathbb Z \backslash \{0\}}.
\end{eqnarray}
%----------------------------------------------------------------------------------------------
They form a quasi-shuffle algebra \cite{HOFF}. Their values for $N \rightarrow \infty$ 
are the multiple zeta values $\zeta_{\vec{a}}$, resp. Euler-Zagier values \cite{EZ} defined by 
$$\zeta_{b,\vec{a}}=\lim_{N\to\infty}S_{b,\vec{a}}(N),\quad b\neq1,$$
see~\cite{MZV}.
In case of massless problems to 3--loop order the results for single scale quantities in Mellin space can be 
written by polynomial expressions in terms of  $S_{\vec{a}}(N)$ and $\zeta_{\vec{a}}$ with coefficients being from the 
rational function field
${\mathbb Q}(N)$, cf. e.g. \cite{Vermaseren:2005qc}.
In this context we consider the Mellin transform
%----------------------------------------------------------------------------------------------
\begin{eqnarray}
\label{eq:MEL1}
\Mvec[f(x)](N) = \int_0^1~dx~x^{N}~f(x)~.
\end{eqnarray}
%----------------------------------------------------------------------------------------------
In most of the applications below we assume $N \in \mathbb{N}_+ = \mathbb{N} \backslash \{0\}$.

%----------------------------------------------------------------------------------------------
In computations at even higher orders in the coupling constant and through finite mass effects, 
however, generalizations of the
nested harmonic sums contribute, at least in intermediary results. One extension concerns
the so-called generalized
harmonic sums \cite{GHS1,GHS2} given by
%----------------------------------------------------------------------------------------------
\begin{eqnarray}
\label{eq:GENSU}
S_{b,\vec{a}}(\zeta, \vec{\xi}; N) &=& \sum_{k=1}^N \frac{\zeta^k}{k^b}
S_{\vec{\xi}}(\vec{r}; k)~,
\end{eqnarray}
%----------------------------------------------------------------------------------------------
with $b, a_i \in
{\mathbb N}_+;
\zeta, \xi_i \in  {\mathbb R}^* = {\mathbb R} \backslash \{0\}$.
Known examples are related to the second index set $\xi_i \in \{1,-1,1/2,-1/2,2,-2\}$,
cf.~\cite{Vermaseren:2005qc,HQ1,HQ2}.

In case of single scale problems with two massive lines and $m_1=m_2$ at 3-loop order summation is
also required over terms
%----------------------------------------------------------------------------------------------
\begin{eqnarray}\label{Equ:CyclotimcSum4}
\frac{(\pm 1)^k}{(2 k +1)^n}~,
\end{eqnarray}
%----------------------------------------------------------------------------------------------
which is a special case of
%----------------------------------------------------------------------------------------------
\begin{eqnarray}
\label{eq:NEWT}
\frac{(\pm 1)^k}{(l \cdot k +m)^n}~,
\end{eqnarray}
%----------------------------------------------------------------------------------------------
with $l,m,n \in {\mathbb N}_+$.
Note that we have deliberately chosen a real representation here, which is of practical importance
in case of fast polynomial operations needed for solving nested summation problems \cite{SIGMA}~\footnote{
Complex representations are related to the so-called colored harmonic sums
$S_{b,\vec{a}}(p, \vec{r}; N) = \sum_{k=1}^N \frac{p^k}{k^b} S_{\vec{a}}(\vec{r}; k)$
with $b, a_i \in {\mathbb N}_+, p, r_i \in \cup_{l = 2}^M \{\exp[2\pi
i (n/l)], n \in \{1,...,l-1\}\}$, cf.~\cite{COLO,GON1}.}. Sums containing fractional terms $m/l$ were
considered
in the context of colored harmonic sums in \cite{DB1,LIL,WZ}.
It is expected that sums of this kind and their iterations will occur in a wide class
of massive calculations in higher order loop calculations in Quantum Electrodynamics, Quantum Chromodynamics, and
other renormalizable Quantum Field Theories, in particular studying single distributions, but also for more
variable differential distributions. Usually objects of this kind emerge first at intermediary steps and,
at even higher orders, they occur in the final results. Therefore, these quantities have to be understood and methods
have to be provided to perform these sums. This does not apply to integer values of the
Mellin variable $N$ only, but
also to the analytic continuation of these quantities to $N \in \mathbb{C}$, since the experimental applications require
the Mellin inversion into momentum-fraction space.

We show that the single sums (\ref{eq:NEWT}) and their nested iterations can be obtained from linear combinations of
Mellin transforms of harmonic polylogarithms over an alphabet of letters containing 
$x^l/\Phi_k(x),~~0< l < \text{deg}(\Phi_k(x))$, where
$\Phi_k(x)$ denotes
the $k$th cyclotomic polynomial \cite{LANG}. One may form words by Poincar\'e-iterated integrals \cite{POINC} over
this alphabet, which leads to the {\sf cyclotomic harmonic polylogarithms} $\mathfrak{H}$,
forming  a shuffle algebra.
This class extends the harmonic polylogarithms \cite{VR}. The Mellin transform of elements of $\mathfrak{H}$ has
support $x \in [0,1]$, but requires an extension in the Mellin variable $N \in \mathbb{N}$,
%----------------------------------------------------------------------------------------------
\begin{eqnarray}
N \rightarrow k \cdot N~,
\end{eqnarray}
%----------------------------------------------------------------------------------------------
where $k$ denotes the index of $\Phi_k(x)$. This assumption allows to associate nested
harmonic sums of the cyclotomic type
by this Mellin transform. Special values are obtained by either the cyclotomic harmonic polylogarithms $H_{\vec{a}}(x)$
at $x=1$ or the associated nested harmonic sums for $N \rightarrow \infty$. They extend the multiple zeta values and
Euler-Zagier values, cf.~Section~\ref{sec:5}. In the present paper we investigate relations and representations of these three
classes of quantities. Special emphasis has been put on the class of nested sums~\eqref{Equ:CyclotimcSum4} where all 
derived algorithms have been incorporated within the computer algebra package {\sf HarmonicSums}~\cite{HASUM}.

The paper is organized as follows. In Section~\ref{sec:2} we establish the connection
between the cyclotomic harmonic sums and
the cyclotomic harmonic polylogarithms through the Mellin transform, at modified argument $k N$. The basic
properties of the cyclotomic harmonic polylogarithms are investigated in
Section~\ref{sec:3}. The cyclotomic harmonic polylogarithms
obey a shuffle algebra. The nested sums at finite values of
$N$ are studied in Section~\ref{sec:4}, including their algebraic and structural
relations, generalizing~\cite{JBSH,STR1,STR2,ABS11}.
Here are also three multiple argument relations of interest. The analytic continuation of the new harmonic sums to complex
values of $N$ is presented based on recursion and asymptotic representations, similar to the case of the nested harmonic
sum \cite{STR1,STR2,ABS11,ANCONT}. The representation of the cyclotomic harmonic sums requires to know
their values at
$N \rightarrow \infty$, which are equivalently given by the cyclotomic harmonic polylogarithms at $x=1$. The set
of special numbers spanning the Euler-Zagier and multiple zeta values \cite{MZV} are extended. We study the
cases of weight {\sf w = 1,2} up to cyclotomy {\sf l = 20} and derive the corresponding relations based on the stuffle
and
shuffle algebra and three multiple argument relations. Furthermore, we investigate the relations of the cyclotomic
harmonic sums for $N \rightarrow \infty$ extending for words resulting from the alphabet $(\pm 1)^k/k, (\pm 1)^k/(2k+1)$,
cf. Section~\ref{sec:5}. For the cyclotomic harmonic polylogarithms, harmonic sums and
their values at $N \rightarrow \infty$
basis representations are derived.
In Section~\ref{sec:6} we study the relations of the infinite nested harmonic sums with
numerators at $l$th root of unity, $\sf l \leq 20$ for weight {\sf w = 1,2}. Here, also the 
distribution relation,
cf.~\cite{GON1}, is considered beyond the relations mentioned before. We also consider a relation valid for finite 
generalized harmonic sums with root numerator weights.
Section~\ref{sec:7} contains the conclusions. Some technical details are given in the
Appendices.

%%%%%%%%%%%%%%%%%%%%%%%%%%%%%%%%%%%%%%%%%%%%%%%%%%%%%%%%%%%%%%%%%%%%%%%%
\section{Basic Formalism}
\label{sec:2}
%%%%%%%%%%%%%%%%%%%%%%%%%%%%%%%%%%%%%%%%%%%%%%%%%%%%%%%%%%%%%%%%%%%%%%%%

\vspace{1mm}
\noindent
We consider \textsf{cyclotomic harmonic sums} defined by
%----------------------------------------------------------------------------------------------
\begin{align}
\label{eq:MSU}
S_{\{a_1,b_1,c_1\}, ...,\{a_l,b_l,c_l\}}(s_1, ...,s_l; N)
&= \sum_{k_1 = 1}^{N} \frac{s_1^k}{(a_1 k_1 + b_1)^{c_1}}
S_{\{a_2,b_2,c_2\}; ...;\{a_l,b_l,c_l\}}(s_2, ...,s_l; k_1), &S_{\emptyset} &= 1,
\end{align}
%----------------------------------------------------------------------------------------------
where $a_i, c_i \in \mathbb{N}_+, b_i \in  \mathbb{N},~~s_i = \pm 1, a_i > b_i$; 
the weight of this sum is defined by $c_1+\dots+c_l$ and $\{a_i, b_i, c_i\}$  denote lists, not sets.
One may generalize
this case further allowing $s_i \in {\mathbb{R}}^*$, \cite{GHS2}.
Of special interest will be the infinite cyclotomic harmonic polylogarithms defined by 
$$\sigma_{\{a_1,b_1,c_1\}, ...,\{a_l,b_l,c_l\}}(s_1, ...,s_l)
=\lim_{N\to\infty}S_{\{a_1,b_1,c_1\}, ...,\{a_l,b_l,c_l\}}(s_1, ...,s_l; N)$$
which diverge if $c_1 = 1, s_1 = 1$. Sometimes we will use the notation  
%----------------------------------------------------------------------------------------------
\begin{eqnarray}
S_{\{a_1,b_1,c_1\}, ...,\{a_l,b_l,c_l\}}(s_1, ...,s_l; N)
&=& S_{\{a_1,b_1,s_1 c_1\}, ...,\{a_l,b_l,s_l c_l\}}(N) \\
\sigma_{\{a_1,b_1,c_1\}, ...,\{a_l,b_l,c_l\}}(s_1, ...,s_l)
&=&\sigma_{\{a_1,b_1,s_1 c_1\}, ...,\{a_l,b_l,s_l c_l\}}
\end{eqnarray}
%----------------------------------------------------------------------------------------------
as shortcut below.

For further considerations, we rely on the following procedure which transforms the sums to integrals.
The denominators have the following integral representation
%----------------------------------------------------------------------------------------------
\begin{eqnarray}
\label{eq:IR1}
\frac{(\pm 1)^k}{a k + b}     &=& \int_0^1~dx~x^{ak+b-1} (\pm 1)^k \\
\label{eq:IR2}
\frac{(\pm 1)^k}{(a k + b)^c} &=& \int_0^1 \frac{dx_1}{x_1} \int_0^{x_1} \frac{dx_2}{x_2} ...
\int_0^{x_{c-2}} \frac{dx_{c-1}}{x_{c-1}}
\int_0^{x_{c-1}}~dx_c~x_c^{ak+b-1} (\pm 1)^k,
\end{eqnarray}
%----------------------------------------------------------------------------------------------
and the sum over $k$ yields
%----------------------------------------------------------------------------------------------
\begin{eqnarray}
\label{eq:SISU}
\sum_{k=1}^{l} (\pm 1)^k x^{ak+b-1}  = x^{a+b-1} \frac{(\pm x^a)^{l+1} -1}{(\pm x^a) -1}~.
\end{eqnarray}
%----------------------------------------------------------------------------------------------
This representation is applied to the innermost sum $(a = a_l, b = b_l, c = c_l)$. One now may perform the next sum in
the
same way, provided $a_{l-1} | a_l$. If this is not the case, one transforms the integration variables in
(\ref{eq:IR1}, \ref{eq:IR2}) such that the next denominator can be generated, etc.
In this way the sums (\ref{eq:MSU})
can be represented in terms of linear combinations of Poincar\'e-iterated integrals. Evidently, the representation of
the cyclotomic harmonic sum (\ref{eq:MSU}) in terms of a (properly regularized) Mellin transform will be related to the
Mellin variable $k N$, with $k$ the least common multiple of $a_1, ... ,a_l$.

Let us illustrate the principle steps in case of the following example~:
%----------------------------------------------------------------------------------------------
\begin{eqnarray}
S_{\{3,2,2\},\{2,1,1\}}(1,-1;N) = \sum_{k=1}^{N} \frac{1}{(3k+2)^2}\sum_{l=1}^{k} \frac{(-1)^l}{(2l+1)}~.
\end{eqnarray}
%----------------------------------------------------------------------------------------------
The first sum yields
%----------------------------------------------------------------------------------------------
\begin{eqnarray}
S_{\{3,2,2\},\{2,1,1\}}(1,-1;N) =  \sum_{k=1}^{N} \int_0^1 dx \frac{x^2}{x^2+1}
\frac{(-x^2)^{k} + 1}{(3k + 2)^2}~.
\end{eqnarray}
%----------------------------------------------------------------------------------------------
Setting $x = y^3$ one obtains
%----------------------------------------------------------------------------------------------
\begin{eqnarray}
\label{eq:ex1}
S_{\{3,2,2\},\{2,1,1\}}(1,-1;N) &=& 12 \int_0^1~dy~\frac{y^8}{y^6+1}
\sum_{k=1}^N \frac{(-y^6)^k-1}{(6k+4)^2}
\nonumber\\
&=& 12\int_0^1~dy~\frac{y^4}{y^6+1} \Biggl\{
\int_0^y
\frac{dz}{z} \int_0^z dt~t^9~\frac{(-t^6)^{N}-1}{t^6+1}
\nonumber\\ && \hspace*{2.9cm}
- y^4 \int_0^1
\frac{dz}{z} \int_0^z dt~t^9~\frac{t^{6N}-1}{t^6-1}
\Biggr\} \\
\label{eq:EX1}
&=& 12\int_0^1~dy~\frac{y^4}{y^6+1}
\int_0^y
\frac{dz}{z} \int_0^z dt~t^9~\frac{(-t^6)^{N}-1}{t^6+1}
\nonumber\\ &&
- \left(4 -\pi \right) \int_0^1
\frac{dz}{z} \int_0^z dt~t^9~\frac{t^{6N}-1}{t^6-1}
\Biggr\}~.
\end{eqnarray}
%----------------------------------------------------------------------------------------------
In general, the polynomials
%----------------------------------------------------------------------------------------------
\begin{eqnarray}
x^a - 1
\end{eqnarray}
%----------------------------------------------------------------------------------------------
in (\ref{eq:SISU}) decompose in a product of cyclotomic polynomials, except for $a=1$ for which the expression is
$\Phi_1(x)$,
see
Section~3. Moreover, the polynomials
%----------------------------------------------------------------------------------------------
\begin{eqnarray}
x^a + 1 = \frac{x^{2a}-1}{x^a-1}
\end{eqnarray}
%----------------------------------------------------------------------------------------------
are either cyclotomic for $a = 2^n, n \in \mathbb{N}$ or decompose into products of cyclotomic
polynomials in other cases, see Appendix~A. All factors divide $(x^a)^l - 1$, resp. $(-x^a)^l - 1$. We remark that Eq.~(\ref{eq:EX1}) is not yet written in terms of a Mellin
transform (\ref{eq:MEL1}). To achieve this in an automatic fashion, the cyclotomic harmonic polylogarithms are introduced in Section~3. Furthermore, their special values at $x=1$
contribute to which we turn in Section~5.

%%%%%%%%%%%%%%%%%%%%%%%%%%%%%%%%%%%%%%%%%%%%%%%%%%%%%%%%%%%%%%%%%%%%%%%%
\section{Cyclotomic Harmonic Polylogarithms}
\label{sec:3}
%%%%%%%%%%%%%%%%%%%%%%%%%%%%%%%%%%%%%%%%%%%%%%%%%%%%%%%%%%%%%%%%%%%%%%%%

\vspace{1mm}
\noindent
To account for the newly emerging sums (\ref{eq:MSU}) in perturbative calculations in
Quantum Field Theory we introduce Poincar\'{e}-iterated integrals over the
alphabet $\mathfrak{A}$
%----------------------------------------------------------------------------------------------
\begin{eqnarray}
\label{eq:alphab}
\mathfrak{A} =
\left\{\frac{1}{x}\right\}\cup\left\{\left. \frac{x^l}{\Phi_k(x)}\right|k\in\mathbb N_+,0\leq 
l<\varphi(k)\right\},
\end{eqnarray}
%----------------------------------------------------------------------------------------------
where $\Phi_k(x)$ denotes the $k$th cyclotomic polynomial \cite{LANG},
and $\varphi(k)$ denotes Euler's totient function \cite{TOTIENT}.
%----------------------------------------------------------------------------------------------
\begin{eqnarray}
\Phi_n(x) = \frac{x^n-1}{\ds \prod_{d|n, d < n} \Phi_d(x)}~,~~~~~d,n \in \mathbb{N}_+,
\end{eqnarray}
%----------------------------------------------------------------------------------------------
and the first cyclotomic polynomials are given by
%----------------------------------------------------------------------------------------------
\begin{eqnarray}
\Phi_1(x) &=& x - 1 \\
\Phi_2(x) &=& x + 1 \\
\Phi_3(x) &=& x^2 + x + 1 \\
\Phi_4(x) &=& x^2 + 1 
\\
\Phi_5(x) &=& x^4 + x^3 + x^2 + x+ 1 \\
\Phi_6(x) &=& x^2 - x + 1 \\
\Phi_7(x) &=& x^6 + x^5 + x^4 + x^3 + x^2 + x+ 1 \\
\Phi_8(x) &=& x^4 + 1 \\
\Phi_9(x) &=& x^6 + x^3 + 1 \\
\Phi_{10}(x) &=& x^4 - x^3 + x^2 - x+ 1 \\
\Phi_{11}(x) &=& x^{10} + x^9 + x^8 + x^7 + x^6 + x^5 + x^4 + x^3 + x^2 + x+ 1 \\
\Phi_{12}(x) &=& x^4 - x^2 + 1,~~\text{etc.}
\end{eqnarray}
%----------------------------------------------------------------------------------------------
The alphabet $\mathfrak{A}$ is an extension of the alphabet
%----------------------------------------------------------------------------------------------
\begin{eqnarray}
\label{eq:alphab1}
\mathfrak{A}_{\rm H} = \left\{\frac{1}{x}, \frac{1}{\Phi_1(x)},\frac{1}{\Phi_2(x)}\right\},
\end{eqnarray}
%----------------------------------------------------------------------------------------------
generating the usual harmonic polylogarithms
\cite{VR}~\footnote{Note that we defined here the second letter by $1/(x-1)$ which differs in sign from the
corresponding letter in \cite{VR}. Numerical implementations were given in
\cite{GR,Vollinga:2004sn}. A
few extensions of iterated integrals
introduced
in \cite{VR} based on linear denominator functions of different kind, which are used in quantum-field theoretic
calculations, were made in \cite{Vollinga:2004sn,EXT1}.}. As a shorthand notation we define the
letters of $\mathfrak{A}$ by
%----------------------------------------------------------------------------------------------
\begin{eqnarray}
\label{eq:f1}
f_0^0(x) &=& \frac{1}{x} \\
\label{eq:f2}
f_k^l(x) &=& \frac{x^l}{\Phi_k(x)},~~~~k \in \mathbb{N}_+, l \in \mathbb{N}, l \leq \varphi(k)~.
\end{eqnarray}
%----------------------------------------------------------------------------------------------
Here the labels $k$ and $l$ form a double index, which always appear sat a {\sf common position}. 
Mellin transformations associated to $1/\Phi_4(x)$ and $1/\Phi_6(x)$ were discussed long ago in \cite{NIELS},
%----------------------------------------------------------------------------------------------
\begin{eqnarray}
\frac{1}{2} \beta\left(\frac{x}{2}\right) &=& \int_0^1 dt \frac{t^{x-1}}{t^2+1} \\
            \beta\left(\frac{x}{3}\right) &=& \beta(x) - \int_0^1 dt~t^{x-1} \frac{t-2}{t^2-t+1}~,
\end{eqnarray}
%----------------------------------------------------------------------------------------------
with Stirling's $\beta$-function \cite{STIRLING}
%----------------------------------------------------------------------------------------------
\begin{eqnarray}
\beta(x) = \frac{1}{2} \left[\Psi\left(\frac{x+1}{2}\right) - \Psi\left(\frac{x}{2}\right)\right]~.
\end{eqnarray}
%----------------------------------------------------------------------------------------------
Integrals of this kind emerge also in particle physics problems in \cite{DB1,DK,STEINH,STGR}.

We form the Poincar\'{e} iterated integrals
%----------------------------------------------------------------------------------------------
\begin{align*}
C_{k_1,...,k_m}^{l_1,...,l_m}(z)  &= \frac{1}{m!}\ln(x)^m&&\text{if } 
(l_1,\dots,l_m)=(0,\dots,0),
(k_1,\dots,k_m)=(0,\dots,0),\\
C_{k_m}^{l_m}(z) &= \int_0^z~dx~f_{k_m}^{l_m}(x)&&\text{if } k_m\neq0,\\
C_{k_1,...,k_m}^{l_1,...,l_m}(z)  &= \int_0^z~dx~f_{k_1}^{l_1}(x)~C_{k_2,...,k_m}^{l_2,...,l_m}(x)&&\text{if } (k_1,\dots,k_m)\neq(0,\dots,0),
\end{align*}
%----------------------------------------------------------------------------------------------
and $C_{\vec{a}}^{\vec{l}}(z)$ denotes  {\sf cyclotomic harmonic polylogarithms}.
They form a shuffle algebra \cite{HOFF,REUT} by multiplication
%----------------------------------------------------------------------------------------------
\begin{eqnarray}
\label{eq:CSH}
C_{\vec{a_1}}^{\vec{a_2}}(z) \cdot C_{\vec{b_1}}^{\vec{b_2}}(z) 
= C_{\vec{a_1}}^{\vec{a_2}}(z) \SH C_{\vec{b_1}}^{\vec{b_2}}(z) 
= \sum_{\tiny
\left[\begin{array}{c}
\vec{c}_2 \\
\vec{c}_1
\end{array}\right]
\in
\left[
\begin{array}{c}
\vec{a}_2 \\
\vec{a}_1
\end{array}\right]
\SH \left[\begin{array}{c}
\vec{b}_2 \\
\vec{b}_1
\end{array}\right]}
C_{\vec{c}_1}^{\vec{c}_2}(z)
\end{eqnarray}
%----------------------------------------------------------------------------------------------
of $M^{\sf w}$ elements at
weight {\sf w}, where $M$ denotes the number of chosen letters from $\mathfrak{A}$. The shuffle symbol $\SH$ implies all
combinations of indices $a_{ij} \in \vec{a}_i$ and  $b_{ij} \in
\vec{b}_j$ leaving the order in both sets unchanged and the  brackets $[~]$ pair the the upper and lower indices 
forming a unity, cf.~(\ref{eq:f2}). 

The number of basis elements spanning the shuffle algebra are given by
%----------------------------------------------------------------------------------------------
\begin{eqnarray}
N^{\rm basic}({\sf w}) = \frac{1}{\sf w} \sum_{d|{\sf w}} \mu\left(\frac{\sf w}{d}\right) M^d, {\sf w} \geq 1
\end{eqnarray}
%----------------------------------------------------------------------------------------------
according to the 1st Witt formula \cite{WITT,REUT}. Here $\mu$ denotes the M\"obius function
\cite{MOEBIUS}. The number of basic cyclotomic harmonic polylogarithms in dependence of {\sf w} and $M$
is given in Table~1.

%----------------------------------------------------------------------------------------------
\begin{table}[h]\centering
\begin{tabular}{|r|r|r|r|r|r|r|r|}
\hline
\multicolumn{1}{|c}{          } &
\multicolumn{7}{|c|}{\sf letters } \\
\cline{2-8}
\multicolumn{1}{|c}{\sf weight} &
\multicolumn{1}{|c}{2} &
\multicolumn{1}{|c}{3} &
\multicolumn{1}{|c}{4} &
\multicolumn{1}{|c}{5} &
\multicolumn{1}{|c}{6} &
\multicolumn{1}{|c}{7} &
\multicolumn{1}{|c|}{8} \\
\hline
      1 &   2 &   3 &    4 &     5 &      6 &      7 &       8  \\
      2 &   1 &   3 &    6 &    10 &     15 &     21 &      28  \\
      3 &   2 &   8 &   20 &    40 &     70 &    112 &     168  \\
      4 &   3 &  18 &   60 &   150 &    315 &    588 &    1008  \\
      5 &   6 &  48 &  204 &   624 &   1554 &   3360 &    6552  \\
      6 &   9 & 116 &  670 &  2580 &   7735 &  19544 &   43596  \\
      7 &  18 & 312 & 2340 & 11160 &  39990 & 117648 &  299592  \\
      8 &  30 & 810 & 8160 & 48750 & 209790 & 729300 & 2096640  \\
\hline
\end{tabular}
\label{TAB1}
\caption[]{\sf Number of basic cyclotomic harmonic polylogarithms in dependence of the
           number of letters and weight.}
\end{table}
%----------------------------------------------------------------------------------------------

Now the cyclotomic harmonic sums (\ref{eq:MSU}) can be represented as Mellin transforms of
cyclotomic harmonic polylogarithms. In the example (\ref{eq:EX1}) the iterated integrals
have to be rewritten. We express respective denominators in terms of products of
cyclotomic polynomials, (\ref{eq:A1}-\ref{eq:A2}), and perform partial fractioning in
the respective integration variable,  (\ref{eq:A3}-\ref{eq:A4}). In our concrete example, the integrand of the
$y$-integral in (\ref{eq:ex1}) has the representation
%----------------------------------------------------------------------------------------------
\begin{eqnarray}
\frac{y^4}{y^6+1} &=& \frac{1}{3} \left[f_4^0(y) - f_{12}^0(y) + 2 f_{12}^2(y)
\right]~.
\end{eqnarray}
%----------------------------------------------------------------------------------------------
With integration  by parts one obtains the following Mellin transforms of argument $6N$
of cyclotomic harmonic
polylogarithms $C_{k_1,...,k_m}^{l_1,...,l_m}(x)$ weighted by the letters
$f^k_l(x)$ of the alphabet $\mathfrak{A}$~:
%----------------------------------------------------------------------------------------------
\begin{eqnarray}
\label{eq:ex2}
S_{\{3,2,2\},\{2,1,1\}}(1,-1;N) &=& \frac{1}{6}(4 - \pi) \int_0^1 dx x^3(x^{6N}-1)
\left[6
+ f_1^0(x)
- f_2^0(x)
- 2 f_3^0(x) \right. \nonumber\\ &&
\hspace*{2cm}
- f_3^1(x) \left.
- 2 f_6^0(x)
+ f_6^1(x)\right] C_0^0(x) \nonumber\\ &&
-2 \int_0^1 dx x^3 \left[(-1)^N x^{6N} - 1\right]\left[3 - f_4^0(x) - 2 f_{12}^0(x) + 2
f_{12}^2(x) \right] C_0^0(x) \nonumber
%\end{eqnarray}
%\begin{eqnarray}
\\ 
&&
-\frac{4}{3} \left[C_{0,4}^{0,0}(1)-C_{0,12}^{0,0}(1)+2C_{0,12}^{0,2}(1) \right]
\int_0^1 dx x^3 \left[(-1)^N x^{6N} - 1\right] \nonumber\\
&&
\hspace*{5cm} \times \left[3 - f_4^0(x) - 2 f_{12}^0(x) + 2
f_{12}^2(x) \right] \nonumber\\ &&
+\frac{4}{3}
\int_0^1 dx x^3 \left[(-1)^N x^{6N} - 1\right]
\left[C_{0,4}^{0,0}(x)-C_{0,12}^{0,0}(x)+2C_{0,12}^{0,2}(x) \right]
\nonumber\\
&&
\hspace*{5cm}
\times \left[3 - f_4^0(x) - 2 f_{12}^0(x) + 2
f_{12}^2(x) \right]~.
\nonumber\\
\end{eqnarray}
%----------------------------------------------------------------------------------------------
The constants $C_{k_1,...,k_m}^{l_1,...,l_m}(1)$ are discussed in Section~5. In
particular one obtains
%----------------------------------------------------------------------------------------------
\begin{eqnarray}
C_{0,4}^{0,0} = - {\bf C},
\end{eqnarray}
%----------------------------------------------------------------------------------------------
with {\bf C} the Catalan number \cite{CATALAN}, and $C_{0,12}^{0,0}$
and $C_{0,12}^{0,2}$ are linear combinations of $\psi'(1/12)$ and
$\psi'(5/12)$. The latter numbers reduce further, cf.~Section~5.
Note that in all cases in which neither $f_1^0(z)$ nor
$C_{1,k_2,...,k_m}^{0,l_2...,l_m}(z)$ are present, the $z$-independent terms in $[(\pm z^k)^N - 1]$
can be integrated, since the other cyclotomic letters $f_l^k(z)$ and $C_{k,k_2,...,k_m}^{l,l_2...,l_m}(z)$ for $k > 1$
are regular at $z=1$.

In general, linear combinations of the Mellin transforms
%---------------------------------------------------------------------------------------------
\begin{eqnarray}
\label{eq:MEL}
\Mvec\left[f_c^d(x) \cdot C_{\vec{a}}^{\vec{b}}(x)\right](l\,N)
&=& \int_0^1~dx~x^{lN}~(f_c^d(x))^u \cdot C_{\vec{a}}^{\vec{b}}(x),\quad u\in\{0,1\}.
\end{eqnarray}
%----------------------------------------------------------------------------------------------
with $l$ being the least common multiple of $a_1, ..., a_k$
allow to represent all cyclotomic harmonic sums.
Summarizing, we can write
%------------------------------------------------------------------------------------------------------------
\begin{eqnarray}\label{Equ:SAsH}
S_{\{a_1,b_1,c_1\},...,\{a_k,b_k,c_k\}}(N) = \sum_{n=1}^s e_n \int_0^1~dx~x^{l N}(f_{\alpha_n}^{\beta_n}(x))^{u_n}
C_{\vec{\gamma_n}}^{\vec{\delta_n}}(x)~,
\end{eqnarray}
%------------------------------------------------------------------------------------------------------------
with $e_n \in \mathbb{R}$ and $u_n\in\{0,1\}$; here $e_n$ is determined by polynomial expressions in terms of cyclotomic harmonic polylogarithms evaluated at $1$ with rational coefficients.
Within this transformation and in the following the letter $1/x$ plays a special role concerning the cyclotomic harmonic sums.
We can exclude the case that the first letter in (\ref{eq:MEL}) is $1/x$ 
since it would just shift the Mellin index by one unit.
In case of $\Phi_1(x)$ and related functions the $+$-regularization
%----------------------------------------------------------------------------------------------
\begin{eqnarray}
\label{eq:MEL3}
\Mvec\left[\left(\frac{f(x)}{x-1}\right)_+  \right](N) = \int_0^1~dx~\frac{x^N-1}{x-1}~f(x)
\end{eqnarray}
%----------------------------------------------------------------------------------------------
is applied to (\ref{eq:MEL}). Only in case that $f_1(x) = 1/(x-1)$ and $C_{\vec{a}}^{\vec{b}}(x)$ do not
vanish in the limit $x \rightarrow 1$ a $+$-function must occur.
Iterations of
cyclotomic letters $f^l_k(x),~~k \geq 1, 0 \leq l < k~~\text{for}~~k \geq 2$ have no singularities
in $x \in [0,1]$.

We remark that the sketched transformation from cyclotomic harmonic sums to their Mellin transforms in terms of cyclotomic harmonic polylogarithms can be reversed, i.e., a given expression in terms of Mellin transforms of cyclotomic harmonic polylogarithms can be expressed in terms of cyclotomic harmonic sums. The explicit algorithms have been worked out in details for the alphabet
%----------------------------------------------------------------------------------------------
\begin{equation}\label{eq:HALP}
\mathfrak{A}':=\left\{f_0^0,f_1^0,f_2^0,f_4^0,f_4^1\right\}
=\left\{\frac{1}{x}, \frac{1}{\Phi_1(x)},\frac{1}{\Phi_2(x)},
\frac{1}{\Phi_4(x)},\frac{x}{\Phi_4(x)}\right\}\subseteq\mathfrak{A},
\end{equation}
%----------------------------------------------------------------------------------------------
which allows one to express the cyclotomic harmonic sums $S_{\{a_1,b_1,c_1\}, ...,\{a_l,b_l,c_l\}}(s_1, ...,s_l; N)$ 
with $a_i\in\{1,2\}$, $b_i\in\{0,1\}$ and $c_i\in\mathbb N_+$ in terms of Mellin transforms of cyclotomic harmonic 
polylogarithms in both directions; the implementation is available within the {\sf HarmonicSums} package \cite{HASUM}. 
This transformation will be used, e.g., in Section~4.1.

The Mellin transforms (\ref{eq:MEL}, \ref{eq:MEL3}) obey difference equations of order $l$ in $N$,
which can be used to define these functions specifying respective initial values for $l$
moments. In the following we illustrate this for the words $x^l/\Phi_k(x)$. One obtains
%----------------------------------------------------------------------------------------------
\begin{eqnarray}
\label{eq:DEQ}
\sum_{n=0}^{N_k} c_{n,k} \phi_k(N+n-l) = \frac{1}{N+1}~,
\end{eqnarray}
%----------------------------------------------------------------------------------------------
where $\Phi_k(x)$ is given by
%----------------------------------------------------------------------------------------------
\begin{eqnarray}
\Phi_k(x) = \sum_{n=0}^{N_k} c_{n,k} x^n~.
\end{eqnarray}
%----------------------------------------------------------------------------------------------
Here we define
%----------------------------------------------------------------------------------------------
\begin{eqnarray}
\label{eq:mel1}
\phi_1(0,N) &=& \int_0^1 dx~\frac{x^N-1}{x-1} \\
\label{eq:mel2}
\phi_k(l,N) &=& \int_0^1 dx~x^N~f_k^l(x),~~~~~~~\phi_k(N)=
\phi_k(0,N),
~~~~k \geq 2 \\
\label{eq:mel3}
\phi_k(l,N)_+ &=& \int_0^1 dx~[x^N-1]~f_k^l(x),~~~~~~~\phi_k(N)_+=
\phi_k(0,N)_+.
\end{eqnarray}
%----------------------------------------------------------------------------------------------
For $ k < 105 $ for all coefficients $c_{n,k} \in \{-1,0,1\}$ holds \cite{CP105}. One derives the following
first order difference equations (\ref{eq:DEQ}) for Mellin transforms associated to the lowest order
cyclotomic polynomials~:
%----------------------------------------------------------------------------------------------
\begin{eqnarray}
\label{eq:rec1}
\phi_1(l,N+1) - \phi_1(l,N) &=& \frac{1}{N+l+1} \\
\phi_2(l,N+1) + \phi_2(l,N) &=& \frac{1}{N+l+1} \\
\phi_3(l,N+3) - \phi_3(l,N) &=& -\frac{1}{(N+l+1)(N+l+2)} \\
\phi_4(l,N+2) + \phi_4(l,N) &=& \frac{1}{N+l+1} \\
\phi_5(l,N+5) - \phi_5(l,N) &=& -\frac{1}{(N+l+1)(N+l+2)} \\
\phi_6(l,N+3) + \phi_6(l,N) &=& \frac{2(N+l)+3}{(N+l+1)(N+l+2)} %\\
\end{eqnarray}
\begin{eqnarray}
\phi_7(l,N+7) - \phi_7(l,N) &=& -\frac{1}{(N+l+1)(N+l+2)}\\
\phi_8(l,N+4) - \phi_8(l,N) &=& \frac{1}{N+l+1}\\
\phi_9(l,N+9) - \phi_9(l,N) &=& -\frac{3}{(N+l+1)(N+l+4)} \\
\phi_{10}(l,N+5) + \phi_{10}(l,N) &=& \frac{3(N+l)+2}{(N+l+1)(N+l+2)} \\
\phi_{11}(l,N+11) - \phi_{11}(l,N) &=& \frac{1}{(N+l+1)(N+l+2)} \\
\label{eq:rec2}
\phi_{12}(l,N+6) + \phi_{12}(l,N) &=& \frac{3(N+l)+2}{(N+l+1)(N+l+2)},~~{\rm etc.}
\end{eqnarray}
%----------------------------------------------------------------------------------------------
%Here we have only introduced those shifts by $l$, which will be needed in the %definition of the corresponding
%single sums up to $l = 6$ in (\ref{eq:NEWT}).
Together with the corresponding initial values, these recurrence relations enable one to compute efficiently the
values for $N$. In particular, due the special form of the recurrences, we get explicit representations in terms of
finite sums. E.g., for $\phi_6(l,N)$, we obtain
%----------------------------------------------------------------------------------------------
\begin{equation}\label{Equ:Phi6Def}
\phi_6(l,3N)=(-1)^N\left(\sum_{i=1}^N\frac{(-1)^i (2(3i+l)-3)}{(3(i+l)-2)(3(i+l)-1)}+\phi_6(l,0)\right),
\end{equation}
%----------------------------------------------------------------------------------------------
and
%----------------------------------------------------------------------------------------------
\begin{eqnarray}
\phi_6(l,3N+1) &=& \phi_6(l+1,3N) \\
\phi_6(l,3N+2  &=& \phi_6(l+2,3N)~.
\end{eqnarray}
%----------------------------------------------------------------------------------------------
Looking at
%----------------------------------------------------------------------------------------------
$$\lim_{N\to\infty}\phi_6(l,N)=\lim_{N\to\infty}\int_{0}^1x^{N+l}\frac{1}{x^2-x-1}dx=
\lim_{N\to\infty}\int_{0}^1
x^{N+l}(1+x-x^3-x^4+\dots)dx=0$$
%----------------------------------------------------------------------------------------------
shows that
%----------------------------------------------------------------------------------------------
$$\phi_6(l,0)=-\sum_{i=1}^{\infty}\frac{(-1)^i [2(3i+l)-3]}{(3i+l-2)(3i+l-1)}.$$
%----------------------------------------------------------------------------------------------
Completely analogously, all the other functions~(\ref{eq:rec1}--\ref{eq:rec2}) can be
written in such a sum representation where the constants are the infinite versions of it multiplied with a minus sign. Note that these constants can be written
as a linear combination of the infinite sums
%----------------------------------------------------------------------------------------------
$$\sigma_{\{a,b,s\}}=\sum_{k=1}^{\infty}\frac{s^k}{a\,k+b}$$
%----------------------------------------------------------------------------------------------
with $s\in\{-1,1\}$ and $a,b\in\mathbb N$ with $a\neq0$.
The relevant values for this article will be worked out explicitly in the Section~5.

The $\phi_k(l,N)_+$ functions are preferred if the
recurrences~(\ref{eq:rec1}--\ref{eq:rec2}) form telescoping equations. In this case, in
particular if $k$ is odd, the $\phi_k(l,r\,N)$ for properly chosen $r$ can be related to sums without any extra constant, e.g.,
%----------------------------------------------------------------------------------------------
\begin{equation}\label{Equ:Phi5Def}
\phi_5(l,5N)_+ = - \sum_{i=1}^N \frac{1}{(5 i+l-4) (5 i+l-3)}.
\end{equation}
%----------------------------------------------------------------------------------------------
$\phi_1$ and $\phi_2$ can be related to the single cyclotomic harmonic sums at
weight {\sf w = 1} as follows
%----------------------------------------------------------------------------------------------
\begin{eqnarray}
\phi_1(l,N)_+  &=&  S_1(N+l),\\
\phi_2(l,N) &=&
%(-1)^N\Big(S_{-1}(N+l)-\sum_{i=1}^{\infty}\frac{(-1)^i}{i+l}\Big)=
(-1)^N \Big[S_{-1}(N+l)-\ln(2)+S_{-1}(l)\Big]~.
\end{eqnarray}
%----------------------------------------------------------------------------------------------
For later considerations we use $S_1(N)$ and $S_{-1}(N)$ instead of $\phi_1(N)$ and $\phi_2(N)$.
%----------------------------------------------------------------------------------------------
In the following Section
the single cyclotomic harmonic sums of weight {\sf w = 1} are expressed in terms of the
Mellin transforms $\phi_k(l,r\,N)$ for properly chosen $r$. In addition, the
cyclotomic harmonic sums of higher weight and depth will be discussed.
%%%%%%%%%%%%%%%%%%%%%%%%%%%%%%%%%%%%%%%%%%%%%%%%%%%%%%%%%%%%%%%%%%%%%%%%
\section{Cyclotomic Harmonic Sums}
\label{sec:4}
%%%%%%%%%%%%%%%%%%%%%%%%%%%%%%%%%%%%%%%%%%%%%%%%%%%%%%%%%%%%%%%%%%%%%%%%

\vspace{1mm}
\noindent
We consider the extension of the finite nested harmonic sums \cite{HS1,HS2}
to those generated by the cyclotomic harmonic polylogarithms discussed in Section~3.
First the single cyclotomic harmonic sums are considered and explicit representations are
given. We derive their analytic continuation to complex values of $N$. Next the
algebraic, differential and three multiple argument relations of the cyclotomic harmonic
sums are discussed. These relations are used to represent these sums over suitable bases.
Finally we consider the nested sums over the alphabet $\{(\pm 1)^k/k,
(\pm 1)^k/(2k+1)\}$ to higher weight deriving explicit relations up to {\sf w = 5}.
%%%%%%%%%%%%%%%%%%%%%%%%%%%%%%%%%%%%%%%%%%%%%%%%%%%%%%%%%%%%%%%%%%%%%%%%
\subsection{The Single Sums}
\label{sec:4.1}
%%%%%%%%%%%%%%%%%%%%%%%%%%%%%%%%%%%%%%%%%%%%%%%%%%%%%%%%%%%%%%%%%%%%%%%%

\vspace{1mm}
\noindent
The single cyclotomic harmonic sums are given by
%------------------------------------------------------------------------------------------------------------
\begin{eqnarray}
\sum_{k=0}^{N} \frac{(\pm 1)^k}{(l \cdot k + m)^n}~.
\end{eqnarray}
%------------------------------------------------------------------------------------------------------------
Here $N$ is either an even or an odd integer. In case one needs representations only
for $N \in \mathbb{N}$ the following representations hold~:
%------------------------------------------------------------------------------------------------------------
\begin{eqnarray}
\sum_{k=0}^{\overline{N}} \frac{(-1)^k}{l \cdot k + m} &=& \left[\sum_{k=0}^N \frac{1}{(2l) \cdot k +m} -
\frac{1}{(2l)
\cdot k + m+l}\right], \\
\sum_{k=0}^{\overline{N}+1} \frac{(-1)^k}{l \cdot k + m} &=& \left[\sum_{k=0}^N \frac{1}{(2l) \cdot k +m} -
\frac{1}{(2l)
\cdot k + m+l}\right] - \frac{1}{(2l) N + l + m}~,
\end{eqnarray}
%------------------------------------------------------------------------------------------------------------
with $\overline{N} = 2N$. However, one is interested in relations for general values of $N$, since for
nested sums more and more cases have to be distinguished. The single sums can be expressed in terms of the
Mellin transforms $\phi_k(l,kN)$, (\ref{eq:mel1}--\ref{eq:mel2}).
Up to {\sf w = 6} one obtains~:
%\allowdisplaybreaks[3]

%------------------------------------------------------------------------------------------------------------
\begin{align}
\label{eq:SUa}
\sum_{k=1}^N \frac{1}{1+2 k}=&-\frac{2 N}{2 N+1}-\frac{S_1(N)}{2}+S_1(2 N),\\
%--
\sum_{k=1}^N \frac{(-1)^k}{1+2 k}=&(-1)^N \left[\frac{1}{2 N+1}-\phi_{4}(2
N)\right]+\sigma_{\{2,1,-1\}},\\
%--
\sum_{k=1}^N \frac{1}{1+3 k}=&-\frac{3 N}{3 N+1}-\frac{S_1(N)}{6}+\frac{1}{2} S_1(3 N)-\frac{1}{2} \phi_{3}(3 N)_+,%\\
\end{align}
\begin{align}
%--
\sum_{k=1}^N \frac{(-1)^k}{1+3 k}=&\frac{1}{6} S_{-1}(N)-\frac{1}{2} S_{-1}(3 N)+(-1)^N
\left[\frac{1}{3 N+1}-\frac{1}{2} \phi_{6}(3 N)\right]+\frac{1}{3}
\sigma_{\{1,0,-1\}}+\sigma_{\{3,1,-1\}},\\
%--
\sum_{k=1}^N \frac{1}{2+3 k}=&-\frac{3 N}{2 (3 N+2)}-\frac{S_1(N)}{6}+\frac{1}{2} S_1(3 N)+\frac{1}{2} \phi_{3}(3 N)_+,\\
%--
\sum_{k=1}^N \frac{(-1)^k}{2+3 k}=&-\frac{1}{6} S_{-1}(N)+\frac{1}{2} S_{-1}(3 N)+(-1)^N
\left[\frac{1}{3 N+2}-\frac{1}{2} \phi_{6}(3 N)\right]
\nonumber\\ &
+\frac{1}{3}
\sigma_{\{1,0,-1\}}+\sigma_{\{3,1,-1\}}+\frac{1}{2},\\
%--
\sum_{k=1}^N \frac{1}{1+4 k}=&-\frac{2 N}{4 N+1}-\frac{1}{4} S_1(2 N)+\frac{1}{2} S_1(4 N)-\frac{1}{2} \phi_{4}(4 N)+\frac{1}{2} \sigma_{\{2,1,-1\}}+\frac{1}{2 (4 N+1)},\\
%--
\sum_{k=1}^N \frac{(-1)^k}{1+4 k}=&(-1)^N \left[\frac{1}{4 N+1}-\phi_{8}(4
N)\right]+\sigma_{\{4,1,-1\}},\\
%--
\sum_{k=1}^N \frac{1}{3+4 k}=&-\frac{10 N}{3 (4 N+3)}-\frac{1}{4} S_1(2 N)+\frac{1}{2} S_1(4 N)+\frac{1}{2} \phi_{4}(4 N)-\frac{1}{2} \sigma_{\{2,1,-1\}}-\frac{3}{2 (4 N+3)},\\
%--
\sum_{k=1}^N \frac{(-1)^k}{3+4 k}=&(-1)^N \left[\frac{1}{4 N+3}-\phi_{8}(2,4
N)\right]+\sigma_{\{4,3,-1\}},\\
%--
\sum_{k=1}^N \frac{1}{1+5 k}=&-\frac{5 N}{5 N+1}-\frac{S_1(N)}{20}+\frac{1}{4} S_1(5 N)-\frac{3}{4} \phi_{5}(5 N)_+-\frac{1}{2} \phi_{5}(1,5 N)_+-\frac{1}{4} \phi_{5}(2,5 N)_+,\\
%--
\sum_{k=1}^N \frac{(-1)^k}{1+5 k}=&(-1)^N \left[-\frac{3}{4} \phi_{10}(5 N)+\frac{1}{2}
\phi_{10}(1,5 N)-\frac{1}{4} \phi_{10}(2,5 N)+\frac{1}{5 N+1}\right]\nonumber\\
&+\frac{1}{20} S_{-1}(N)-\frac{1}{4} S_{-1}(5 N)+\frac{1}{5} \sigma_{\{1,0,-1\}}+\sigma_{\{5,1,-1\}},\\
%--
\sum_{k=1}^N \frac{1}{2+5 k}=&-\frac{5 N}{2 (5 N+2)}-\frac{S_1(N)}{20}+\frac{1}{4} S_1(5 N)+\frac{1}{4} \phi_{5}(5 N)_+-\frac{1}{2} \phi_{5}(1,5 N)_+-\frac{1}{4} \phi_{5}(2,5 N)_+,\\
%--
\sum_{k=1}^N \frac{(-1)^k}{2+5 k}=&(-1)^N \left[-\frac{1}{4} \phi_{10}(5 N)-\frac{1}{2}
\phi_{10}(1,5 N)+\frac{1}{4} \phi_{10}(2,5 N)+\frac{1}{5 N+2}\right]\\
%--
&-\frac{1}{20} S_{-1}(N)+\frac{1}{4} S_{-1}(5 N)-\frac{1}{5} \sigma_{\{1,0,-1\}}+\sigma_{\{5,2,-1\}},\nonumber\\
\sum_{k=1}^N \frac{1}{3+5 k}=&-\frac{5 N}{3 (5 N+3)}-\frac{S_1(N)}{20}+\frac{1}{4} S_1(5 N)+\frac{1}{4} \phi_{5}(5 N)_++\frac{1}{2} \phi_{5}(1,5 N)_+-\frac{1}{4} \phi_{5}(2,5 N)_+,\\
%--
\sum_{k=1}^N \frac{(-1)^k}{3+5 k}=&(-1)^N \left[\frac{1}{4} \phi_{10}(5 N)-\frac{1}{2}
\phi_{10}(1,5 N)-\frac{1}{4} \phi_{10}(2,5 N)+\frac{1}{5 N+3}\right]\nonumber\\
&+\frac{1}{20} S_{-1}(N)-\frac{1}{4} S_{-1}(5 N)+\frac{1}{5} \sigma_{\{1,0,-1\}}+\sigma_{\{5,3,-1\}},%\\
%--
\end{align}
\begin{align}
\sum_{k=1}^N \frac{1}{4+5 k}=&-\frac{5 N}{4 (5 N+4)}-\frac{S_1(N)}{20}
+\frac{1}{4} S_1(5 N)+\frac{1}{4} \phi_{5}(5 N)_+ \nonumber\\ &
+\frac{1}{2} \phi_{5}(1,5 N)_++\frac{3}{4} \phi_{5}(2,5 N)_+,\\
%--
\sum_{k=1}^N \frac{(-1)^k}{4+5 k}=&(-1)^N \left[-\frac{1}{4} \phi_{10}(5 N)+\frac{1}{2}
\phi_{10}(1,5 N)-\frac{3}{4} \phi_{10}(2,5 N)+\frac{1}{5 N+4}\right]\nonumber\\
&-\frac{1}{20} S_{-1}(N)+\frac{1}{4} S_{-1}(5 N)+\frac{3}{5} \sigma_{\{1,0,-1\}}+\sigma_{\{5,1,-1\}}-\sigma_{\{5,2,-1\}}+\sigma_{\{5,3,-1\}}+\frac{7}{12},\\ \sum_{k=1}^N \frac{1}{1+6 k}=&-\frac{6 N}{6 N+1}+\frac{S_1(N)}{12}-\frac{1}{6} S_1(2 N)-\frac{1}{4} S_1(3 N)+\frac{1}{2} S_1(6 N)-\frac{1}{4} \phi_{3}(3 N)_+-\frac{1}{2} \phi_{3}(6 N)_+,\\
%--
\sum_{k=1}^N \frac{(-1)^k}{1+6 k}=&(-1)^N \left[-\phi_{12}(6 N)+\frac{1}{3} \phi_{4}(2
N)+\frac{1}{6 N+1}\right]+\sigma_{\{6,1,-1\}},\\
%--
\sum_{k=1}^N \frac{1}{5+6 k}=&-\frac{6 N}{5 (6 N+5)}+\frac{S_1(N)}{12}
-\frac{1}{6} S_1(2 N)-\frac{1}{4} S_1(3 N)+\frac{1}{2} S_1(6 N)
\nonumber\\ &
+\frac{1}{4} \phi_{3}(3 N)_++\frac{1}{2} \phi_{3}(6 N)_+,\\
%--
\label{eq:SUb}
\sum_{k=1}^N \frac{(-1)^k}{5+6 k}=&(-1)^N \left[\phi_{12}(6 N)-\frac{2}{3} \phi_{4}(2
N)-\phi_{4}(6 N)+\frac{1}{6 N+5}\right]+\frac{4}{3}
\sigma_{\{2,1,-1\}}-\sigma_{\{6,1,-1\}}+\frac{2}{15}.
\end{align}
%------------------------------------------------------------------------------------------------------------
Taking the sum representations such as~\eqref{Equ:Phi5Def}
and~(\ref{Equ:Phi6Def})
given by the recurrence relations~(\ref{eq:rec1}--\ref{eq:rec2})
and the corresponding initial values $\phi_k(l,0)$
(which are expressible by the infinite sums $\sigma_{\{a,b,\pm1\}}$, cf. Section~5), we
used the
summation package {\sf Sigma} \cite{SIGMA} to perform this transformation. As a
consequence, the single cyclotomic
harmonic sums can be expressed in terms of the sums
%------------------------------------------------------------------------------------------------------------
\begin{align*}
&S_{-1}(N),S_{-1}(3 N),S_{-1}(5 N),S_1(N),S_1(2 N),S_1(3 N),S_1(4 N),S_1(5 N),S_1(6 N),\\
&\phi_{3}(3 N)_+,\phi_{3}(6 N)_+,\phi_{4}(2 N),\phi_{4}(4 N),\phi_{4}(6 N),\phi_{5}(5 N)_+,\phi_{5}(1,5 N)_+,\\
&\phi_{5}(2,5 N)_+,\phi_{6}(3 N),\phi_{8}(4 N),\phi_{8}(2,4 N),\phi_{10}(5 N),\phi_{10}(1,5 N),\phi_{10}(2,5
N),\phi_{12}(6 N).
\end{align*}
%------------------------------------------------------------------------------------------------------------
In particular, in the way how this construction is carried out it follows by the summation theory of~\cite{SIGMA} that
the sequences produced by these sums
form an algebraic independent basis over the ring of sequences generated by
the elements from $\mathbb R(N)[(-1)^N]$. Note that we exploited the algebraic
relations~(\ref{Equ:Sigma1}--\ref{Equ:Sigma2}) which implies that only the following
constants
%------------------------------------------------------------------------------------------------------------
$$\sigma_{\{1,0,-1\}},\sigma_{\{2,1,-1\}},\sigma_{\{3,1,-1\}},\sigma_{\{4,1,-1\}},
\sigma_{\{4,3,-1\}},\sigma_{\{5,1,-1\}},\sigma_{\{5,2,-1\}},\sigma_{\{5,3,-1\}}
\sigma_{\{6,1,-1\}}$$
%------------------------------------------------------------------------------------------------------------
from $\mathbb R$ appear in the representation found for the single cyclotomic harmonic sums with weight {\sf w = 1}.

Due to (\ref{eq:mel2}, \ref{eq:mel3}) the functions $\Phi_k(l,N), k
\geq 1$ can be represented as factorial series~\cite{NIELS,LANDAU}. Therefore they are
meromorphic functions in $N$ with poles at $-n$, $n \in \mathbb{N}$. This also applies
to $\phi_k(0,N)$, (\ref{eq:mel1}). The latter function grows $\propto \ln(N)$ for $N
\rightarrow \infty,~|{\rm arg}(N)| < \pi$. The recursion relations (\ref{eq:DEQ}) allow one to
shift $\phi_k(l,N)$ in $N \rightarrow N+1$. To represent a function $\phi_k(l,N)$ for
$N \in \mathbb{C}$ one needs to know its asymptotic representation in addition.

It is given in analytic form in terms of series involving the Stirling numbers of the 2nd
kind \cite{NIELS,STIRLING}. The corresponding representations read~:
%---------------------------------------------------------------------------------
\begin{eqnarray}
%--
 \phi_1(0,N) &\sim& \gamma +\ln(N)+\frac{1}{2 N}-\frac{1}{12
   N^2}+\frac{1}{120 N^4}-\frac{1}{252 N^6}+\frac{1}{240 N^8}\NN\\&&-\frac{1}{132
   N^{10}}+\frac{691}{32760 N^{12}}+O\left(\frac{1}{N^{13}}\right) \\
%--
 \phi_2(0,N) &\sim& \frac{1}{2 N}-\frac{1}{4 N^2}+\frac{1}{8 N^4}-\frac{1}{4
   N^6}+\frac{17}{16 N^8}-\frac{31}{4 N^{10}}+\frac{691}{8
   N^{12}}+O\left(\frac{1}{N^{13}}\right) \\
%--
 \phi_3(0,N) &\sim& \frac{1}{3 N}-\frac{2}{9 N^3}+\frac{2}{3 N^5}-\frac{14}{3
   N^7}+\frac{1618}{27 N^9}-\frac{3694}{3
   N^{11}}+O\left(\frac{1}{N^{13}}\right) \\
%--
 \phi_4(0,N) &\sim& \frac{1}{2 N}-\frac{1}{2 N^3}+\frac{5}{2 N^5}-\frac{61}{2
   N^7}+\frac{1385}{2 N^9}-\frac{50521}{2
   N^{11}}+O\left(\frac{1}{N^{13}}\right) %\\
\end{eqnarray}
\begin{eqnarray}
%--
 \phi_5(0,N) &\sim& \frac{1}{5 N}+\frac{1}{5 N^2}-\frac{1}{5
   N^3}-\frac{1}{N^4}+\frac{31}{25 N^5}+\frac{67}{5 N^6}-\frac{109}{5
   N^7}-\frac{361}{N^8}+\frac{3779}{5 N^9}\NN\\&&+\frac{412751}{25 N^{10}}-\frac{214093}{5
   N^{11}}-\frac{1150921}{N^{12}}+O\left(\frac{1}{N^{13}}\right) \\
%--
 \phi_6(0,N) &\sim&
   \frac{1}{N}-\frac{2}{N^3}+\frac{22}{N^5}-\frac{602}{N^7}+\frac{30742}{N^9}-\frac{25
   23002}{N^{11}}+O\left(\frac{1}{N^{13}}\right) \\
%--
 \phi_7(0,N) &\sim& \frac{1}{7 N}+\frac{2}{7 N^2}-\frac{16}{7 N^4}-\frac{12}{7
   N^5}+\frac{56}{N^6}+\frac{3900}{49 N^7}-\frac{20296}{7
   N^8}-\frac{5796}{N^9}\NN\\&&+\frac{1809992}{7 N^{10}}+\frac{4582500}{7
   N^{11}}-\frac{35282968}{N^{12}}+O\left(\frac{1}{N^{13}}\right) \\
%--
 \phi_8(0,N) &\sim& \frac{1}{2 N}+\frac{1}{2 N^2}-\frac{3}{2 N^3}-\frac{11}{2
   N^4}+\frac{57}{2 N^5}+\frac{361}{2 N^6}-\frac{2763}{2 N^7}-\frac{24611}{2
   N^8}+\frac{250737}{2 N^9}\NN\\&&+\frac{2873041}{2 N^{10}}-\frac{36581523}{2
   N^{11}}-\frac{512343611}{2 N^{12}}+O\left(\frac{1}{N^{13}}\right) \\
%--
 \phi_9(0,N) &\sim& \frac{1}{3 N}+\frac{2}{3 N^2}-\frac{2}{3 N^3}-\frac{28}{3
   N^4}+\frac{34}{3 N^5}+\frac{1172}{3 N^6}-\frac{1862}{3 N^7}-\frac{101428}{3
   N^8}+\frac{207394}{3 N^9}\NN\\&&+\frac{14999012}{3 N^{10}}-\frac{37996022}{3
   N^{11}}-\frac{3386034628}{3 N^{12}}+O\left(\frac{1}{N^{13}}\right) \\
%--
 \phi_{10}(0,N) &\sim&
   \frac{1}{N}+\frac{1}{N^2}-\frac{5}{N^3}-\frac{17}{N^4}+\frac{151}{N^5}
   +\frac{871}{N^6}-\frac{11465}{N^7}-\frac{92777}{N^8}+\frac{1626151}{N^9}\NN\\&&+\frac{1692
   2791}{N^{10}}-\frac{370714025}{N^{11}}-\frac{4715323337}{N^{12}}+O\left(\frac{1}{N^{13}}\right) \\
%--
 \phi_{11}(0,N) &\sim& \frac{1}{11 N}+\frac{4}{11 N^2}+\frac{6}{11
   N^3}-\frac{56}{11 N^4}-\frac{282}{11 N^5}+\frac{3064}{11 N^6}+\frac{26646}{11
   N^7}-\frac{382616}{11 N^8}\NN\\&&-\frac{4592442}{11
   N^9}+\frac{7618184}{N^{10}}+\frac{13945859346}{121 N^{11}}-\frac{28200213176}{11
   N^{12}}+O\left(\frac{1}{N^{13}}\right) \\
%--
 \phi_{12}(0,N) &\sim&
   \frac{1}{N}+\frac{1}{N^2}-\frac{7}{N^3}-\frac{23}{N^4}+\frac{305}{N^5}
   +\frac{1681}{N^6}-\frac{33367}{N^7}-\frac{257543}{N^8}+\frac{6815585}{N^9}\NN\\&&+\frac{67
   637281}{N^{10}}-\frac{2237423527}{N^{11}}-\frac{27138236663}{N^{12}}+O\left(\frac{1}{N^{13}}\right)%\\
%--
\end{eqnarray}
\begin{eqnarray}
 \phi_{5}(2,N) &\sim&\frac{1}{5 N}-\frac{2}{5 N^3}+\frac{86}{25 N^5}
   -\frac{338}{5 N^7}+\frac{12094}{5 N^9}-\frac{690866}{5
N^{11}}+O\left(\frac{1}{N^{13}}\right)\\
%--
 \phi_{8}(2,N) &\sim&\frac{1}{2 N}-\frac{2}{N^3}+\frac{40}{N^5}-\frac{1952}{N^7}
   +\frac{177280}{N^9}-\frac{25866752}{N^{11}}+O\left(\frac{1}{N^{13}}\right)\\
%--
 \phi_{10}(1,N)
&\sim&\frac{1}{N}+\frac{1}{N^2}-\frac{5}{N^3}-\frac{17}{N^4}+\frac{151}{N^5}
+\frac{871}{N^6}-\frac{11465}{N^7}-\frac{92777}{N^8}+\frac{1626151}{N^9}\NN\\
&&+\frac{16922791}{N^{10}}
   -\frac{370714025}{N^{11}}-\frac{4715323337}{N^{12}}+O\left(\frac{1}{N^{13}}\right)\\
%--
 \phi_{10}(2,N) &\sim& \frac{1}{N}-\frac{6}{N^3}+\frac{186}{N^5}-\frac{14166}{N^7}
   +\frac{2009946}{N^9}-\frac{458225526}{N^{11}}+O\left(\frac{1}{N^{13}}\right)~.
\end{eqnarray}
%---------------------------------------------------------------------------------
The numerical accuracy of the asymptotic representations at a given suitably
large value of $N$ lowers with growing $k$, i.e., one has to choose larger values
of $N$ correspondingly to apply the asymptotic formulae. The above expansions can
easily be extended to higher inverse powers of $N$.
The recursion for $\phi_k(l,kN)$ is given in (\ref{eq:DEQ}),
resp.~~(\ref{eq:rec1}--\ref{eq:rec2})
in a more compact form. Due to these for any $N \in \mathbb{C}$ at which $\phi_k(l,N)$ is analytic
one may map $\phi_k(l,N)$ to values $|N| \gg 1, \text{arg}(N) < \pi$ and use the asymptotic representations.

The single cyclotomic harmonic sums of higher weight obey the representations
%---------------------------------------------------------------------------------
\begin{eqnarray}
\label{eq:suhw1}
\sum_{k=0}^{N-1} \frac{1}{(lk+m)^n}
&=& \frac{(-1)^{n-1}}{(n-1)!} \int_0^1~dx~ \ln^{n-1}(x)~x^{m-1}
\frac{x^{lN}-1}{x^l-1},~~~m < l
\\
%---
\label{eq:suhw2}
\sum_{k=0}^{N-1} \frac{(-1)^k}{(lk+m)^n} &=& \frac{(-1)^{n}}{(n-1)!} \int_0^1~dx~
\ln^{n-1}(x)~x^{m-1} \frac{(-x^l)^N - 1}{x^l+1},~~~m < l,
\end{eqnarray}
%---------------------------------------------------------------------------------
for $l, m, n \in \mathbb{N}_+$, which may be expressed in terms of
cyclotomic letters again. We note that
%---------------------------------------------------------------------------------
\begin{eqnarray}
\frac{\partial^n}{\partial m^n}
\int_0^1~dx~x^{m-1}~f(x)
= \int_0^1~dx~x^{m-1}~~\ln^n(x)~f(x)~.
\end{eqnarray}
%---------------------------------------------------------------------------------
Therefore, (\ref{eq:suhw1}, \ref{eq:suhw2}) can be expressed by the corresponding
derivatives of $\phi_k(n,lN)$ for $N$ and corresponding constants.

%%%%%%%%%%%%%%%%%%%%%%%%%%%%%%%%%%%%%%%%%%%%%%%%%%%%%%%%%%%%%%%%%%%%%%%%
\subsection{\boldmath Cyclotomic harmonic polylogarithms at $x =1$}
%%%%%%%%%%%%%%%%%%%%%%%%%%%%%%%%%%%%%%%%%%%%%%%%%%%%%%%%%%%%%%%%%%%%%%%%

\vspace*{1mm}
\noindent
We consider the cyclotomic harmonic polylogarithm $C_{a,\vec{b}}^{c, \vec{d}}(x)$.
Its value at $x=1$ is given by
%---------------------------------------------------------------------------------
\begin{eqnarray}
\label{eq:Hval2}
C_{a,\vec{b}}^{c, \vec{d}}(1) = \int_0^1 dx f^c_a(x) C_{\vec{b}}^{\vec{d}}(x)~.
\end{eqnarray}
%---------------------------------------------------------------------------------
Let $f^c_a(x) = x^l/\Phi_k(x)$, with $l < \text{deg}(\Phi_k(x))$ and $n$ be the smallest
integer such that $\Phi_k(x) | (x^n-1$), with
%---------------------------------------------------------------------------------
\begin{eqnarray}
\frac{1}{\Phi_k(x)} = \frac{\prod_j \Phi_j(x)}{x^n-1}~.
\end{eqnarray}
%---------------------------------------------------------------------------------
Since $x \in [0,1]$, the representation 
%---------------------------------------------------------------------------------
\begin{eqnarray}
\frac{x^l}{\Phi_k(x)} = - x^l \prod_k \Phi_k(x) \sum_{j=0}^\infty x^{jn}
= \sum_{m=0}^w a_m(l) x^m \sum_{j=0}^\infty x^{jn},~~ a_m(l) \in \mathbb{Z}
\end{eqnarray}
%---------------------------------------------------------------------------------
holds. Thus we get
%---------------------------------------------------------------------------------
\begin{eqnarray}
\label{eq:Hval1}
\int_0^1 dx \frac{x^l}{\Phi_k(x)} = 
\sum_{m=0}^w a_m(l) \sum_{j=0}^\infty \frac{1}{jn + m + 1}~,
\end{eqnarray}
%---------------------------------------------------------------------------------
representing the value of the depth {\sf d=1} cyclotomic harmonic polylogarithms at $x=1$
as a linear combination of the corresponding infinite cyclotomic sums, see Section~\ref{sec:5}.

We return now to Eq.~(\ref{eq:Hval2}). Integration by parts yields
%---------------------------------------------------------------------------------
\begin{eqnarray}
\label{eq:HP2}
C_{a,\vec{b}}^{c, \vec{d}}(1) = 
C^c_a(1) C_{\vec{b}}^{\vec{d}}(1)
-\sum_{j=0}^\infty \sum_{m=0}^w a_m 
\int_0^1 dx \frac{x^{jn+m+1}}{jn+m+1} C_{\vec{b}}^{\vec{d}}(x)~.
\end{eqnarray}
%---------------------------------------------------------------------------------
Since the Mellin transform of a cyclotomic harmonic polylogarithm can be represented by a linear combination of
finite cyclotomic harmonic sums, the weighted infinite sum of the latter ones in (\ref{eq:HP2}) is thus given 
as a polynomial of infinite cyclotomic harmonic sums, Section~5.
%%%%%%%%%%%%%%%%%%%%%%%%%%%%%%%%%%%%%%%%%%%%%%%%%%%%%%%%%%%%%%%%%%%%%%%%
\subsection{Relations of Cyclotomic Harmonic Sums}
\label{sec:4.2}
%%%%%%%%%%%%%%%%%%%%%%%%%%%%%%%%%%%%%%%%%%%%%%%%%%%%%%%%%%%%%%%%%%%%%%%%

\vspace*{1mm}
\noindent
The cyclotomic harmonic sums obey differentiation relations, cf.~\cite{HS1,JBSH},
stuffle relations due to their quasi--shuffle algebra, cf. \cite{HOFF}, and multiple
argument relations.~\footnote{For the relations given in this Section we mostly
present the results, giving for a few cases the proofs in Appendix~B. The other proofs
proceed in a similar manner.}

%%%%%%%%%%%%%%%%%%%%%%%%%%%%%%%%%%%%%%%%%%%%%%%%%%%%%%%%%%%%%%%%%%%%%%%%
\subsubsection{Differentiation}
%%%%%%%%%%%%%%%%%%%%%%%%%%%%%%%%%%%%%%%%%%%%%%%%%%%%%%%%%%%%%%%%%%%%%%%%

\vspace*{1mm}
\noindent
As has been illustrated in Section~3, cyclotomic harmonic sums can be represented as linear
combinations of Mellin transforms~\eqref{Equ:SAsH} of cyclotomic harmonic polylogarithms.
Based on this representation, the differentiation of these sums is defined by
%------------------------------------------------------------------------------------------------------------
\begin{eqnarray}
\label{eq:DIF1}
\frac{\partial^m}{\partial N^m}
S_{\{a_1,b_1,c_1\},...,\{a_k,b_k,c_k\}}(N) = \sum_{n=1}^s e_n \int_0^1~dx~x^{l N}
~l^m~
\ln^m(x) (f_{\alpha_n}^{\beta_n}(x))^{u_n}
C_{\vec{\gamma_n}}^{\vec{\delta_n}}(x).
\end{eqnarray}
%------------------------------------------------------------------------------------------------------------
The product $\ln^m(x) C_{\vec{\alpha_n}}^{\vec{\beta_n}}(x)=m!C_{0,\dots,0}^{0,\dots,0} C_{\vec{\alpha_n}}^{\vec{\beta_n}}(x)$ with $m$ consecutive zeros may be transformed into a linear
combination of cyclotomic harmonic polylogarithms using the shuffle
relation (\ref{eq:CSH}). Finally, using the inverse Mellin transform, the derivative (\ref{eq:DIF1}) of a cyclotomic
harmonic sum w.r.t.\ $N$ is given as a polynomial expression in terms of cyclotomic harmonic sums and cyclotomic harmonic polylogarithms at $x=1$. Together with the previous section, the derivative (\ref{eq:DIF1}) can be expressed as a polynomial expression with rational coefficients in terms of cyclotomic harmonic sums and their values at
$N \rightarrow \infty$. The corresponding relations are denoted by $(D)$. Further details on cyclotomic harmonic sums 
at infinity are given in~Section~5.

A given finite cyclotomic harmonic sum is determined for $N \in \mathbb{C}$ by its asymptotic 
representation and 
the corresponding recursion from $N \rightarrow (N-1)$. Both the asymptotic representation and the recursion
can be easily differentiated analytically. Therfore any differentiation of a cyclotomic harmonic sum w.r.t. $N$
is closely related to the original sum. For this reason one may collect these derivatives in classes 
%------------------------------------------------------------------------------------------------------------
\begin{eqnarray}
S_{\{a_1,b_1,c_1\},...,\{a_k,b_k,c_k\}}^{(D)}(N)
= \left\{\frac{\partial^n}{\partial N^n}S_{\{a_1,b_1,c_1\},...,\{a_k,b_k,c_k\}}(N);
n \in \mathbb{N}\right\}.
\end{eqnarray}
%------------------------------------------------------------------------------------------------------------
%%%%%%%%%%%%%%%%%%%%%%%%%%%%%%%%%%%%%%%%%%%%%%%%%%%%%%%%%%%%%%%%%%%%%%%%
\subsubsection{Stuffle Algebra}
%%%%%%%%%%%%%%%%%%%%%%%%%%%%%%%%%%%%%%%%%%%%%%%%%%%%%%%%%%%%%%%%%%%%%%%%

\vspace*{1mm}
\noindent
To derive the stuffle relations, cf. \cite{JBSH,MZV}, we consider the product of two
denominator terms. For $a_1,a_2,b_1,b_2,c_1,c_2,i \in \N$
they are given by
%------------------------------------------------------------------------------------------------------------
\begin{eqnarray}
\frac{1}{(a_1 i +b_1)^{c_1}(a_2 i +b_2)^{c_2}}&=&
%\hspace{-2cm}
(-1)^{c_1}\sum_{j=1}^{c_1}{ (-1)^j
\binom{c_1+c_2-j-1}{c_2-1}\frac{a_1^{c_2}a_2^{c_1-j}}{(a_1 b_2-a_2 b_1)^{c_1+c_2-j}}\frac{1}{(a_1 i+b_1)^j}}\NN\\
&+&(-1)^{c_2}\sum_{j=1}^{c_2}{ (-1)^j
\binom{c_1+c_2-j-1}{c_1-1}\frac{a_1^{c_2-j}a_2^{c_1}}
{(a_2 b_1-a_1 b_2)^{c_1+c_2-j}}\frac{1}{(a_2 i+b_2)^j}},
\NN\\
\end{eqnarray}
%------------------------------------------------------------------------------------------------------------
and
%------------------------------------------------------------------------------------------------------------
\begin{eqnarray}
\frac{1}{(a_1 i +b_1)^{c_1}(a_2 i +b_2)^{c_2}}&=&
\left(\frac{a_1}{a_2}\right)^{c_2}\frac{1}{(a_1 i+b_1)^{c_1+c_2}},
\end{eqnarray}
%------------------------------------------------------------------------------------------------------------
if $a_1 b_2= a_2 b_1$.

The product of two cyclotomic harmonic sums has the following representation. Let
$a_i,b_i,d_i,e_i,k,l,n \in \N_+$ and $c_i,f_i \in \Z\setminus{\{0\}}.$
If $a_1 e_1 \neq d_1 b_1$ one has
%------------------------------------------------------------------------------------------------------------
\begin{eqnarray}
%\hspace*{-1cm}
S_{\{a_1,b_1,c_1\},\ldots,\{a_k,b_k,c_k\}}(n)
S_{\{d_1,e_1,f_1\},\ldots,\{d_l,e_l,f_l\}}(n)&=& \NN\\
&& \hspace*{-3.6cm}
\sum_{i=1}^n\frac{\sign{c_1}^i}{(a_1 i+b_1)^{\abs{c_1}}}
S_{\{a_2,b_2,c_2\},\ldots,\{a_k,b_k,c_k\}}(i)
S_{\{d_1,e_1,f_1\},\ldots,\{d_l,e_l,f_l\}}(i)\NN\\
&& \hspace*{-4cm}
+\sum_{i=1}^n\frac{\sign{f_1}^i}{(d_1 i+e_1)^{\abs{f_1}}}
S_{\{a_1,b_1,c_1\},\ldots,\{a_k,b_k,c_k\}}(i)
S_{\{d_2,e_2,f_2\},\ldots,\{d_l,e_l,f_l\}}(i)\NN\\
&& \hspace*{-4cm}
-\sum_{i=1}^n \left( (-1)^{\abs{c_1}}\sum_{j_1=1}^{\abs{c_1}}  (-1)^j \binom{\abs{c_1}+\abs{f_1}-j-1}
{\abs{f_1}-1}\frac{a_1^{\abs{f_1}}d_1^{\abs{c_1}-j}}{a_1 e_1-d_1 b_1}
\frac{1}{(a_1 i+b_1)^j}\right.\NN\\
&&\hspace{-4cm}\left.+(-1)^{\abs{f_1}}\sum_{j=1}^{\abs{f_1}}
{ (-1)^j \binom{\abs{c_1}+\abs{f_1}-j-1}
{\abs{f_1}-1}\frac{a_1^{\abs{f_1}-j}d_1^{\abs{c_1}}}{d_1 b_1-a_1 e_1}
\frac{1}{(d_1 i+e_1)^j}}\right)\NN\\
&&\hspace{-4cm}\times
S_{\{a_2,b_2,c_2\},\ldots,\{a_k,b_k,c_k\}}(i)
S_{\{d_2,e_2,f_2\},\ldots,\{d_l,e_l,f_l\}}(i),
\label{eq:SHA1}
\end{eqnarray}
%------------------------------------------------------------------------------------------------------------
resp. for $a_1 e_1= d_1 b_1$ one has
%------------------------------------------------------------------------------------------------------------
\begin{eqnarray}
S_{\{a_1,b_1,c_1\},\ldots,\{a_k,b_k,c_k\}}(n)
S_{\{d_1,e_1,f_1\},\ldots,\{d_l,e_l,f_l\}}(n)&=&\NN\\
&& \hspace*{-3.6cm}
\sum_{i=1}^n\frac{\sign{c_1}^i}{(a_1 i+b_1)^{\abs{c_1}}}
S_{\{a_2,b_2,c_2\},\ldots,\{a_k,b_k,c_k\}}(i)
S_{\{d_1,e_1,f_1\},\ldots,\{d_l,e_l,f_l\}}(i)\NN\\
\NN\\
%\end{eqnarray}
%\begin{eqnarray}
&& \hspace*{-4cm}
+\sum_{i=1}^n\frac{\sign{f_1}^i}{(d_1 i+e_1)^{\abs{f_1}}}
S_{\{a_1,b_1,c_1\},\ldots,\{a_k,b_k,c_k\}}(i)
S_{\{d_2,e_2,f_2\},\ldots,\{d_l,e_l,f_l\}}(i)\NN\\
&& \hspace*{-4cm}
-\left(\frac{a_1}{d_1}\right)^{\abs{f_1}}\sum_{i=1}^n\frac{(\sign{c_1}
\sign{f_1})^i}{(a_1 i+b_1)^{\abs{c_1}+\abs{f_1}}}
\NN\\
&& \hspace*{-4cm} \times
S_{\{a_2,b_2,c_2\},\ldots,\{a_k,b_k,c_k\}}(i)S_{\{d_2,e_2,f_2\},\ldots,\{d_l,e_l,f_l\}}(i)~.
\label{eq:SHA2}
\end{eqnarray}
%------------------------------------------------------------------------------------------------------------
Subsequently, the relations given by (\ref{eq:SHA1}, \ref{eq:SHA2}) are denoted by $(A)$.

%%%%%%%%%%%%%%%%%%%%%%%%%%%%%%%%%%%%%%%%%%%%%%%%%%%%%%%%%%%%%%%%%%%%%%%%
\subsubsection{Synchronization}
%%%%%%%%%%%%%%%%%%%%%%%%%%%%%%%%%%%%%%%%%%%%%%%%%%%%%%%%%%%%%%%%%%%%%%%%

\vspace*{1mm}
\noindent
A first multiple argument relation is implied as follows. Let $a, b, k \in \N$, $c\in \Z
\setminus \{0\}$, $k\geq 2$. Then
%------------------------------------------------------------------------------------------------------------
\begin{eqnarray}
\X{\{a,b,c\}}{k \cdot N}=\sum_{i=0}^{k-1}\sign{c}^i\X{\{k\cdot a,b-a \cdot i,
\sign{c}^k\abs{c}\}}{N}~.
\label{CSmultint1}
\end{eqnarray}
%------------------------------------------------------------------------------------------------------------
For $a_i, b_i, m, k \in \N$, $c_i\in \Z \setminus \{0\}$, $k \geq 2$, the general
cyclotomic harmonic sums obey
%------------------------------------------------------------------------------------------------------------
\begin{eqnarray}
\X{\{a_m,b_m,c_m\},\{a_{m-1},b_{m-1}, c_{m-1}\},\ldots,\{a_1,b_1,c_1\}}{k \cdot
N}&=&\NN\\
&&\hspace{-2cm}\sum_{i=0}^{m-1}\sum_{j=1}^{N}
\frac{\X{\{a_{m-1},b_{m-1},c_{m-1}\},\ldots,\{a_1,b_1,c_1\}}{k \cdot
j-i}\sign{c_m}^{k\cdot j -i}} {(a_m (k\cdot j-i)+b_1)^{\abs{c_m}}}~.
\NN\\
\label{CSmultint}
\end{eqnarray}
%------------------------------------------------------------------------------------------------------------
Repeated application of (\ref{CSmultint1}, \ref{CSmultint}) allows to represent
cyclotomic harmonic sums of argument $k n$ by those of argument $n$. 
Subsequently, the resulting relations are denoted by $(M)$.

%%%%%%%%%%%%%%%%%%%%%%%%%%%%%%%%%%%%%%%%%%%%%%%%%%%%%%%%%%%%%%%%%%%%%%%%
\subsubsection{Duplication Relations}
%%%%%%%%%%%%%%%%%%%%%%%%%%%%%%%%%%%%%%%%%%%%%%%%%%%%%%%%%%%%%%%%%%%%%%%%

\vspace*{1mm}
\noindent
The usual duplication relation \cite{MZV} holds also for cyclotomic harmonic sums~:
%------------------------------------------------------------------------------------------------------------
\begin{eqnarray}
\sum_{\pm}{\X{\{a_m,b_m,\pm c_m\},\{a_{m-1},b_{m-1}, \pm c_{m-1}\},\ldots,\{a_1,b_1,
\pm c_1\}}{2N}} =2^m \X{\{2 a_m,b_m, c_m\},\ldots,\{2 a_1,b_1, c_1\}}{N}~.
\label{halfint}
\end{eqnarray}
%------------------------------------------------------------------------------------------------------------
The proof of (\ref{halfint}) is given in Appendix~B.

Similar to (\ref{halfint}) one obtains a second duplication relation~:
%------------------------------------------------------------------------------------------------------------
\begin{eqnarray}
\sum_{d_i \in \{-1,1\}}
{d_m d_{m-1}\cdots d_1 \X{\{a_m,b_m,d_m c_m\},\{a_{m-1},b_{m-1}, d_{m-1}
c_{m-1}\},\ldots,\{a_1,b_1,d_1
c_1\}}{2N}} && \NN\\
	&&\hspace{-3cm} = 2^m \X{\{2 a_m,b_m-a_m, c_m\},\ldots,\{2 a_1,b_1-a_1, c_1\}}{N}
\NN;\\
\label{halfint2}
\end{eqnarray}
%------------------------------------------------------------------------------------------------------------
its proof is given in Appendix~B. The resulting algebraic relations of~\eqref{halfint} and~\eqref{halfint2} are 
denoted by $(H_1)$ and $(H_2)$, respectively. We remark that more general relations arise for generalized cyclotomic 
harmonic sums presented in Section~6, (\ref{eq:DISTR}, \ref{eq:DISTR1}). 
%%%%%%%%%%%%%%%%%%%%%%%%%%%%%%%%%%%%%%%%%%%%%%%%%%%%%%%%%%%%%%%%%%%%%%%%
\subsection{Sums of Higher Depth and Weight}
%%%%%%%%%%%%%%%%%%%%%%%%%%%%%%%%%%%%%%%%%%%%%%%%%%%%%%%%%%%%%%%%%%%%%%%%

\vspace{1mm}
\noindent
As an example we consider the class of cyclotomic sums, which occur in the physical
application mentioned in Section~1. They are given by iteration of the summands
%---------------------------------------------------------------------------------
\begin{eqnarray}
\label{eqSel}
\frac{1}{k^{l_1}},~~~~~~
\frac{(-1)^k}{k^{l_2}},~~~~~~
\frac{1}{(2k+1)^{l_3}},~~~~~~
\frac{(-1)^k}{(2k+1)^{l_4}}
\end{eqnarray}
%---------------------------------------------------------------------------------
with $N \geq k \geq 1,~~l_i, k \in \mathbb{N}_+$. As pointed out earlier, this class of cyclotomic harmonic sums can be expressed by Mellin transforms in terms of cyclotomic harmonic polylogarithms generated by the alphabet~\eqref{eq:HALP}.
 
We apply the relations in Section~4.2 to derive the corresponding bases
for given weight {\sf w}. The cyclotomic harmonic sums can be represented over the
corresponding bases. Applying the relations using computer algebra we find the
pattern for the number of basis elements up to {\sf w = 5} given in Table~2.
%---------------------------------------------------------------------------------
\restylefloat{table}
\begin{table}[H]
\scalebox{0.91}{%
\begin{tabular}{| r | r | r | r | r | r | r | r | r | r | r | r | }
\hline	
{\sf w}
& $N_S$
& $H_1$
& $H_1,H_2$
& $H_1,M$
& $H_1,H_2,M$
& $D$
& $H_1,H_2,M,D$
& $A$
& $H_1,H_2,M,A$
& $A,D$
& {\sf all}\\
\hline
     1 &    4 &    3 &    3 &    2 &    2 &    4 &    2 &   4 &   2 &   4 &   2 \\
     2 &   20 &   18 &   17 &   16 &   15 &   16 &   13 &  10 &   8 &   6 &   6 \\
     3 &  100 &   96 &   93 &   92 &   89 &   80 &   74 &  40 &  35 &  30 &  27 \\
     4 &  500 &  492 &  485 &  484 &  477 &  400 &  388 & 150 & 142 & 110 & 107 \\
     5 & 2500 & 2484 & 2469 & 2468 & 2453 & 2000 & 1976 & 624 & 607 & 474 & 465 \\
\hline
\end{tabular}
\caption[]{
\label{TAB2}
\small \sf
Reduction of the number of cyclotomic harmonic sums $N_S$ over the elements (\ref{eqSel})
at given weight w by applying the three multiple argument relations $(H_1, H_2, M)$,
differentiation w.r.t.\ to the external sum index $N$, $(D)$, and
the algebraic relations $(A)$. A sequence of symbols corresponds to the combination
of these relations.
}
}
\end{table}
%---------------------------------------------------------------------------------

Due to the arguments given in Section~4.3.1 above, for any occurrence of a differential 
operator $(\partial^m/\partial N^m) S_{a_1,b_1,c_1}, ...,{a_k,b_k,c_k}(N)$
only one representative  is counted.

The total number of sums of weight {\sf w} for the alphabet (\ref{eqSel}) is
%---------------------------------------------------------------------------------
\begin{eqnarray}
N_S(w)  &=& 4\cdot5^{w-1}~.
\end{eqnarray}
%---------------------------------------------------------------------------------
We mention that the single application of any of the multiple argument relations $(M,
H_1, H_2)$ leads to the same number of basis sums
%---------------------------------------------------------------------------------
\begin{eqnarray}
N_{H_1}   &=& N_{H_2} = N_M~.
\end{eqnarray}
%---------------------------------------------------------------------------------
This also applies to the combinations $H_1 M$ and $H_2 M$
%---------------------------------------------------------------------------------
\begin{eqnarray}
N_{H_1,M} &=& N_{H_2,M}~.
\end{eqnarray}
%---------------------------------------------------------------------------------
Explicit counting relations for the number of basis elements  given Table~2 can be derived~:
%---------------------------------------------------------------------------------
\begin{eqnarray}
N_A(w)  &=& \frac{1}{w}\sum_{d|w}{\mu\left(\frac{w}{d}\right)5^d}
\\
N_D(w)  &=& N_S(w)-N_S(w-1)=16\cdot5^{w-2}
\\
N_{H_1}(w) &=& N_S(w)-2^{w-1}=4\cdot5^{w-1}-2^{w-1}
\\
N_{H_1H_2}(w) &=& N_S(w)-(2\cdot2^{w-1}-1)=4\cdot5^{w-1}-(2\cdot2^{w-1}-1)
\\
N_{H_1M}(w) &=& N_S(w)-2\cdot2^{w-1}=4\cdot5^{w-1}-2\cdot2^{w-1}
\\
N_{H_1H_2M}(w) &=& N_S(w)-(3\cdot2^{w-1}-1)=4\cdot5^{w-1}-(3\cdot2^{w-1}-1)
\\
N_{AD}(w)&=& N_A(w)-N_A(w-1)\nonumber\\
	 &=& \frac{1}{w}\sum_{d|w}{\mu\left(\frac{w}{d}\right)5^d}-\frac{1}{w-1}\sum_{d|w-1}{\mu\left(\frac{w-1}{d}\right)5^d} \\
N_{AH_1H_2M}(w)  &=&
\frac{1}{w}\sum_{d|w}{\mu\left(\frac{w}{d}\right)5^d}-\left(3\cdot\frac{1}{w}\sum_{d|w}{\mu\left(\frac{w}{d}\right)2^d}-1\right) \\
N_{DH_1H_2M}(w)  &=&N_{H_1H_2M}(w)-N_{H_1H_2M}(w-1)=16\cdot5^{w-2}-3\cdot2^{w-2} \\
N_{ADH_1H_2M}(w) &=&N_{AH_1H_2M}(w)-N_{AH_1H_2M}(w-1) \\
	&=&\frac{1}{w}\sum_{d|w}{\mu\left(\frac{w}{d}\right)
(5^d-3\cdot2^d)}-\frac{1}{w-1}\sum_{d|w-1}{\mu\left(\frac{w-1}{d}\right)
(5^d-3\cdot2^d)}~.
\NN\\
\end{eqnarray}
%---------------------------------------------------------------------------------
Here $\mu$ denotes the M\"obius function \cite{MOEBIUS}.

The analytic continuation of the cyclotomic harmonic sums can be performed
as outlined for the case of the single sums in Section~4.1. Their representation as
a Mellin transform of the cyclotomic harmonic polylogarithms~(\ref{eq:MEL}) relates
them to factorial series, except the case $c_1 = 1, s_1 = 1$, cf.~(\ref{eq:MSU}).  
If a sequential set of first indices $c_i = 1, s_i = 1$ occurs, one may reduce the corresponding 
sum algebraically to convergent sums, separating factors, cf.~\cite{STRUCT5}. I.e., also in the general 
case the poles of the cyclotomic harmonic sums are located at
$-k, k \in \N$. The recursion relations of the cyclotomic harmonic sums
(\ref{eq:MSU}) imply the shift relations $N \rightarrow N+1$ in a hierarchic manner,
referring to the sums of lower depth.

To accomplish the analytic continuation in $N$, the asymptotic representations of
the cyclotomic harmonic sums have to be computed. Since the cyclotomic harmonic sums are represented
over respective bases, only the asymptotic representations for the basis elements have to
be derived. One way consists in using iterated integration by parts
%---------------------------------------------------------------------------------
\begin{eqnarray}
\int_0^1~dx~x^N~f(x) = \frac{f(1)}{N+1} - \frac{1}{N+1} \int_0^1~dx~x^N~[x f'(x)]~,
\label{eq:PINT}
\end{eqnarray}
%---------------------------------------------------------------------------------
with $f(x)$ a linear combination of cyclotomic polylogarithms. Here $f(x)$ is
conveniently expressed in a power series to which we turn now.

We illustrate the principle steps considering the alphabet~\eqref{eq:HALP}
related to (\ref{eqSel}), cf.~(\ref{eq:f1}, \ref{eq:f2}). In general the cyclotomic
harmonic polylogarithms $C_{\vec a}^{\vec b}(x)$ over the alphabet (\ref{eq:HALP}) do not have a
regular Taylor series expansion, cf.~\cite{VR}. This is due to the
effect that trailing zeroes in the index set may cause powers of $\ln(x)$. Hence
the proper expansion is one in terms of both $x$ and $\ln(x)$. For depth one and $0 < x <1$ one obtains
%---------------------------------------------------------------------------------
\begin{eqnarray}
\label{eq:cycl1}
C_0^0(x)&=& \ln(x)
\\
C_2^0(x)&=&-\sum_{i=1}^{\infty}\frac{(-x)^i}{i}=-\sum_{i=1}^{\infty}\frac{(-x)^{2
i}}{2
i}+\sum_{i=1}^{\infty}\frac{(-x)^{2 i+1}}{2 i+1}
\\
C_1^0(x)&=&-\sum_{i=1}^{\infty}\frac{x^i}{i}=\sum_{i=1}^{\infty}\frac{(-x)^{2 i}}{2
i}+\sum_{i=1}^{\infty}\frac{(-x)^{2 i+1}}{2 i+1}
\\
C_4^0(x)&=&-\sum_{i=1}^{\infty}\frac{(-1)^ix^{2 i-1}}{2 i-1}
\\
\label{eq:cycl4}
C_4^1(x)&=&\sum_{i=1}^{\infty}\frac{(-1)^i x^{2 i}}{2 i}~.
\end{eqnarray}
%---------------------------------------------------------------------------------
Let $C_{\vec{a}}^{\vec{b}}(x)$ be a cyclotomic harmonic polylogarithm with depth $d.$ Assume that
its power series expansion is of the form
%---------------------------------------------------------------------------------
\begin{eqnarray}
C_{\vec{a}}^{\vec{b}}(x) = \sum_{j=1}^w\sum_{i=1}^{\infty}\frac{\sigma^ix^{2 i+c_j}}{(2
i+c_j)^a}\X{\vec{n}_j}i
\end{eqnarray}
%---------------------------------------------------------------------------------
with the index sets ${\vec{a}}$ and  ${\vec{b}}$ according to the iteration of the 
letters (\ref{eq:cycl1}--\ref{eq:cycl4}) and $\vec{n}_j$ a corresponding index 
set of the cyclotomic harmonic sum, $x \in (0,1)$, $w\in \N$ and $c_j\in \Z$.

Then the expansion of the cyclotomic harmonic polylogarithms of depth $d+1$ is obtained by using
%---------------------------------------------------------------------------------
\begin{eqnarray}
\label{eq:Cf00}
C_{0,\vec{a}}^{0,\vec{b}}(x) &=& \sum_{j=1}^w{\sum_{i=1}^{\infty}\frac{\sigma^i x^{2 i+c_j}}{(2
i+c_j)^{a+1}}\X{\vec n_j}i}
\end{eqnarray}
%---------------------------------------------------------------------------------
\begin{eqnarray}
C_{0,\vec{a}}^{1,\vec{b}}(x) &=& \sum_{j=1}^w{\sum_{i=1}^{\infty}\frac{x^{2 i+c_j+1}}{(2
i+c_j+1)}\X{\{2,c_j,\sigma a\},\vec{n}_j}i}
		+\sum_{j=1}^w{\sum_{i=1}^{\infty}\frac{x^{2 i+c_j+2}}{(2
i+c_j+2)}\X{\{2,c_j,\sigma a\},\vec{n}_j}i}\NN\\
\\
C_{2,\vec{a}}^{0,\vec{b}}(x) &=& \sum_{j=1}^w{\sum_{i=1}^{\infty}\frac{x^{2 i+c_j+1}}{(2
i+c_j+1)}\X{\{2,c_j,\sigma a\},\vec{n}_j}i}
		-\sum_{j=1}^w{\sum_{i=1}^{\infty}\frac{x^{2 i+c_j+2}}{(2
i+c_j+2)}\X{\{2,c_j,\sigma a\},\vec{n}_j}i}\NN\\
\\
C_{4,\vec{a}}^{0,\vec{b}}(x) &=& \sum_{j=1}^w{\sum_{i=1}^{\infty}\frac{(-1)^i x^{2 i+c_j+1}}{(2
i+c_j+1)}\X{\{2,c_j,-\sigma a\},\vec{n}_j}i}
\label{eq:Cf40}
\\
C_{4,\vec{m}}^{1,\vec{m}}(x) &=& \sum_{j=1}^w{\sum_{i=1}^{\infty}\frac{(-1)^i x^{2 i+c_j+2}}{(2
i+c_j+2)}\X{\{2,c_j,-\sigma a\},\vec{n}_j}i}~.
\label{eq:Cf41}
\end{eqnarray}
%---------------------------------------------------------------------------------
Sample proofs of these relations are given in Appendix~B.

The analytic continuation of the cyclotomic harmonic sums at larger depths to $N \in \mathbb{C}$ is
performed analogously to the case discussed in Section~4.1. Shifts parallel to the real axis are
performed with the recurrence relation induced by (\ref{eq:MSU}). The asymptotic relations for
$N \rightarrow \infty,~~|\text{arg}(N)| < \pi$ can be derived analytically to arbitrary precision.
Examples are~:
%---------------------------------------------------------------------------------
\begin{align}
\X{\{2,1,2\},\{1,0,-2\}}N \sim &
4 C_{f_4^0,f_0^0}(1)^2+4 C_{f_4^0}(1) C_{f_0^0,f_4^0,f_0^0}(1)-4 
C_{f_4^0,f_0^0,f_4^0,f_0^0}(1)+\left(\frac{\pi ^2}{2}-\frac{1}{N}\right.\nonumber\\
& +\frac{1}{N^2}-\frac{11}{12 N^3}+\frac{3}{4 N^4}-\frac{127}{240 N^5}+\frac{5}{16 N^6}-\frac{221}{1344 
N^7}+\frac{7}{64 N^8}\nonumber\\
&\left.-\frac{367}{3840 N^9}+\frac{9}{256 N^{10}}\right) C_{f_4^1,f_0^0}(1)+(-1)^N \left(\frac{1}{16 
N^4}-\frac{1}{4 N^5}+\frac{27}{64 N^6}\right.\nonumber\\
&\left.-\frac{1}{32 N^7}-\frac{269}{256 N^8}-\frac{11}{32 N^9}+\frac{8699}{1024 N^{10}}\right)+O\left(\frac{1}{N^{11}}\right)\\
\X{\{2,1,2\},\{2,1,-2\}}N \sim &-\frac{\pi 
^2}{8}+C_{f_4^0}(1)C_{f_0^0,f_4^1,f_0^0}(1)-C_{f_4^0,f_0^0,f_4^1,f_0^0}(1)+\frac{1}{4 N}-\frac{1}{4 
N^2}+\frac{11}{48 N^3}\nonumber\\
&-\frac{3}{16 N^4}+\frac{127}{960N^5}-\frac{5}{64 N^6}+\frac{221}{5376 N^7}-\frac{7}{256 
N^8}+\frac{367}{15360 N^9}\nonumber\\
&-\frac{9}{1024 N^{10}}+(-1)^N \left(\frac{1}{64 N^4}-\frac{5}{64 N^5}+\frac{25}{128 
N^6}-\frac{61}{256N^7}\right.\nonumber\\
&\left.-\frac{77}{1024 N^8}+\frac{221}{512 N^9}+\frac{1545}{1024 N^{10}}\right)+C_{f_4^0,f_0^0}(1) 
\left(C_{_4^1,f_0^0}(1)-\frac{\pi ^2}{8}\right.\nonumber\\
&\left.+\frac{1}{4 N}-\frac{1}{4 N^2}+\frac{11}{48 N^3}-\frac{3}{16 N^4}+\frac{127}{960N^5}-\frac{5}{64 
N^6}+\frac{221}{5376 N^7}\right.\nonumber\\
&\left.-\frac{7}{256 N^8}+\frac{367}{15360 N^9}-\frac{9}{1024 
N^{10}}\right)+O\left(\frac{1}{N^{11}}\right)~.
\end{align}
%---------------------------------------------------------------------------------
Given a cyclotomic harmonic sum with the iterative denominators~\eqref{eqSel} 
and given the number of desired terms, the corresponding expansion can be computed 
on demand by the {\sf HarmonicSums} package.

%%%%%%%%%%%%%%%%%%%%%%%%%%%%%%%%%%%%%%%%%%%%%%%%%%%%%%%%%%%%%%%%%%%%%%%%
\section{Special Values}
\label{sec:5}
%%%%%%%%%%%%%%%%%%%%%%%%%%%%%%%%%%%%%%%%%%%%%%%%%%%%%%%%%%%%%%%%%%%%%%%%

\vspace{1mm} \noindent
The values of the cyclotomic harmonic polylogarithms at argument
$x=1$ and, related to it, the associated cyclotomic harmonic sums at $N \rightarrow
\infty$ occur in various relations of the finite cyclotomic harmonic sums
and the Mellin transforms of cyclotomic harmonic polylogarithms. In this
Section we investigate their relations and basis representations. The
infinite cyclotomic harmonic sums extend the Euler-Zagier and multiple
zeta values \cite{MZV} and are related at lower weight and depth to other
known special numbers. We first consider the single non-alternating and
alternating sums up to cyclotomy {\sf l = 6} at general weight {\sf w}.
Next the relations of the infinite cyclotomic harmonic sums associated to
the summands (\ref{eqSel}) up to weight {\sf w = 6} are worked out. Finally we investigate
the sums of weight {\sf w = 1} and {\sf 2} up to cyclotomy {\sf l = 20}.
%%%%%%%%%%%%%%%%%%%%%%%%%%%%%%%%%%%%%%%%%%%%%%%%%%%%%%%%%%%%%%%%%%%%%%%%
\subsection{Single Infinite Sums}
%%%%%%%%%%%%%%%%%%%%%%%%%%%%%%%%%%%%%%%%%%%%%%%%%%%%%%%%%%%%%%%%%%%%%%%%

\vspace{1mm}
\noindent
We consider the single sums of the type
%----------------------------------------------------------------------------------------------
\begin{eqnarray}
\sum_{k=0}^\infty \frac{(\pm 1)^k}{(lk + m)^n}~,
\end{eqnarray}
%----------------------------------------------------------------------------------------------
with $l,m,n \in {\mathbb N}_+, l > m, n \geq 1$. These sums are linearly related to the
colored harmonic sums by
%----------------------------------------------------------------------------------------------
\begin{eqnarray}
\sum_{k=1}^\infty \frac{e_l^k}{k^n}
= \sum_{m=1}^l e^m_l \sum_{k=0}^\infty \frac{1}{(lk+m)^n},~~~~~e_l = \exp\left(\frac{2
\pi i}{l}\right)~,
\end{eqnarray}
%----------------------------------------------------------------------------------------------
and similar linear relations for the nested sums. We will address the latter sums in Section~\ref{sec:6}.
%%%%%%%%%%%%%%%%%%%%%%%%%%%%%%%%%%%%%%%%%%%%%%%%%%%%%%%%%%%%%%%%%%%%%%%%
\subsubsection{Non-alternating Single Sums at {\sf w = 1}}
\label{sec:5.1.1}
%%%%%%%%%%%%%%%%%%%%%%%%%%%%%%%%%%%%%%%%%%%%%%%%%%%%%%%%%%%%%%%%%%%%%%%%

\vspace{1mm}
\noindent
For the non-alternating sums one obtains
%----------------------------------------------------------------------------------------------
\begin{eqnarray}
\sum_{k=0}^{\infty} \frac{1}{lk+m} &=& \frac{1}{l} \left[ \sigma_0 - \gamma_E
                                      - \psi\left(\frac{m}{l}\right) \right]~,
\label{eq:EVINF1}
\end{eqnarray}
%----------------------------------------------------------------------------------------------
with
%----------------------------------------------------------------------------------------------
\begin{eqnarray}
\sigma_0 = \sum_{k=1}^\infty \frac{1}{k}
\end{eqnarray}
%----------------------------------------------------------------------------------------------
denoting the divergence in (\ref{eq:EVINF1}). The corresponding sums, except $\sigma_0$ can always be
regularized.

The reflection symmetry of the
$\psi$-function interchanging the arguments $x$ and $(1-x)$, \cite{NIELS}, implies
%----------------------------------------------------------------------------------------------
\begin{eqnarray}
\label{eq:EVINF2}
\sum_{k=0}^\infty\left[
  \frac{1}{nk+l}
- \frac{1}{nk+(n-l)} \right] &=& \frac{\pi}{n} \cot\left(\frac{l}{n}\pi\right)~.
\end{eqnarray}
%----------------------------------------------------------------------------------------------
The digamma-function at positive rational arguments obeys \cite{GAUJEN,NIELS}
%----------------------------------------------------------------------------------------------
\begin{eqnarray}
\label{eq:GAUSS1}
\psi\left(\frac{p}{q}\right) &=& - \gamma_E - \ln(2 q) - \frac{\pi}{2} \cot\left(\frac{p \pi}{q}\right)
+ 2 \sum_{k=1}^{[(q-1)/2]} \cos\left(\frac{2 \pi k p}{q}\right) \ln\left[\sin\left(\frac{\pi k}{q}\right)
\right]\\
\psi\left(\frac{1}{n}\right) &=& - n\left( \gamma_E + \ln(n)\right)
- \sum_{k=2}^n \psi\left(\frac{k}{n}\right)~.
\end{eqnarray}
%----------------------------------------------------------------------------------------------
Eq.~(\ref{eq:GAUSS1}) is used to remove dependencies in (\ref{eq:EVINF1}, \ref{eq:EVINF2}). If the
regular $q$--polygon is constructible, the trigonometric functions in (\ref{eq:GAUSS1}) are algebraic numbers.
This is the case for 
%----------------------------------------------------------------------------------------------
\begin{eqnarray}
q \in \{2, 3, 4, 5, 6, 8, 10, 12, 15, 16, 17, 20, 24, 30, 32, 34, 40, 48, 51, 60, 64,...\}~,
\end{eqnarray}
%----------------------------------------------------------------------------------------------
\cite{CONSTR}~\footnote{
See \cite{SINCOS} for special values of the
trigonometric functions occurring in (\ref{eq:GAUSS1}).}.
Due to (\ref{eq:GAUSS1}) the constants $\ln(d_i)$ with $d_i \neq 1 $, the divisors of $q$, and
logarithms of further algebraic numbers will occur. Theses are all transcendental numbers
\cite{BAKER}.

Let us list the first few of these relations, see also~\cite{SC}~:
%----------------------------------------------------------------------------------------------
\begin{eqnarray}
-\frac{1}{2}\left[\gamma_E+\psi\left(\frac{1}{2}\right)\right] &=& \ln(2) \\
-\frac{1}{3}\left[\gamma_E+\psi\left(\frac{1}{3}\right)\right] &=& \frac{1}{2}\ln(3) + \frac{\pi}{18} \sqrt{3} \\
-\frac{1}{3}\left[\gamma_E+\psi\left(\frac{2}{3}\right)\right] &=& \frac{1}{2}\ln(3) - \frac{\pi}{18} \sqrt{3} \\
%\end{eqnarray}
%\begin{eqnarray}
-\frac{1}{4}\left[\gamma_E+\psi\left(\frac{1}{4}\right)\right] &=& \frac{3}{4}\ln(2) + \frac{\pi}{8}  \\
-\frac{1}{4}\left[\gamma_E+\psi\left(\frac{3}{4}\right)\right] &=& \frac{3}{4}\ln(2) - \frac{\pi}{8}
\\
-\frac{1}{5}\left[\gamma_E+\psi\left(\frac{1}{5}\right)\right] &=&
\frac{\sqrt{5}}{10} \left[ \ln(2) - \ln(\sqrt{5}-1) \right]
+ \frac{1}{4} \ln(5)
+ \frac{\sqrt{25 +10 \sqrt{5}}}{50} \pi
\\
-\frac{1}{5}\left[\gamma_E+\psi\left(\frac{2}{5}\right)\right] &=&
- \frac{\sqrt{5}}{10}\left[\ln(2) - \ln(\sqrt{5}-1)\right] + \frac{1}{4} \ln(5)
+\frac{\sqrt{10 - 2 \sqrt{5}}}{40} \left[1 - \frac{\sqrt{5}}{5}\right] \pi
\nonumber\\
\\
-\frac{1}{5}\left[\gamma_E+\psi\left(\frac{3}{5}\right)\right] &=&
- \frac{\sqrt{5}}{10}\left[\ln(2) - \ln(\sqrt{5}-1)\right] + \frac{1}{4} \ln(5)
-\frac{\sqrt{10 - 2 \sqrt{5}}}{40} \left[1 - \frac{\sqrt{5}}{5}\right] \pi
\nonumber\\
\\
-\frac{1}{5}\left[\gamma_E+\psi\left(\frac{4}{5}\right)\right] &=&
\frac{\sqrt{5}}{10} \left[ \ln(2) - \ln(\sqrt{5}-1) \right]
+ \frac{1}{4} \ln(5)
- \frac{\sqrt{25 +10 \sqrt{5}}}{50} \pi \\
%\end{eqnarray}
%\begin{eqnarray}
-\frac{1}{6}\left[\gamma_E+\psi\left(\frac{1}{6}\right)\right] &=& \frac{1}{3}\ln(2) + \frac{1}{4} \ln(3)
+ \frac{\pi}{12} \sqrt{3}  \\
-\frac{1}{6}\left[\gamma_E+\psi\left(\frac{5}{6}\right)\right] &=& \frac{1}{3}\ln(2) + \frac{1}{4} \ln(3)
- \frac{\pi}{12} \sqrt{3},~~{\rm etc.}
\end{eqnarray}
%----------------------------------------------------------------------------------------------
%%%%%%%%%%%%%%%%%%%%%%%%%%%%%%%%%%%%%%%%%%%%%%%%%%%%%%%%%%%%%%%%%%%%%%%%
\subsubsection{Alternating Single Sums at {\sf w = 1}}
\label{sec:5.1.2}
%%%%%%%%%%%%%%%%%%%%%%%%%%%%%%%%%%%%%%%%%%%%%%%%%%%%%%%%%%%%%%%%%%%%%%%%

\vspace{1mm}
\noindent
The alternating sums at {\sf w = 1} have the representation~:
%----------------------------------------------------------------------------------------------
\begin{eqnarray}
\label{eq:EVINF3}
\sum_{k=0}^{\infty} \frac{(-1)^k}{lk+m} &=& \frac{1}{2l}
\left[\psi\left(\frac{m+l}{2l}\right) - \psi\left(\frac{m}{2l}\right)\right]~,
\end{eqnarray}
%----------------------------------------------------------------------------------------------
with the reflection relation
%----------------------------------------------------------------------------------------------
\begin{eqnarray}
\sum_{k=0}^\infty\left[
  \frac{(-1)^k}{nk+l}
- \frac{(-1)^k}{nk+(n-l)} \right] &=& \frac{1}{2n} \Biggl[
\psi\left(\frac{1}{2} -\frac{l}{2n}\right)
+ \psi\left(\frac{1}{2} +\frac{l}{2n}\right)
%\nonumber\\ & &
- \psi\left(1 -\frac{l}{2n}\right)
- \psi\left(\frac{l}{2n}\right) \Biggr]~.
\nonumber\\
\end{eqnarray}
%----------------------------------------------------------------------------------------------

For the sums at {\sf w = 1} one obtains~:
%----------------------------------------------------------------------------------------------
\begin{eqnarray}
\sum_{k=0}^{\infty} \frac{(-1)^k}{2k+1} &=& \frac{\pi}{4} \\
%--
\sum_{k=0}^{\infty} \frac{(-1)^k}{3k+1} &=& \frac{1}{3} \left[\frac{\pi}{\sqrt{3}}  +  \ln(2) \right] \\
%--
\sum_{k=0}^{\infty} \frac{(-1)^k}{3k+2} &=& \frac{1}{3} \left[\frac{\pi}{\sqrt{3}}  -  \ln(2) \right] \\
%--
\sum_{k=0}^{\infty} \frac{(-1)^k}{4k+1} &=& \frac{1}{2 \sqrt{2}} \left[\frac{\pi}{2}  -  \ln(\sqrt{2}-1) \right]
\\
%--
\sum_{k=0}^{\infty} \frac{(-1)^k}{4k+3} &=& \frac{1}{2 \sqrt{2}} \left[\frac{\pi}{2}  +  \ln(\sqrt{2}-1) \right]
\\
%\end{eqnarray}
%\begin{eqnarray}
%--
\sum_{k=0}^{\infty} \frac{(-1)^k}{5k+1} &=& \frac{1}{5} \left[1+ \sqrt{5}\right] \ln(2) -
\frac{1}{\sqrt{5}}\ln(\sqrt{5}-1) + \frac{1+\sqrt{5}}{5\sqrt{10 + 2 \sqrt{5}}} \pi
\\
%--
\sum_{k=0}^{\infty} \frac{(-1)^k}{5k+2} &=&
-\frac{1}{5}\left[1-\sqrt{5}\right] \ln(2) - \frac{1}{\sqrt{5}} \ln(\sqrt{5}-1)
+ \frac{\sqrt{5}-1}{5\sqrt{10 - 2 \sqrt{5}}} \pi \\
%--
\sum_{k=0}^{\infty} \frac{(-1)^k}{5k+3} &=&
\frac{1}{5}\left[1-\sqrt{5}\right] \ln(2) + \frac{1}{\sqrt{5}} \ln(\sqrt{5}-1)
+ \frac{\sqrt{5}-1}{5\sqrt{10 - 2 \sqrt{5}}} \pi \\
%--
\sum_{k=0}^{\infty} \frac{(-1)^k}{5k+4} &=&
-\frac{1}{5} \left[1+ \sqrt{5}\right] \ln(2) +
\frac{1}{\sqrt{5}}\ln(\sqrt{5}-1) + \frac{1+\sqrt{5}}{5\sqrt{10 + 2 \sqrt{5}}} \pi
\\
%--
\sum_{k=0}^{\infty} \frac{(-1)^k}{6k+1} &=&
\frac{\pi}{6} + \frac{1}{\sqrt{3}}\left[\frac{1}{2} \ln(2)
                                                          - \ln(\sqrt{3}-1)\right] \\
%--
\sum_{k=0}^{\infty} \frac{(-1)^k}{6k+5} &=&
\frac{\pi}{6} - \frac{1}{\sqrt{3}}\left[\frac{1}{2} \ln(2)
                                                          -
\ln(\sqrt{3}-1)\right],~~\text{etc.}
\end{eqnarray}
%----------------------------------------------------------------------------------------------
%%%%%%%%%%%%%%%%%%%%%%%%%%%%%%%%%%%%%%%%%%%%%%%%%%%%%%%%%%%%%%%%%%%%%%%%
\subsubsection{Single Infinite Sums of Higher Weight}
%%%%%%%%%%%%%%%%%%%%%%%%%%%%%%%%%%%%%%%%%%%%%%%%%%%%%%%%%%%%%%%%%%%%%%%%

\vspace{1mm}
\noindent
These sums obey representations which are obtained by repeated differentiation of
(\ref{eq:EVINF1}) and (\ref{eq:EVINF3}) for $m$~:
%----------------------------------------------------------------------------------------------
\begin{eqnarray}
\sum_{k=0}^\infty \frac{1}{(lk+m)^n} &=& \frac{1}{l^n} \zeta_{\rm H}\left(n,
\frac{m}{l}\right) = \frac{1}{\Gamma(n)}\left(- \frac{1}{l}\right)^n
\psi^{(n-1)}\left(\frac{m}{l}\right) \\
\sum_{k=0}^\infty \frac{(-1)^k}{(lk+m)^n} &=& \frac{(-1)^{n-1}}{\Gamma(n)}
\frac{1}{(2l)^n}
\left[\psi^{(n-1)}\left(\frac{m+l}{2l}\right) -
\psi^{(n-1)}\left(\frac{m}{2l}\right)\right]~.
\end{eqnarray}
%----------------------------------------------------------------------------------------------
Here $\zeta_{\rm H}$ is the Hurwitz $\zeta$--function \cite{HURWITZ} with the serial
representation
%----------------------------------------------------------------------------------------------
\begin{eqnarray}
\label{eq:HURW}
\zeta_{\rm H}(s,a) = \sum_{n=0}^\infty \frac{1}{(n+a)^s}~.
\end{eqnarray}
%----------------------------------------------------------------------------------------------
We consider the representations of the polygamma functions at rational arguments for
$p/q$, $p \nmid  q, q \in \mathbb{N}_+, q \leq 12$.\footnote{Special examples were also
considered in  \cite{SPEC}.}
One obtains~:
%----------------------------------------------------------------------------------------------
\begin{eqnarray}
\psi^{(l)}\left(\frac{1}{2}\right) &=&  (-1)^{l+1} l! \left(2^{l+1}
- 1 \right) \zeta_{l+1}~,
\label{eq:half1}
\end{eqnarray}
%----------------------------------------------------------------------------------------------
with
%----------------------------------------------------------------------------------------------
\begin{eqnarray}
\zeta_{2n} = (-1)^{n-1} \frac{(2 \pi)^{2n}}{2(2n)!} B_{2n},
\end{eqnarray}
%----------------------------------------------------------------------------------------------
where $B_n$ denote the Bernoulli numbers \cite{BERNOULLI1,BERNOULLI2}. They are generated by
%----------------------------------------------------------------------------------------------
\begin{eqnarray}
\frac{x}{\exp(x) - 1} &=& \sum_{n=0}^\infty \frac{B_n}{n!} x^n,~~~B_{2n+1} =
0~\text{for}~n > 1~.
\end{eqnarray}
%----------------------------------------------------------------------------------------------
For odd values of $l$ no new basis elements occur due to (\ref{eq:half1}).

The reflection formula \cite{NIELS}
%----------------------------------------------------------------------------------------------
\begin{eqnarray}
\label{eq:REFL}
\psi^{(n)}(1-z) = (-1)^n \left[\psi^{n}(z) + \pi \frac{d^n}{dz^n} \cot(\pi z) \right]
\end{eqnarray}
%----------------------------------------------------------------------------------------------
implies the relations for the argument $p/q$ and $(q-p)/q$. Likewise,
the multiplication formula for the Hurwitz zeta function \cite{HURWITZ} holds,
%----------------------------------------------------------------------------------------------
\begin{eqnarray}
\zeta_{\rm H}(s, kz) &=& \frac{1}{k^s} \sum_{n=0}^{k-1} \zeta_{\rm H}
\left(s, z + \frac{n}{k}\right),~~~k \in \mathbb{N}_+, \\
\label{eq:HUR2}
m^{l+1} \psi^{(l)}(m z) &=& \sum_{k=0}^{m-1} \psi^{(l)}\left(z +\frac{k}{m}\right),
\end{eqnarray}
%----------------------------------------------------------------------------------------------
with the special relation
%----------------------------------------------------------------------------------------------
\begin{eqnarray}
n! (-1)^{n+1} \zeta_{n+1} \left[m^{n+1} - 1 \right] &=& \sum_{k=1}^{m-1}
\psi^{(n)}\left(\frac{k}{m}\right)~.
\end{eqnarray}
%----------------------------------------------------------------------------------------------
One obtains, cf. also \cite{MA},
%----------------------------------------------------------------------------------------------
\begin{eqnarray}
\psi^{(l)}\left(\frac{2}{3}\right) = (-1)^{l+1} l! \left(3^{l+1}-1\right) \zeta_{l+1}
-
\psi^{(l)}\left(\frac{1}{3}\right)~.
\end{eqnarray}
%----------------------------------------------------------------------------------------------
For even values of $l$, $\psi^{(l)}(1/3)$ and $\psi^{(l)}(2/3)$ are linear
in $\pi^{l+1} \sqrt{3}$ and $\zeta_{l+1}$, with
%----------------------------------------------------------------------------------------------
\begin{eqnarray}
\psi^{(2)}\left(\frac{1}{3}\right) &=&   -\frac{4}{9} \sqrt{3} \pi^3 - 26 \zeta_3  \\
\psi^{(2)}\left(\frac{2}{3}\right) &=&   ~~\frac{4}{9} \sqrt{3} \pi^3 - 26 \zeta_3  \\
\psi^{(4)}\left(\frac{1}{3}\right) &=&   -\frac{16}{3} \sqrt{3} \pi^5 - 2904 \zeta_5 \\
\psi^{(4)}\left(\frac{2}{3}\right) &=&   ~~\frac{16}{3} \sqrt{3} \pi^5 - 2904
\zeta_5,~~\text{etc.}
\end{eqnarray}
%----------------------------------------------------------------------------------------------
Therefore, only for odd values of $l$ one new basis element due to $\psi^{(2l+1)}(1/3)$
contributes.

The two values at argument $1/4$ and $3/4$ are related, see e.g.~\cite{KOLB},
%----------------------------------------------------------------------------------------------
\begin{eqnarray}
\psi^{(2l-1)}\left(\frac{1}{4}\right) &=&  \frac{4^{2l-1}}{2l}\left[\pi^{2l} ( 2^{2l} -1 ) |B_{2l}| +
2(2l)! \beta_D(2l)\right]\\
\psi^{(2l-1)}\left(\frac{3}{4}\right) &=&  \frac{4^{2l-1}}{2l}\left[\pi^{2l} ( 2^{2l} -1 ) |B_{2l}| -
2(2l)! \beta_D(2l)\right]\\
\psi^{(2l)}\left(\frac{1}{4}\right) &=&  - 2^{2l-1} \left[\pi^{2l+1} |E_{2l}| +
 2(2l)! (2^{2l+1}-1) \zeta_{2l+1}\right]\\
\psi^{(2l)}\left(\frac{3}{4}\right) &=&  + 2^{2l-1} \left[\pi^{2l+1} |E_{2l}| -
 2(2l)! (2^{2l+1}-1) \zeta_{2l+1}\right]~.
\end{eqnarray}
%----------------------------------------------------------------------------------------------
Here,  $E_{2l}$ denote the
Euler numbers \cite{EULER-NU1,EULER-NU2,BERN1,BERN2}. They are generated by
%----------------------------------------------------------------------------------------------
\begin{eqnarray}
\frac{2}{\exp(x) + \exp(-x)} &=& \sum_{n=0}^\infty \frac{E_n}{n!} x^n,~~~E_{2n+1} =
0~\text{for}~n > 1~.
\end{eqnarray}
%----------------------------------------------------------------------------------------------
$\beta_D$ is the Dirichlet $\beta$--function
\cite{HURWITZ,DIRICHLET}
%----------------------------------------------------------------------------------------------
\begin{eqnarray}
\label{eq:DIRL}
\beta_D(l) = \sum_{k=0}^{\infty} \frac{(-1)^k}{(2k+1)^l} &=& \Ti_l(1), l \in {\mathbb
N}_+,
\end{eqnarray}
%----------------------------------------------------------------------------------------------
which is also given by the inverse tangent integral ${\rm Ti}_l(x)$ \cite{SPENCE,LEWIN1},
being related to Clausen integrals \cite{CLAUSEN}.
The special value for $l=2$ yields Catalan's constant \cite{CATALAN}
%----------------------------------------------------------------------------------------------
\begin{eqnarray}
\sum_{k=0}^{\infty} \frac{(-1)^k}{(2k+1)^2} &=& {\bf C}~.
\end{eqnarray}
%----------------------------------------------------------------------------------------------
At even values of $l$ no new constants appear. Odd values contribute with ${\rm Ti}(2l)$.

For $z = 1/5, 2/5, 3/5, 4/5$ the relations
%----------------------------------------------------------------------------------------------
\begin{eqnarray}
\psi^{(l)}\left(\frac{3}{5}\right) &=& (-1)^l \left[\psi^{(l)}\left(\frac{2}{5}\right)
+ \pi \frac{d^l}{dz^l} \left. \cot(\pi z)\right|_{z=2/5} \right] \\
\psi^{(l)}\left(\frac{4}{5}\right) &=& (-1)^l \left[\psi^{(l)}\left(\frac{1}{5}\right)
+ \pi \frac{d^l}{dz^l} \left. \cot(\pi z)\right|_{z=1/5} \right] \\
l! (-1)^{l+1} \zeta_{l+1} \left[5^{l+1}-1\right]
&=&
\psi^{(l)}\left(\frac{1}{5}\right)
+\psi^{(l)}\left(\frac{2}{5}\right)
+\psi^{(l)}\left(\frac{3}{5}\right)
+\psi^{(l)}\left(\frac{4}{5}\right)
\end{eqnarray}
%----------------------------------------------------------------------------------------------
hold. For even values of $l$, $\psi^{(l)}(2/5)$ is thus dependent on $\psi^{(l)}(1/5)$.

For $z = 1/6, 5/6$ the relation
%----------------------------------------------------------------------------------------------
\begin{eqnarray}
\psi^{(l)}\left(\frac{5}{6}\right) = l! (-1)^{l+1} \zeta_{l+1} \left[6^{l+1} - 3^{l+1}
-2^{l+1} +1 \right] - \psi^{(l)}\left(\frac{1}{6}\right)
\end{eqnarray}
%----------------------------------------------------------------------------------------------
holds. The reflection formula (\ref{eq:REFL}) relates
$\psi^{(l)}(1/6)$ and $\psi^{(l)}(5/6)$ for even values of $l$ to linear combinations
of $\pi^{l+1} \sqrt{3}$ and $\zeta_{2l+1}$.
For odd $l$ the $\psi$-values can be expressed by $\psi^{(l)}\left({1}/{6}\right)$.
The shift-formula by $1/2$
%----------------------------------------------------------------------------------------------
\begin{eqnarray}
\label{eq:shift}
2^{l+1} \psi^{(l)}(2z) =  \psi^{(l)}(z) + \psi^{(l)}\left(z+\frac{1}{2}\right)
\end{eqnarray}
%----------------------------------------------------------------------------------------------
implies
%----------------------------------------------------------------------------------------------
\begin{eqnarray}
\psi^{(l)}\left(\frac{1}{6}\right) =  \left(2^{l+1} + 1 \right)
\psi^{(l)}\left(\frac{1}{3} \right) + (-1)^l l! \left(3^{l+1} - 1\right) \zeta_{l+1}~,
\end{eqnarray}
%----------------------------------------------------------------------------------------------
and relates all values to $\psi^{(l)}(1/3)$.

For $z = 1/8, 3/8, 5/8, 7/8$ one obtains
%----------------------------------------------------------------------------------------------
\begin{eqnarray}
  \psi^{(l)}\left(\frac{1}{8}\right)
+ \psi^{(l)}\left(\frac{3}{8}\right)
+ \psi^{(l)}\left(\frac{5}{8}\right)
+ \psi^{(l)}\left(\frac{7}{8}\right)
= l! (-1)^{l+1} \zeta_{l+1} 4^{l+1} \left( 2^{l+1} - 1 \right)~.
\end{eqnarray}
%----------------------------------------------------------------------------------------------
Due to (\ref{eq:HUR2}) and (\ref{eq:REFL}) the corresponding
$\psi$-values can be expressed by $\psi^{(l)}(1/8)$
%----------------------------------------------------------------------------------------------
\begin{eqnarray}
\psi^{(l)}\left(\frac{3}{8}\right) &=&
2^{l+1} \psi^{(l)}\left(\frac{3}{4}\right) - (-1)^l
\left[\psi^{(l)}\left(\frac{1}{8}\right) + \pi \frac{d^l}{dz^l} \left.
\cot(\pi z)\right|_{z=1/8}\right]
\\
\psi^{(l)}\left(\frac{5}{8}\right) &=&
2^{l+1}
  \psi^{(l)}\left(\frac{1}{4} \right)
- \psi^{(l)}\left(\frac{1}{8} \right)\\
%
%(-1)^l\left[\psi^{(l)}\left(\frac{3}{8}\right)
%                                        + \pi \frac{d^l}{dz^l} \left. \cot(\pi z)
%\right|_{z=3/8}\right] \\
\psi^{(l)}\left(\frac{7}{8}\right) &=&
(-1)^l\left[\psi^{(l)}\left(\frac{1}{8}\right)
                                        + \pi \frac{d^l}{dz^l} \left. \cot(\pi z)
\right|_{z=1/8} \right]~.
\end{eqnarray}
%----------------------------------------------------------------------------------------------

For $z = 1/10, 3/10, 7/10, 9/10$ two reflection relations and
%----------------------------------------------------------------------------------------------
\begin{eqnarray}
\psi^{(l)}\left(\frac{1}{10}\right)
+ \psi^{(l)}\left(\frac{3}{10}\right)
+ \psi^{(l)}\left(\frac{7}{10}\right)
+ \psi^{(l)}\left(\frac{9}{10}\right)
= l! (-1)^{l+1} \zeta_{l+1} \left[10^{l+1} - 5^{l+1} - 2^{l+1} +1 \right]
\nonumber\\
\end{eqnarray}
%----------------------------------------------------------------------------------------------
hold. By the shift relation (\ref{eq:shift}) one obtains
%----------------------------------------------------------------------------------------------
\begin{eqnarray}
\psi^{(l)}\left(\frac{1}{10}\right) &=& 2^{l+1}
\psi^{(l)}\left(\frac{1}{5}\right)
- \psi^{(l)}\left(\frac{3}{5}\right)
\\
\psi^{(l)}\left(\frac{3}{10}\right) &=& 2^{l+1}
\psi^{(l)}\left(\frac{3}{5}\right)
- \psi^{(l)}\left(\frac{4}{5}\right)
\\
\psi^{(l)}\left(\frac{7}{10}\right) &=& 2^{l+1}
\psi^{(l)}\left(\frac{2}{5}\right)
- \psi^{(l)}\left(\frac{1}{5}\right)
\\
\psi^{(l)}\left(\frac{9}{10}\right) &=& 2^{l+1}
\psi^{(l)}\left(\frac{4}{5}\right)
- \psi^{(l)}\left(\frac{2}{5}\right)~.
\end{eqnarray}
%----------------------------------------------------------------------------------------------
No new constants contribute.

For $z = 1/12, 5/12, 7/12, 11/12$ two reflection relations and
%----------------------------------------------------------------------------------------------
\begin{eqnarray}
\label{eq:12A}
  \psi^{(l)}\left(\frac{1}{12}\right)
+ \psi^{(l)}\left(\frac{5}{12}\right)
+ \psi^{(l)}\left(\frac{7}{12}\right)
+ \psi^{(l)}\left(\frac{11}{12}\right)
&=& l! (-1)^{l+1} \zeta_{l+1} 2^{l+1}
\nonumber\\ &&  \times
\left(
  6^{l+1}
- 3^{l+1}
- 2^{l+1}
+ 1 \right)
\end{eqnarray}
%----------------------------------------------------------------------------------------------
hold. All values can be expressed by $\psi^{(l)}(1/12)$.
%----------------------------------------------------------------------------------------------
\begin{eqnarray}
\psi^{(l)}\left(\frac{5}{12}\right) &=&
(-1)^l \left[2^{l+1}
\psi^{(l)}\left(\frac{1}{6}\right) -
\psi^{(l)}\left(\frac{1}{12}\right)
+ \pi \frac{d^l}{dz^l} \left. \cot(\pi z)\right|_{z=7/12} \right]
\\
\psi^{(l)}\left(\frac{7}{12}\right) &=&  2^{l+1} \psi^{(l)} \left(\frac{1}{6}\right)
- \psi^{(l)} \left(\frac{1}{12}\right)
\\
\psi^{(l)}\left(\frac{11}{12}\right) &=&
(-1)^l \left[
\psi^{(l)}\left(\frac{1}{12}\right)
+ \pi \frac{d^l}{dz^l} \left. \cot(\pi z)\right|_{z=1/12} \right],
~~\text{etc.}
\end{eqnarray}
%----------------------------------------------------------------------------------------------
Eq.~(\ref{eq:HUR2}) yields
%----------------------------------------------------------------------------------------------
\begin{eqnarray}
\psi^{(l)}\left(\frac{5}{12}\right) = 3^{l+1} \psi^{(l)}\left(\frac{1}{4}\right)
- \psi^{(l)}\left(\frac{3}{4}\right) - \psi^{(l)}\left(\frac{1}{12}\right)~.
\end{eqnarray}
%----------------------------------------------------------------------------------------------
For odd values of $l$ one obtains
%----------------------------------------------------------------------------------------------
\begin{eqnarray}
\psi^{(l)}\left(\frac{1}{12} \right)= 2^l \psi^{(l)}\left(\frac{1}{6}\right)
+ \frac{1}{2}\left[3^{l+1} \psi^{(l)}\left(\frac{1}{4}\right) -
\psi^{(l)}\left(\frac{3}{4}\right) + \pi \frac{d^l}{dz^l} \left. \cot(\pi
z)\right|_{z=7/12} \right]~.
\end{eqnarray}
%----------------------------------------------------------------------------------------------
Eq.~(\ref{eq:12A}) does not imply a further relation. $\psi^{(l)}(1/12)$
contributes as new constant for even values of $l$.

For even values of $l = 2q$ the Hurwitz $\zeta$-function (\ref{eq:EVINF2}) obeys the representation
\cite{REP1}
%----------------------------------------------------------------------------------------------
\begin{eqnarray}
\label{eq:LEG0}
\sum_{k=0}^{\infty} \frac{1}{(2q k + 2p-1)^n} &=&
\frac{1}{(2q)^n} \zeta_{\rm H}\left(n, \frac{2p-1}{2q}\right)
\nonumber\\ &=&
\frac{1}{q} \sum_{k=1}^q
\left[ C_n\left(\frac{k}{q}\right) \cos \left( \frac{(2p-1) k \pi}{q}\right)
     + S_n\left(\frac{k}{q}\right) \sin \left( \frac{(2p-1) k \pi}{q}\right) \right]~,
\nonumber\\
\end{eqnarray}
%----------------------------------------------------------------------------------------------
where $C_\nu$ and $S_\nu$ are represented by the Legendre $\chi$-function
\cite{LEGENDRE-CHI}
(\ref{eq:LEG1}),
%----------------------------------------------------------------------------------------------
\begin{eqnarray}
\label{eq:LEG1}
\chi_\nu(z) = \frac{1}{2} \left[ \Li_{\nu}(z) -  \Li_{\nu}(-z)\right]
\end{eqnarray}
%----------------------------------------------------------------------------------------------
with~\footnote{We corrected typos in Eq.~(10) of Ref.~\cite{REP1}.}
%----------------------------------------------------------------------------------------------
\begin{eqnarray}
\label{eq:LEG2}
C_\nu(x) &=& {\sf Re} \chi_\nu\left(\exp(i \pi x)\right)  \nonumber\\
S_\nu(x) &=& {\sf Im} \chi_\nu\left(\exp(i \pi x)\right)~.
\end{eqnarray}
%----------------------------------------------------------------------------------------------
For $\nu \in {\mathbb N}$, (\ref{eq:LEG2}) can be represented by Euler polynomials \cite{EULERP}
and powers of $\pi$,
%----------------------------------------------------------------------------------------------
\begin{eqnarray}
C_{2n}\left(\frac{p}{q}\right)   &=& \frac{(-1)^n}{4(2n-1)!} \pi^{2n}
E_{2n-1}\left(\frac{p}{q}\right) \\
S_{2n+1}\left(\frac{p}{q}\right) &=& \frac{(-1)^n}{4(2n)!} \pi^{2n+1}
E_{2n}\left(\frac{p}{q}\right),~p,q \in \mathbb{N}_+, p \leq q,
\end{eqnarray}
%----------------------------------------------------------------------------------------------
with
%----------------------------------------------------------------------------------------------
\begin{eqnarray}
\frac{2 \exp(xt)}{\exp(t)+1} &=& \sum_{n=0}^\infty E_n(x) \frac{t^n}{n!}~.
\end{eqnarray}
%----------------------------------------------------------------------------------------------
Since
%----------------------------------------------------------------------------------------------
\begin{eqnarray}
\psi^{(m-1)}\left(\frac{1}{l}\right) = (-1)^m l^m (m-1)! \sum_{k=0}^\infty
\frac{1}{(lk+1)^m},
\end{eqnarray}
%----------------------------------------------------------------------------------------------
in particular $\psi^{(l)}(1/8)$ and $\psi^{(l)}(1/12), l \in \mathbb{N}_+, l \geq 1$
consist of one term containing an integer power of $\pi$ and a second term $\propto
{\sf Re(Im)}(\chi_{l+1}(r)), r \in \mathbb{Q}$ according to
(\ref{eq:LEG0}, \ref{eq:LEG2}).

In conclusion,  the representation of the single sums of weight {\sf w
$\geq$ 2} and $l \leq 6$ require the additional constants~:
%----------------------------------------------------------------------------------------------
\begin{eqnarray}
\label{eq:num1}
\zeta_{2k+1}, \psi^{(2k+1)}\left(\frac{1}{3}\right), {\rm Ti}_{2k}(1),
\psi^{(k)}\left(\frac{1}{5}\right),
\psi^{(2k+1)}\left(\frac{2}{5}\right),
\psi^{(k)}\left(\frac{1}{8}\right),
\psi^{(2k)}\left(\frac{1}{12}\right)~, k \in \mathbb{N}_+~.
\end{eqnarray}
%----------------------------------------------------------------------------------------------

Finally we would like to comment on a relation in Ramanujan's notebooks
\cite{RAMA} Chapter 9, (11.3), which was claimed to involve a cyclotomic harmonic sum,
%----------------------------------------------------------------------------------------------
\begin{eqnarray}
G(1) &=& \frac{1}{8} \sum_{r=1}^\infty \frac{1}{r^3} \sum_{s=1}^r \frac{1}{2l-1}
\\
H(1) &=&   \frac{\pi}{4} \sum_{r=0}^\infty \frac{(-1)^r}{(4r+1)^3}
         - \frac{\pi}{3 \sqrt{3}} \sum_{r=0}^\infty \frac{1}{(2r+1)^3}~,
\end{eqnarray}
%----------------------------------------------------------------------------------------------
with
%----------------------------------------------------------------------------------------------
\begin{eqnarray}
G(1) &=& H(1)~.
\end{eqnarray}
%----------------------------------------------------------------------------------------------
This relation is unfortunately incorrect, cf. \cite{RAMA}, p.~257.
$G(1)$ may be given by the following representation in multiple zeta values
introducing nested harmonic sums \cite{SITA}
%----------------------------------------------------------------------------------------------
\begin{eqnarray}
\label{eq:G1a}
G(1) &=& \frac{7}{16} \sigma_{3,1} + \frac{1}{2} \sigma_{-3,1} =
- \frac{53}{160} \zeta_2^2
         - \frac{1}{4} \zeta_2 \ln^2(2)
         + \frac{7}{8} \zeta_3 \ln(2)
         + \frac{1}{24} \ln^4(2) + \Li_4\left(\frac{1}{2}\right)
\nonumber\\
     &\simeq& 0.16227193947148339072~
\end{eqnarray}
%----------------------------------------------------------------------------------------------
using the  representations in \cite{MZV}.
At present, no relation between the special constants used in (\ref{eq:G1a})
are known. On the other hand, $H(1)$ is given by
%----------------------------------------------------------------------------------------------
\begin{eqnarray}
H(1) &=&
-\frac{\pi}{2048}\left\{8\left[\pi^3+28\left(1+\frac{8}{9}\sqrt{3}\right)\zeta_3\right]
+\psi^{(2)}\left(\frac{1}{8}\right)\right\}
\nonumber\\
    &\simeq& 0.14402290986880995023~.
\end{eqnarray}
%----------------------------------------------------------------------------------------------
%%%%%%%%%%%%%%%%%%%%%%%%%%%%%%%%%%%%%%%%%%%%%%%%%%%%%%%%%%%%%%%%%%%%%%%%
\subsection{\boldmath Infinite multiple sums at higher depth}
%%%%%%%%%%%%%%%%%%%%%%%%%%%%%%%%%%%%%%%%%%%%%%%%%%%%%%%%%%%%%%%%%%%%%%%%

\vspace{1mm}
\noindent
As a further extension of the infinite cyclotomic harmonic sums \cite{MZV}
we consider the iterated summation of the terms (\ref{eqSel}). The corresponding sums 
diverge if the first indices have the pattern $c_i = 1, s_i = 1$, (\ref{eq:MSU}). However, these 
divergences
can be regulated by polynomials in $\sigma_0$ and cyclotomic harmonic sums, which are convergent for
$N \rightarrow \infty$, very similar to the case of the usual harmonic sums \cite{HS1, HS2}.
We study the
relations given in Section~\ref{sec:4.2} supplemented by those of the shuffle algebra
(SH) of the cyclotomic harmonic polylogarithms (at argument $x = 1$), Eq.~(\ref{eq:CSH}).
In Table~3 we present the number of basis elements obtained applying the
respective relations up to weight {\sf w = 6}. The representation of all sums were
computed by means of computer algebra in explicit form.
We derive the cumulative basis, quoting only the new elements in the next
weight. Up to {\sf w = 4} we derive also suitable integral representations
over known functions.
One possible choice of basis elements is~:
\begin{eqnarray}
\text{\sf w = 1:}
\hspace*{5cm}
&&   \nonumber\\
\sigma_{\{1, 0, 1\}} &=& \sigma_0 \\
\sigma_{\{1, 0, -1\}} &=& -\ln(2) \\
\sigma_{\{2, 1, -1\}} &=& -1 + \frac{\pi}{4}
\end{eqnarray}
%----------------------------------------------------------------------------------------------
\begin{table}
\begin{center}
{\small
\begin{tabular}{| r | r | r | r | r | r | r | r | r|}
\hline
{\sf weight} & $\sf N_S$ & $\sf A$  & $\sf SH$ &  $\sf A + SH $ &  $\sf A+sh+H_1$   &
$\sf A + SH +H_1 + H_2$ & $\sf A + SH + H_1 + H_2 + M$\\
\hline
   1 &     4 &    4 &    4 &   4 &   4 &   3 &   3  \\
   2 &    20 &   10 &   13 &   3 &   3 &   2 &   1  \\
   3 &   100 &   40 &   46 &   6 &   6 &   5 &   3  \\
   4 &   500 &  150 &  163 &  10 &  10 &   9 &   6  \\
   5 &  2500 &  624 &  650 &  21 &  21 &  19 &  13  \\
   6 & 12500 & 2580 & 2635 &  36 &  36 &  34 &  25  \\
 \hline
\end{tabular}
}
\end{center}
\label{TT:3}
\caption[]{\sf Basis representations of the infinite cyclotomic harmonic sums over the
alphabet $\{(\pm 1)^k/k, (\pm 1)^k/(2k+1)\}$ after applying the stuffle (A), shuffle
(SH) relations, their combination, and their application together with the three
multiple argument relations ($\sf H_1, H_2, M$), as far these lead not to new quantities.
In the latter case we quote the cumulative number of basis elements appearing at the new
weight.
} \end{table}
%----------------------------------------------------------------------------------------------

\vspace*{-5mm}
\noindent
%-----------------------------------------------------------------------------------------------
\begin{eqnarray}
%---
\text{\sf w = 2:} \hspace*{5cm}
&&   \nonumber\\
\sigma_{\{2, 1, -2\}} &=& -1 + {\bf C} \\
%---
\text{\sf w = 3:} \hspace*{5cm}
&&   \nonumber\\
\sigma_{\{1, 0, 3\}} &=& \zeta_3 \\
\sigma_{\{1, 0, -2\}, \{2, 1, -1\}} &=&
\frac{\pi^2}{12} -\frac{\pi^3}{48} + \frac{1}{2} \int_0^1 dx \frac{\sqrt{x}}{x+1} \Li_2(x) 
\\
\sigma_{\{2, 1, -2\}, \{1, 0, -1\}} &=& -{\bf C} \ln(2)
+ \int_0^1 dx \frac{1}{1+x} \frac{\chi_2(\sqrt{x})}{\sqrt{x}} \\
\text{\sf w = 4:} \hspace*{5cm}
&&   \nonumber\\
%---
\sigma_{\{1, 0, -1\}, \{1, 0, 1\}, \{1, 0, 1\},\{1, 0, 1\}}
&=& - \Li_4\left(\frac{1}{2}\right)\\
\sigma_{\{2, 1, -4\}} &=& -1 - i \chi_4(i)\\
\sigma_{\{2, 1, -3\}, \{2, 1, -1\}} &=& i\left(1- \frac{\pi}{4}\right) \chi_3(i)
+\frac{1}{2}\int_0^1 dx \frac{1}{x+1} \chi_3(\sqrt{x}) \\
\sigma_{\{1, 0, -2\}, \{1, 0, -1\}, \{2, 1, -1\}} &=&
- \left(\frac{1}{4} \pi^2 \ln(2) - \frac{5}{8}
\zeta_3\right)\left(1-\frac{\pi}{4}\right) \nonumber\\ &&
+ \frac{1}{2} \int_0^1 \frac{\sqrt{x}}{1+x}\Biggl[\left(
\Li_2(-x)+\frac{\pi^2}{12}\right) \ln(1-x)
\nonumber\\ && \hspace*{1.2cm}
+ \frac{1}{2} {\rm S}_{1,2}(x^2)
-{\rm S}_{1,2}(x)
-{\rm S}_{1,2}(-x)\Biggr]
\\
\sigma_{\{1, 0, -2\}, \{2, 1, -1\}, \{1, 0, 1\}} &=&
-\frac{1}{2} \int_0^1 dx \frac{\sqrt{x}}{1+x} \left[ \ln\left(1-\sqrt{x}\right)
-\ln\left(1+\sqrt{x}\right)\right] \Li_2(x) \nonumber\\
&&
- \ln(2)\int_0^1 dx \frac{\sqrt{x}}{1+x} \Li_2(x)
+\frac{\pi^2}{24}\left[\ln(2) \pi -2 {\bf C}\right] \\
\sigma_{\{1, 0, -2\}, \{2, 1, -1\}, \{2, 1, 1\}} &=&
- \frac{1}{4} \int_0^1 dx
\frac{\sqrt{x}}{1+x} \ln(1-x) \Li_2(x) \nonumber\\ 
&&
- \frac{1}{2} \left[1 - \ln(2)\right] \int_0^1 dx
\frac{\sqrt{x}}{1+x} \Li_2(x)
\nonumber
\end{eqnarray}
\begin{eqnarray}
%\\ 
&&
+ \frac{\pi^2}{24}\left[\frac{\pi}{2}\left[1 -\frac{1}{2} \ln(2)\right] - {\bf
C}\right]~.
\end{eqnarray}
%-----------------------------------------------------------------------------------------------
Here $S_{1,2}(x)$ denotes a Nielsen integral \cite{NIELS1},
%-----------------------------------------------------------------------------------------------
\begin{eqnarray}
S_{1,2}(x) = \frac{1}{2}\int_0^x \frac{dz}{z} \ln^2(1-z)~.
\end{eqnarray}
%-----------------------------------------------------------------------------------------------

At weight {\sf w = 5,6} we give only a few integral representations. They can in general be obtained
form the Mellin transforms setting the kernel $x^{N} \rightarrow 1$.
The following basis elements are obtained~:
%-----------------------------------------------------------------------------------------------
\begin{eqnarray}
\text{\sf w = 5:}
\hspace*{5cm}
&&   \nonumber\\
&& \hspace*{-6cm}
\sigma_{\{1, 0, 5\}} = \zeta_5,\\
&& \hspace*{-6cm}
\sigma_{\{1, 0, -1\}, \{1, 0, 1\}, \{1, 0, 1\}, \{1, 0, 1\}, \{1, 0, 1\}} = - \Li_5\left(\frac{1}{2}\right), \\
&& \hspace*{-6cm}
\sigma_{\{1, 0, -4\}, \{2, 1, -1\}} = \frac{7}{720} \pi^4 - \frac{7}{2880} \pi^5  + \frac{1}{2} \int_0^1 dx
\frac{\sqrt{x}}{1+x} \Li_4(x),\\
&& \hspace*{-6cm}
\sigma_{\{1, 0, 4\}, \{2, 1, -1\}} = -\frac{\pi^4}{90} + \frac{\pi^5}{360} + \frac{1}{2} \int_0^1 dx
\frac{\sqrt{x}}{1+x} \Li_4(-x),
\\
&& \hspace*{-6cm}
\sigma_{\{2, 1, -4\}, \{1, 0, -1\}},
\sigma_{\{1, 0, -3\}, \{1, 0, -1\}, \{2, 1, 1\}},
\sigma_{\{1, 0, -3\}, \{2, 1, -1\}, \{2, 1, -1\}},
\sigma_{\{1, 0, 3\}, \{2, 1, -1\}, \{2, 1, -1\}}, \NN\\
&& \hspace*{-6cm}
\sigma_{\{2, 1, -3\}, \{2, 1, -1\}, \{2, 1, 1\}},
\sigma_{\{1, 0, -2\}, \{1, 0, -1\}, \{1, 0, -1\}, \{2, 1, -1\}},
\sigma_{\{1, 0, -2\}, \{1, 0, -1\}, \{2, 1, -1\}, \{1, 0, -1\}}, \NN\\
&& \hspace*{-6cm}
\sigma_{\{1, 0, -2\}, \{2, 1, -1\}, \{1, 0, 1\}, \{1, 0, 1\}},
\sigma_{\{1, 0, -2\}, \{2, 1, -1\}, \{2, 1, 1\}, \{1, 0, 1\}}~.	
\\
%-----------------------------------------------------------------------------------------------
&&   \nonumber
%\\
\end{eqnarray}
\begin{eqnarray}
\text{\sf w = 6:}
\hspace*{5cm}
&&   \nonumber\\
&& \hspace*{-6cm}
\sigma_{\{1, 0, -5\}, \{1, 0, -1\}} = \sigma_{-5,-1},
\\
&& \hspace*{-6cm}
\sigma_{\{1, 0, -1\}, \{1, 0, 1\}, \{1, 0, 1\}, \{1, 0, 1\}, \{1, 0, 1\},\{1, 0, 1\}} = - \Li_6\left(\frac{1}{2}\right),
\\
&& \hspace*{-6cm}
\sigma_{\{2, 1, -6\}} = -1 - i \chi_6(i),
\\
&& \hspace*{-6cm}
\sigma_{\{1, 0, -2\}, \{2, 1, -1\}, \{2, 1, 1\}, \{1, 0, 1\}, \{1, 0, 1\}},
\sigma_{\{1, 0, -2\}, \{2, 1, -1\}, \{1, 0, 1\}, \{1, 0, 1\}, \{1, 0, 1\}},
\nonumber\\
&& \hspace*{-6cm}
\sigma_{\{1, 0, -2\}, \{1, 0, -1\}, \{1, 0, -1\}, \{2, 1, -1\}, \{2, 1, 1\}},
\sigma_{\{1, 0, -2\}, \{1, 0, -1\}, \{1, 0, -1\}, \{2, 1, -1\}, \{1, 0, 1\}},
\nonumber\\
&& \hspace*{-6cm}
\sigma_{\{1, 0, -2\}, \{1, 0, -1\}, \{1, 0, -1\}, \{1, 0, -1\}, \{2, 1, -1\}},
\sigma_{\{1, 0, 3\}, \{2, 1, -1\}, \{2, 1, -1\}, \{1, 0, 1\}},
\nonumber\\
&& \hspace*{-6cm}
\sigma_{\{1, 0, -3\}, \{1, 0, -1\}, \{2, 1, 1\}, \{1, 0, 1\}},
\sigma_{\{1, 0, -3\}, \{1, 0, -1\}, \{1, 0, 1\}, \{2, 1, 1\}},
\sigma_{\{2, 1, -3\}, \{2, 1, -1\}, \{2, 1, -1\}, \{2, 1, -1\}},
\nonumber\\
&& \hspace*{-6cm}
\sigma_{\{1, 0, -3\}, \{2, 1, -1\}, \{2, 1, -1\}, \{1, 0, -1\}},
\sigma_{\{1, 0, -3\}, \{2, 1, -1\}, \{1, 0, -1\}, \{2, 1, -1\}},
\sigma_{\{1, 0, -3\}, \{1, 0, -1\}, \{2, 1, -1\}, \{2, 1, -1\}},
\nonumber\\
&& \hspace*{-6cm}
\sigma_{\{2, 1, 4\}, \{1, 0, -1\}, \{2, 1, -1\}},
\sigma_{\{1, 0, 4\}, \{2, 1, -1\}, \{1, 0, -1\}},
\sigma_{\{1, 0, 4\}, \{1, 0, -1\}, \{2, 1, -1\}},
\nonumber\\
&& \hspace*{-6cm}
\sigma_{\{2, 1, -4\}, \{1, 0, -1\}, \{2, 1, 1\}},
\sigma_{\{2, 1, -4\}, \{1, 0, -1\}, \{1, 0, 1\}},
\sigma_{\{1, 0, -4\}, \{2, 1, -1\}, \{2, 1, 1\}},
\nonumber\\
&& \hspace*{-6cm}
\sigma_{\{1, 0, -4\}, \{2, 1, -1\}, \{1, 0, 1\}},
\sigma_{\{1, 0, -4\}, \{1, 0, -1\}, \{2, 1, -1\}},
\sigma_{\{2, 1, -4\}, \{2, 1, -2\}},
\nonumber\\
&& \hspace*{-6cm}
\sigma_{\{2, 1, -5\}, \{2, 1, -1\}}~.
\end{eqnarray}
%-----------------------------------------------------------------------------------------------

Recently, infinite sums of a type proposed in \cite{JORDAN,MILGRAM}~\footnote{For similar sums see \cite{OSL}.} were
studied in
\cite{LAURENZI}, Eqs.~(11a-c). They can be expressed in terms of sums studied in this Section~:
%----------------------------------------------------------------------------------------------
\begin{eqnarray}
{\cal J}_1(r) &=& \frac{1}{2} \sum_{k=0}^\infty \frac{[\psi(k+1/2) - \psi(1/2)]}{(2k+1)^r}
= \sigma_{\{2,1,r\}} - \sigma_{\{2,1,r+1\}} +\sigma_{\{2,1,r+1\},\{2,1,1\}}
\end{eqnarray}
\begin{eqnarray}
{\cal J}_2(r) &=& \frac{1}{2} \sum_{k=0}^\infty \frac{[\psi(k+1/2) - \psi(1/2)]}{(2k)^r}
= \frac{1}{2^r} \sigma_{\{1,0,r\},\{2,-1,1\}}
\\
{\cal M}(r) &=& \frac{1}{2} \sum_{k=0}^\infty \frac{[\psi(k+1) + \gamma]}{(2k+1)^r}
= \frac{1}{2} \sigma_{\{2,1,r\},\{1,0,1\}}~.
\end{eqnarray}
%----------------------------------------------------------------------------------------------
Moreover, values of usual harmonic polylogarithms \cite{VR} at $x=1$, $\int_0^1 dt [\Li_p(\pm t) - \Li_p(\pm 1)]
/(1 \pm t)$, are discussed. In part these integrals refer to all three letters of the corresponding alphabet.
The corresponding representations
involve infinite harmonic sums of depth {$\sf d > 1$} naturally, as e.g. for
%----------------------------------------------------------------------------------------------
\begin{eqnarray}
\int_0^1 dx \frac{\Li_5(x)}{1+x}
= H_{-1,0,0,0,1}(1) = -\frac{15}{16} \zeta_5 \ln(2) + \sigma_{-5,-1}
\end{eqnarray}
%----------------------------------------------------------------------------------------------
and in similar cases of higher weight, see Ref.~\cite{HS2,MZV,VR}.
%%%%%%%%%%%%%%%%%%%%%%%%%%%%%%%%%%%%%%%%%%%%%%%%%%%%%%%%%%%%%%%%%%%%%%%%%%
\subsection{\boldmath Infinite multiple sums with more cyclotomic letters}
%%%%%%%%%%%%%%%%%%%%%%%%%%%%%%%%%%%%%%%%%%%%%%%%%%%%%%%%%%%%%%%%%%%%%%%%%%

\vspace{1mm}
\noindent
Let us now consider more cyclotomic letters.
We study the sums up to weight {\sf w = 2} and cyclotomy {\sf l = 20}~\footnote{Relations between colored
nested infinite harmonic sums have been investigated also in Refs.~\cite{DELIGNE,DG}
recently.}, based on
the sets of the non-alternating and alternating sums using the letters
%----------------------------------------------------------------------------------------------
\begin{eqnarray} \label{eq:def1} \frac{(\pm 1)^k}{(lk+m)^n},~~~~1
\leq n \leq 2, 1 \leq l \leq 20, m
< l~. \end{eqnarray}
%----------------------------------------------------------------------------------------------
We use the stuffle (quasi-shuffle) relations for the sums, the shuffle relations on the side of the
associated cyclotomic harmonic polylogarithms, and the multiple argument relations for these
sums, cf. Section~\ref{sec:4.2}.
In some
cases the latter request to include sums which are outside the above pattern. In this case the
corresponding relations are not accounted for. At {\sf w = 1} the respective numbers of
basis elements is summarized in Table~4.

%----------------------------------------------------------------------------------------------
\begin{table}[H]\centering
%\begin{tabular}{l*{6}{c}r}
{\small
\begin{tabular}{|r|r|r|r|r|r|r|r|r|r|r|r|r|r|r|r|r|r|r|r|r|}
\hline
\multicolumn{1}{|c}{\sf l}             &
\multicolumn{1}{|c}{1}                 &
\multicolumn{1}{|c}{2}                 &
\multicolumn{1}{|c}{3}                 &
\multicolumn{1}{|c}{4}                 &
\multicolumn{1}{|c}{5}                 &
\multicolumn{1}{|c}{6}                 &
\multicolumn{1}{|c}{7}                 &
\multicolumn{1}{|c}{8}                 &
\multicolumn{1}{|c}{9}                 &
\multicolumn{1}{|c}{10}                &
\multicolumn{1}{|c}{11}                &
\multicolumn{1}{|c}{12}                &
\multicolumn{1}{|c}{13}                &
\multicolumn{1}{|c}{14}                &
\multicolumn{1}{|c}{15}                &
\multicolumn{1}{|c}{16}                &
\multicolumn{1}{|c}{17}                &
\multicolumn{1}{|c}{18}                &
\multicolumn{1}{|c}{19}                &
\multicolumn{1}{|c|}{20}               \\
\hline
{\sf sums} &  2 & 4 & 6 & 8 & 10 & 12 & 14 & 16 & 18 & 20 & 22 & 24 & 26 & 28 & 30 & 32 & 34 & 36 & 38 &
40
\\
\hline
{\sf basis}
& 2 & 3 & 4 & 5 &  6 &  6 &  8 &  9 & 8 & 10 & 12 & 10 &14 & 14 & 11 & 17 & 18 & 14 & 20 & 18\\
\hline
{\sf new basis}
& 2 & 1 & 2 & 2 &  4 &  1 &  6 &  4 & 4 & 3 & 10 & 2 &12 & 5 & 3 & 8 & 16 & 4 & 18 & 6\\
{\sf sums}
& & & & & & & & & &
& & & & & & & & & & \\
\hline
\end{tabular}
}
\label{TT:1a}
\caption[]{\sf The number of the {\sf w = 1} cyclotomic harmonic sums (\ref{eq:def1}) up to {\sf l =
20}, the basis elements at fixed value of {\sf l}, and the new basis elements in ascending sequence.}
\end{table}
%----------------------------------------------------------------------------------------------

\noindent
The reflection relation (\ref{eq:REFL}) for the $\psi$-functions
for $x \leftrightarrow (1-x)$ implies that there are at most $l+1$ basis elements. We showed that the use 
of the above analytic 
representations and the shuffle, stuffle, and multiple argument relations lead to the same number of basis elements 
for $l \leq 6$ in the non-alternating and alternating case.

The independent sums at {\sf w = 1} up to {\sf l = 6} are~:
%----------------------------------------------------------------------------------------------
\begin{eqnarray}
&&\sigma_{\{1, 0, 1\}},
\sigma_{\{1, 0, -1\}},
\sigma_{\{2, 1, -1\}},
\sigma_{\{3, 1,  1\}},
\sigma_{\{3, 1, -1\}},
\sigma_{\{4, 1, -1\}},
\sigma_{\{4, 3, -1\}},
\sigma_{\{5, 1,  1\}},
\sigma_{\{5, 1, -1\}},
\nonumber\\ &&
\sigma_{\{5, 2, -1\}},
\sigma_{\{5, 3, -1\}},
\sigma_{\{6, 1, -1\}}~,
\label{eq:basw1}
\end{eqnarray}
%----------------------------------------------------------------------------------------------
see Sections~\ref{sec:5.1.1}, \ref{sec:5.1.2} for equivalent representations.

The dependent sums up to {\sf l = 6} are
%----------------------------------------------------------------------------------------------
\begin{eqnarray}
\label{Equ:Sigma1}
\sigma_{\{2, 1, 1\}} &=& -1 - \sigma_{\{1, 0, -1\}} + \frac{1}{2} \sigma_{\{1, 0, 1\}}
\\
%----
\sigma_{\{3, 2, 1\}} &=& - \frac{1}{2} - \frac{1}{3}\sigma_{\{1, 0, -1\}}
                         - \sigma_{\{3, 1, -1\}} + \sigma_{\{3, 1, 1\}}
\\
\sigma_{\{3, 2, -1\}} &=&   \frac{1}{2} + \frac{2}{3} \sigma_{\{1, 0, -1\}}
                         + \sigma_{\{3, 1, -1\}}
\\
%----
\sigma_{\{4, 1, 1\}} &=&
   - \frac{1}{2} - \frac{3}{4} \sigma_{\{1, 0, -1\}} + \frac{1}{4} \sigma_{\{1, 0, 1\}}
   + \frac{1}{2} \sigma_{\{2, 1, -1\}}
\\
\sigma_{\{4, 3, 1\}} &=&
-\frac{5}{6} - \frac{3}{4} \sigma_{\{1, 0, -1\}} + \frac{1}{4} \sigma_{\{1, 0, 1\}}
- \frac{1}{2} \sigma_{\{2, 1, -1\}}
\\
%----
\sigma_{\{5, 2, 1\}} &=&
\frac{1}{5} \sigma_{\{1, 0, -1\}}
+ \sigma_{\{5, 1, 1\}}
- \sigma_{\{5, 2, -1\}}
\\
\sigma_{\{5, 3, 1\}} &=&
- \frac{1}{3}
- \frac{1}{5} \sigma_{\{1, 0, -1\}}
- \sigma_{\{5, 1, -1}\}
+ \sigma_{\{5, 1, 1\}}
\\
\sigma_{\{5, 4, 1\}} &=& -\frac{7}{12}
- \frac{2}{5} \sigma_{\{1, 0, -1\}}
- \sigma_{\{5, 1, -1\}}
+ \sigma_{\{5, 1, 1\}}
- \sigma_{\{5, 3, -1\}}
\\
\sigma_{\{5, 4, -1}\} &=& \frac{7}{12}
+ \frac{4}{5} \sigma_{\{1, 0, -1\}}
+ \sigma_{\{5, 1, -1\}}
- \sigma_{\{5, 2, -1\}}
+ \sigma_{\{5, 3, -1\}}
\\
%----
\sigma_{\{6, 1, 1\}} &=&
   - \frac{1}{6} \sigma_{\{1, 0, - 1\}}
   + \frac{1}{2} \sigma_{\{3, 1, -1\}}
   + \frac{1}{2} \sigma_{\{3, 1, 1\}}
\\
\sigma_{\{6, 5, 1\}} &=&
- \frac{7}{10} - \frac{2}{3} \sigma_{\{1, 0, -1}\}
- \sigma_{\{3, 1, -1\}}
+ \frac{1}{2} \sigma_{\{3, 1, 1\}}
\\
\sigma_{\{6, 5, -1\}} &=&
\frac{2}{15}
+ \frac{4}{3} \sigma_{\{2, 1, -1\}}
- \sigma_{\{6, 1, -1\}},~\text{etc.}
\label{Equ:Sigma2}
\end{eqnarray}
%----------------------------------------------------------------------------------------------
The remaining sums are related to those given in (\ref{eq:basw1}--\ref{Equ:Sigma2}).
The following counting relations for the basis elements were tested up to {\sf l = 700} using computer algebra
methods.
Let $p,p_i,q$ be pairwise distinct primes $>2,$ and let $k,k_i$ be positive integers.
The number of basis elements at {\sf w = 1} and cyclotomy {\sf l} are given by
%------------------------------------------------------------------------------------------------------------
\begin{eqnarray}
\varphi(l)&=&\left\{
		  	\begin{array}{lll}
						l+1,\  & l=1 \text{ or } l=2^k\\
						(p-1)p^{k-1}+2,\  & l=p^k\\
						2 \varphi\left(2^{k-1}\prod_{i=1}^np_i^{k_i}\right)-n-1,\  & l=2^k\prod_{i=1}^np_i^{k_i}\\
						(q-1) \varphi\left(\prod_{i=1}^np_i^{k_i}\right)-n(q-2)-q+3,\  & l=q\prod_{i=1}^np_i^{k_i}\\
						q\; \varphi\left(q^{k-1}\prod_{i=1}^np_i^{k_i}\right)
-(n+2)(q-1),\  & l=q^k\prod_{i=1}^np_i^{k_i},\ k>1~.\\
			\end{array}
		\right.
\end{eqnarray}
%------------------------------------------------------------------------------------------------------------

Let us now consider the case {\sf w = 2}. Applying the relations given in
Section~\ref{sec:4.2} and the shuffle algebra of the cyclotomic harmonic polylogarithms
at argument $x=1$ the results given in Table~5 are obtained for the number of basis
elements. Again we solved the corresponding linear systems using computer algebra methods
and derived the representations for the dependent sums analytically.
%----------------------------------------------------------------------------------------------
\begin{table}[h]\centering
\begin{tabular}{|r|r|r|r|r|r|r|r|r|}
\hline
\multicolumn{1}{|c}{\sf l }                &
\multicolumn{1}{|c}{$\sf N_S$}              &
\multicolumn{1}{|c}{\sf SH}                &
\multicolumn{1}{|c}{\sf A}                &
\multicolumn{1}{|c}{\sf A + SH}           &
\multicolumn{1}{|c}{$\sf A + SH + H_1$}      &
\multicolumn{1}{|c}{$\sf A  + SH + H_1 +H_2$}      &
\multicolumn{1}{|c|}{$\sf A + SH + H_1 + H_2 +M$}\\
\hline
  1 &   6 &   4 &   3 &   1 &   1 &   1 &    1 \\
  2 &  20 &  13 &  10 &   3 &   3 &   2 &    1 \\
  3 &  42 &  27 &  21 &   7 &   6 &   6 &    5 \\
  4 &  72 &  46 &  36 &  12 &  11 &  10 &    3 \\
  5 & 110 &  70 &  55 &  19 &  17 &  17 &   16 \\
  6 & 156 &  99 &  78 &  27 &  25 &  24 &    5 \\
  7 & 210 & 133 & 105 &  37 &  34 &  34 &   33 \\
  8 & 272 & 172 & 136 &  48 &  45 &  44 &   12 \\
  9 & 342 & 216 & 171 &  61 &  57 &  57 &   52 \\
 10 & 420 & 265 & 210 &  75 &  71 &  70 &   22 \\
 11 & 506 & 319 & 253 &  91 &  86 &  86 &   85 \\
 12 & 600 & 378 & 300 & 108 & 103 & 102 &   21 \\
 13 & 702 & 442 & 351 & 127 & 121 & 121 &  120 \\
 14 & 812 & 551 & 406 & 147 & 141 & 140 &   49 \\
 15 & 930 & 585 & 465 & 169 & 162 & 162 &  145 \\
 16 &1056 & 664 & 528 & 192 & 185 & 184 &   50 \\
 17 &1190 & 748 & 595 & 217 & 209 & 209 &  208 \\
 18 &1332 & 837 & 666 & 243 & 235 & 234 &   63 \\
 19 &1482 & 931 & 741 & 271 & 262 & 262 &  261 \\
 20 &1640 &1030 & 820 & 300 & 291 & 290 &   74 \\
\hline
\end{tabular}
\label{TT:2}
\caption[]{\sf Number of basis elements of the {\sf w = 2} cyclotomic harmonic sums
(\ref{eq:def1}) up to cyclotomy {\sf l = 20} after applying the quasi-shuffle algebra of
the sums (A), the shuffle algebra of the cyclotomic harmonic polylogarithms (SH), and the
three multiple argument relations ($\sf H_1, H_2, M$) of the sums.}
\end{table}

%----------------------------------------------------------------------------------------------

The number of the weight {\sf w = 2} infinite sums for cyclotomy {\sf l} is
%------------------------------------------------------------------------------------------------------------
\begin{eqnarray}
{\sf N_S} = 2l (2l+1)~.
\end{eqnarray}
%------------------------------------------------------------------------------------------------------------
One may guess, based on the results for {$\sf l \leq 20$},  counting relations for the
length of the bases listed in Table~5. We found for all but the last column~:
%------------------------------------------------------------------------------------------------------------
\begin{eqnarray}
{\sf N_A}(l)  &=& l(2l+1)
\end{eqnarray}
\begin{eqnarray}
%\\
{\sf N_{SH}}(l)  &=& \frac{(5l+3)l}{2}
\\
{\sf N_{A,SH}}(l) &=& \frac{6l^2+1-(-1)^l}{8}
\\
{\sf N_{A,SH,H_1}}(l) &=& \frac{6l^2-4l+7-(-1)^l}{8}
\\
\label{eq:BROAD}
{\sf N_{A,SH,H_1,H_2}}(l) &=& \frac{6l^2-4l+3(1-(-1)^l)}{8}~.
\end{eqnarray}
%----------------------------------------------------------------------------------------------
The latter relation (\ref{eq:BROAD}) has been derived prior to us by D. Broadhurst~\footnote{We would like to
thank D. Broadhurst for communicating this relation to us.}~:
%------------------------------------------------------------------------------------------------------------
\begin{eqnarray}
\label{eq:DB1}
{\sf N_{A,SH,H_1,H_2}}(l)
 = \frac{3}{4} l^2  - \frac{1}{2} l + \text{if(modp(l,2)}=0, 1, 3/4)
\end{eqnarray}
%------------------------------------------------------------------------------------------------------------
and the corresponding generating function
%------------------------------------------------------------------------------------------------------------
\begin{eqnarray}
\label{eq:DB2}
f(x) = \left[1 + \frac{x^3}{1+x} \right] \frac{1}{(1-x)^3} = \sum_{l=0}^\infty a(l) x^l~.
\end{eqnarray}
%------------------------------------------------------------------------------------------------------------
We conjecture that in case of $\sf N_{A,SH,H_1,H_2}(l)$ the $(M)$-relations lead to
a reduction of one in the basis for {\sf l} being a prime. Otherwise quite
significant reductions are obtained for which we do not know an explicit counting
relation. The corresponding sequence is also not recorded yet in the data base \cite{SLOAN}.
%%%%%%%%%%%%%%%%%%%%%%%%%%%%%%%%%%%%%%%%%%%%%%%%%%%%%%%%%%%%%%%%%%%%%%%%%%%%%%%%%%%%%%%%%%%%%%%
\section{Generalized Harmonic Sums at Roots of Unity}
\label{sec:6}

\vspace{1mm}
\noindent
%%%%%%%%%%%%%%%%%%%%%%%%%%%%%%%%%%%%%%%%%%%%%%%%%%%%%%%%%%%%%%%%%%%%%%%%%%%%%%%%%%%%%%%%%%%%%%%
In Section~\ref{sec:5} we considered real representations for the infinite cyclotomic harmonic 
sums. These are related to the infinite generalized harmonic sums at the roots of unity.
We define
%--------------------------------------------------------------------------------
\begin{eqnarray}
\lim_{N \rightarrow \infty} S_{k_1,...,k_m}(x_1, ..., x_m; N)
= \sigma_{k_1,...,k_m}(x_1, ..., x_m),
\end{eqnarray}
%--------------------------------------------------------------------------------
with $S_{k_1,...,k_m}(x_1, ..., x_m; N)$ a generalized harmonic sum (\ref{eq:GENSU}), see also 
\cite{GHS1,GHS2}, and $x_j 
\in 
{\cal C}_n, n \geq 1$, with
%--------------------------------------------------------------------------------
${\cal C}_n  \in \left \{e_n| e_n^n = 1, e_n \in \mathbb{C} \right\}$; $k_1 \neq 1 $ for $x_1 = 1$.
%--------------------------------------------------------------------------------

We seek the relations between the sums of {\sf w = 1,2}. They can be expressed in terms
of polylogarithms by~: 
%--------------------------------------------------------------------------------
\begin{eqnarray}
\sigma_{\sf w}(x) &=& \Li_{\sf w}(x),~~~~~\text{\sf w $\in \mathbb{N}$, w
$\geq$ 1} \\
\sigma_1(x) =\Li_1(x) &=& - \ln(1-x)\\ 
\sigma_{1,1}(x,y) 
&=& \Li_2(x) + \frac{1}{2} \ln^2(1-x)
+ \Li_2\left(- \frac{x(1-y)}{1-x}\right)\\  
\sigma_{1,1}(x,x^*) &=& \Li_2(x) + \frac{1}{2} \ln^2(1-x)\\
\Li_{\sf w}(x)  &=& \Li_{\sf w}^*(x^*)
\end{eqnarray}
%--------------------------------------------------------------------------------
and $^*$ denotes complex conjugation.
%--------------------------------------------------------------------------------
Furthermore,
the symmetric combination $\sigma_{1,1}(x,y) + \sigma_{1,1}(y,x)$ is given by \cite{GHS1,GHS2}
%--------------------------------------------------------------------------------
\begin{eqnarray}
\label{eq:SIGSYM}
\sigma_{1,1}(x,y) + \sigma_{1,1}(y,x) = \ln(1-x) \ln(1-y) +\Li_2(xy)~.
\end{eqnarray}
%--------------------------------------------------------------------------------
Knowing the representations at {\sf w = 1} and the dilogarithms at the corresponding roots of unity
one term $\sigma_{1,1}(y,x)$ may be expressed by (\ref{eq:SIGSYM}).

In analyzing the functions $\sigma_{k_1,...,k_m}(x_1, ..., x_m)$ with $x_k^l = 1$ one may use the 
real representations of nested cyclotomic sums for $N \rightarrow \infty$,
accounting for the respective sub-cycles at $d,~d|l$ and complex conjugation.

An example is given by
%--------------------------------------------------------------------------------
\begin{eqnarray}
\left\{\left. e_{12}^k \right|_{k=1}^{12} \right\} \equiv \left\{
e_{12}, e_6, e_4, e_3, e_{12}^5, e_2, {e_{12}^5}^*, e_3^*, e_4^*, e_6^*, e_{12}^*, 
e_1\right\}~.
\end{eqnarray}
%--------------------------------------------------------------------------------
Here we labeled the elements occurring in sub-cycles accordingly.

The polylogarithms $\Li_1(e_l^k)$ and $\Li_2(e_l^k)$ obey~: 
%--------------------------------------------------------------------------------
\begin{eqnarray}
\label{eq:IM1}
{\sf Im}\left[\Li_1(e_l^k)\right] &=&  \frac{l-2k}{2l} \pi \\
%------
\label{eq:RE1}
{\sf Re}\left[\Li_2(e_l^k)\right] &=& \frac{6k(k-l)+l^2}{6 l^2} \pi^2~.
\end{eqnarray}
%--------------------------------------------------------------------------------
More generally, {\tt PSLQ} \cite{PSLQ} tests let conjecture that 
${\sf Im}\left[\Li_n(e_l^k)\right] = r_{n,l,k} \pi^n~\text{for}~n~\text{odd}$ and
${\sf Re}\left[\Li_n(e_l^k)\right] = r_{n,l,k} \pi^n~\text{for}~n~\text{odd}$ with
$r_{n,l,k} \in \mathbb{Q}$.

Now we extend Proposition 2.3 of Ref.~\cite{GON1}, where we  consider generalized harmonic sums
$S_{k_1,...,k_m}(x_1, ..., x_m; N)$ with  $N \in \mathbb{N}, k_i \in \mathbb{N}_+, 
x_i \in 
\mathbb{C}, |x_i| \leq 1$. 
%--------------------------------------------------------------------------------
Let $l \in \mathbb{N}_+$ and 
%--------------------------------------------------------------------------------
\begin{eqnarray}
\label{eq:POW}
y_i^{l} = x_i 
\end{eqnarray}
%--------------------------------------------------------------------------------
then
%--------------------------------------------------------------------------------
\begin{eqnarray}
\label{eq:DISTR}
S_{k_1,...,k_m}(x_1, ..., x_m; N) = \prod_{j=1}^m l^{k_j-1} \sum_{y_i^l = x_i} 
S_{k_1,...,k_m}(y_1, ..., y_m; l N)~.
\end{eqnarray}
%--------------------------------------------------------------------------------
Here the sum is over the $l$th roots of $x_i$ for $i \in [1,m]$.
Eq.~(\ref{eq:DISTR}) is called {\sf Distribution  Relation}. It  follows from the Vieta's theorem 
\cite{VIETA} for (\ref{eq:POW}) and properties of symmetric polynomials \cite{ALG1}. 
Eq.~(\ref{eq:DISTR}) contains the well--known duplication relation, cf. Eq.~(2.15) \cite{MZV}.

If also $x_1 \neq 1$ for $k_1 = 1$ the limit
%--------------------------------------------------------------------------------
\begin{eqnarray}
\label{eq:DISTR1}
\sigma_{k_1,...,k_m}(x_1, ..., x_m) 
= \lim_{N \rightarrow \infty}
S_{k_1,...,k_m}(x_1, ..., x_m;  N)
\end{eqnarray}
%--------------------------------------------------------------------------------
exists.
One may apply (\ref{eq:DISTR},\ref{eq:DISTR1}) to roots of unity $x_i$ and $y_j$, 
i.e. $x_i = \exp(2 \pi i n_i/m_i)$ and $y_{jk} = \exp(2 \pi i k n_i/(m_i l)), k = 
1 ... (l-1), n_i, m_i \in \mathbb{N}_+$. 
%-----------------------------------------------------------------------------------------
Let us now consider the cases {\sf w = 1, 2 } in more detail. 

%%%%%%%%%%%%%%%%%%%%%%%%%%%%%%%%%%%%%%%%%%%%%%%%%%%%%%%%%%%%%%%%%%%%%%%%%%%%%%%%%%%%%%%%%%%%%%%
\subsection{\sf w = 1}

\vspace*{1mm}
\noindent
%%%%%%%%%%%%%%%%%%%%%%%%%%%%%%%%%%%%%%%%%%%%%%%%%%%%%%%%%%%%%%%%%%%%%%%%%%%%%%%%%%%%%%%%%%%%%%%
The first element is real and occurs at {\sf l = 2} 
%--------------------------------------------------------------------------------
\begin{eqnarray}
\Li_1(e_2)                 &=& - \ln(2),
\end{eqnarray}
%--------------------------------------------------------------------------------
representing the simplest alternating multiple zeta value, cf. e.g. \cite{MZV}.
At {\sf l = 3} we get the complex conjugate numbers 
%--------------------------------------------------------------------------------
\begin{eqnarray}
\Li_1(e_3)  &=& 
- \frac{1}{2} \ln(3) + \frac{\pi i}{6} \\ \Li_1(e_3^2)  &=& 
- \frac{1}{2} \ln(3) - \frac{\pi i}{6}~. 
\end{eqnarray}
%--------------------------------------------------------------------------------
Due to (\ref{eq:IM1}), $\ln(3)$ and $i \pi$ are considered as basis elements from this
level on. For all higher values of {\sf l} one thus needs only to consider the real part
of the {\sf w = 1} sums and one may work with the real representations given in Section~\ref{sec:5}.
Let us consider the example
%--------------------------------------------------------------------------------
\begin{eqnarray}
\Li_1(e_4) &=& - \ln(1 - e_4) \nonumber\\
           &=& 
  e_4   \sum_{k=1}^\infty \frac{1}{4k-3}
+ e_2   \sum_{k=1}^\infty \frac{1}{4k-2}
+ e_4^* \sum_{k=1}^\infty \frac{1}{4k-1}
+ e_1   \sum_{k=1}^\infty \frac{1}{4k} \nonumber\\
\label{eq:4A}
           &=& 
  e_4 \sum_{k=1}^\infty \left( \frac{1}{4k-3} - \frac{1}{4k}\right)
+ e_2 \sum_{k=1}^\infty \left( \frac{1}{4k-2} - \frac{1}{4k}\right)
+ e_4^* \sum_{k=1}^\infty \left( \frac{1}{4k-1} - \frac{1}{4k}\right)~.
\nonumber\\
\end{eqnarray}
%--------------------------------------------------------------------------------
Eq.~(\ref{eq:4A}) follows from
%--------------------------------------------------------------------------------
\begin{eqnarray}
\sum_{k=1}^{n-1} e_n^k = 0~.
\end{eqnarray}
%--------------------------------------------------------------------------------
The type of sums occurring in (\ref{eq:4A}) leads to digamma-functions and one may use
their relations given before to find the corresponding basis representations. The first terms are given 
by~:
%--------------------------------------------------------------------------------
\begin{eqnarray}
\Li_1(e_4)                 &=& - \frac{1}{2} \ln(2) + \frac{\pi i}{4} \\
\Li_1(e_4^3) &=& \Li_1^*(e_4) \\
\Li_1(e_5)    &=& \frac{1}{2} \ln\left(\frac{\sqrt{5}+1}{2}\right) - \frac{1}{4} \ln(5)    
+ i  \frac{3}{10}  \pi %\\ 
\end{eqnarray}
\begin{eqnarray}
\Li_1(e_5^2)  &=& \frac{1}{2} \ln\left(\frac{\sqrt{5}-1}{2}\right)  -\frac{1}{4} \ln(5)  
+ i \frac{1}{10} \pi  
\\
{\sf Re}(\Li_1(e_5))  +
{\sf Re}(\Li_1(e_5^2)) 
&=& - \frac{1}{2} \ln(5) 
\\
\Li_1(e_6)    &=&     \frac{\pi i}{3} \\
{\sf Re}(\Li_1(e_8))    &=&     -\frac{1}{4} \ln(2) -\frac{1}{2} \ln(\sqrt{2}-1)\\
{\sf Re}(\Li_1(e_{12}))    &=&     \frac{1}{2} \ln(2) - \ln(\sqrt{3}-1)~. 
\end{eqnarray}
%--------------------------------------------------------------------------------

\noindent
In Table~6 we summarize the number of basis elements.
%-----------------------------------------------------------------------------------------
\begin{table}[H]\centering
%\begin{tabular}{l*{6}{c}r}
{\small
\begin{tabular}{|c|r|r|r|r|r|r|r|r|r|r|r|r|r|r|r|r|r|r|r|r|}
\hline
\multicolumn{1}{|c}{\sf l}             &
\multicolumn{1}{|c}{1}                 &
\multicolumn{1}{|c}{2}                 &
\multicolumn{1}{|c}{3}                 &
\multicolumn{1}{|c}{4}                 &
\multicolumn{1}{|c}{5}                 &
\multicolumn{1}{|c}{6}                 &
\multicolumn{1}{|c}{7}                 &
\multicolumn{1}{|c}{8}                 &
\multicolumn{1}{|c}{9}                 &
\multicolumn{1}{|c}{10}                &
\multicolumn{1}{|c}{11}                &
\multicolumn{1}{|c}{12}                &
\multicolumn{1}{|c}{13}                &
\multicolumn{1}{|c}{14}                &

\multicolumn{1}{|c}{15}                &
\multicolumn{1}{|c}{16}                &
\multicolumn{1}{|c}{17}                &
\multicolumn{1}{|c}{18}                &
\multicolumn{1}{|c}{19}                &
\multicolumn{1}{|c|}{20}               \\
\hline
%{\sf w = 1} 
%&   &  &  &  &  &  &  &  &  &  &  &  &  &  &  &  &  &  & & \\
{\sf basis} 
&  0  & 1  & 2  & 2  & 3  & 3  & 4  & 3 & 4 & 4 &  6 & 4 & 7 & 5 & 6 & 5 & 9 & 5 & 10 & 6  
\\
{\sf Ref.~\cite{RAC}} 
&  0  & 1  & 2  & 2  & 3  & 3  & 4  & 3 & 5 & 4 &  6  &  &  &  &  &  &  &  &  &  
\\
{\sf new} 
&   &  &  &  &  &  &  &  &  &  &  &  &  &  &  &  &  &  & & \\
{\sf elements} 
&  0  & 1  & 2  & 0  & 2  & 0  & 3  & 1 & 2 & 0 &  5 & 1 & 6 & 0 & 2 & 2 & 8 & 0 & 9 & 2 
\\
\hline
\end{tabular}
}
\label{TT:XXa}
\caption[]{\sf The number of the basis elements spanning the {\sf w=1} cyclotomic harmonic polylogarithms at  
$l$th roots of unity  up to {\sf  20}.}.
\end{table}
\noindent
At cyclotomy {\sf l = 9} we find one basis element less than reported in \cite{RAC}.
The new elements contributing at the respective level of cyclotomy
for {\sf l $\leq$ 20} are~:
%--------------------------------------------------------------------------------
\begin{eqnarray}
{\sf l = 2}   & & \ln(2) \\
{\sf l = 3}   & & \ln(3), \pi  \\
{\sf l = 4}   & & -  \\
{\sf l = 5}   & &{\sf Re}(\Li_1(e_{5})),  {\sf Re}(\Li_1(e_{5}^2))
   \\
{\sf l = 6}   & & -  \\
{\sf l = 7}   & & \left. {\sf Re}(\Li_1(e_7^k))\right|_{k=1}^3\\
{\sf l = 8}   & & {\sf Re}(\Li_1(e_8)) \\
{\sf l = 9}   & &  {\sf Re}(\Li_1(e_{9})),
{\sf Re}(\Li_1(e_{9}^2)) \\
{\sf l = 10}  & & -  \\
{\sf l = 11}  & & \left.
{\sf Re}(\Li_1(e_{11}^k))\right|_{k=1}^5\\  
{\sf l = 12}  & & {\sf Re}(\Li_1(e_{12}))\\
{\sf l = 13}  & & \left.
{\sf Re}(\Li_1(e_{13}))\right|_{k=1}^6 \\  
{\sf l = 14}   & & - \\
{\sf l = 15}   & &  {\sf Re}(\Li_1(e_{15})),  {\sf Re}(\Li_1(e_{15}^2))  
\\
{\sf l = 16}   & & {\sf Re}(\Li_1(e_{16})),  {\sf Re}(\Li_1(e_{16}^3))\\
%\end{eqnarray}
%\begin{eqnarray}
{\sf l = 17}  & & \left.
{\sf Re}(\Li_1(e_{17}))\right|_{k=1}^8\\  
{\sf l = 18}   & & - \\
{\sf l = 19}  & & \left.
{\sf Re}(\Li_1(e_{19}))\right|_{k=1}^9  \\
{\sf l = 20}   & & {\sf Re}(\Li_1(e_{20})),  {\sf Re}(\Li_1(e_{20}^3))~.
\end{eqnarray}
%--------------------------------------------------------------------------------

\noindent
%%%%%%%%%%%%%%%%%%%%%%%%%%%%%%%%%%%%%%%%%%%%%%%%%%%%%%%%%%%%%%%%%%%%%%%%%%%%%%%%%%%%%%%%%%%%%%%
\subsection{\sf w = 2}

\vspace*{1mm}
\noindent
%%%%%%%%%%%%%%%%%%%%%%%%%%%%%%%%%%%%%%%%%%%%%%%%%%%%%%%%%%%%%%%%%%%%%%%%%%%%%%%%%%%%%%%%%%%%%%%

We first consider the relations for $\Li_2(e_n^k)$. The following well--known representations 
for the function  holds, cf.~\cite{Newman, KIR}~:
%--------------------------------------------------------------------------------
\begin{eqnarray}
\Li_2\left(e^{i\theta}\right) = \pi^2 \bar{B}_2\left(\frac{\theta}{2\pi}\right) +
i {\rm Cl}_2(\theta)~,
\end{eqnarray}
%--------------------------------------------------------------------------------
with 
%--------------------------------------------------------------------------------
\begin{eqnarray}
{\rm Cl}_2(\theta) &=& \sum_{k=1}^\infty
\frac{\sin(k \theta)}{k^2} \\
\bar{B}_2(x) &=& - \frac{1}{\pi^2} \sum_{k=1}^\infty 
\frac{\cos(2 \pi k x)}{k^2}~.
\end{eqnarray}
%--------------------------------------------------------------------------------
$\bar{B}_2$ denotes the second modified Bernoulli polynomial. Due to (\ref{eq:RE1})
only the imaginary parts have to be considered. The number $\pi$ occurs at {\sf w = 1,
l = 3} only. The first terms are given by~:
%--------------------------------------------------------------------------------
\begin{eqnarray}
\Li_2(e_1) &=&  \frac{\pi^2}{6} \\
\label{eq:LI2a}
\Li_2(e_2) &=&  - \frac{\pi^2}{12} \\
{\sf Im}(\Li_2(e_3)) &=&  \frac{\sqrt{3}}{9} \left[ \psi^{(1)}\left(\frac{1}{3}\right) - \frac{2}{3} \pi^2\right]\\
{\sf Im}(\Li_2(e_4)) &=&   {\bf C} 
%\end{eqnarray}
%\begin{eqnarray}
\\
\Li_2(e_4^3) &=& \Li_2^*(e_4) \\
{\sf Im}(\Li_2(e_5)) &=&  {5 \sqrt{10}}\Biggl\{
\sqrt{1+\frac{1}{\sqrt{5}}} \left[\psi^{(1)}\left(\frac{1}{5}\right) - 
\pi^2\left(1+\frac{1}{\sqrt{5}}\right)\right] \nonumber\\ &&
\hspace*{1cm}
+
\sqrt{1-\frac{1}{\sqrt{5}}} \left[\psi^{(1)}\left(\frac{2}{5}\right) - 
\pi^2\left(1-\frac{1}{\sqrt{5}}\right)\right]\Biggr\} 
\end{eqnarray}
\begin{eqnarray}
%\\
{\sf Im}(\Li_2(e_5^2)) &=& \frac{i}{5 
\sqrt{10}}\Biggl\{
\sqrt{1-\frac{1}{\sqrt{5}}} \left[\psi^{(1)}\left(\frac{1}{5}\right) - 
\pi^2\left(1+\frac{1}{\sqrt{5}}\right)\right] \nonumber\\ &&
\hspace*{1.1cm}
-
\sqrt{1+\frac{1}{\sqrt{5}}} \left[\psi^{(1)}\left(\frac{2}{5}\right) - 
\pi^2\left(1-\frac{1}{\sqrt{5}}\right)\right]\Biggr\} 
\\
\Li_2(e_5^3) &=& \Li_2^*({e_5^2})\\
\Li_2(e_5^4) &=& \Li_2^*(e_5) \\
{\sf Im}(\Li_2(e_6)) &=&  \frac{3}{2} 
{\sf Im}(\Li_2(e_3)) \\
\Li_2(e_6^5) &=& \Li_2^*(e_6) \\ 
{\sf Im}\Li_2(e_8) &=& \frac{\sqrt{2}}{32} \psi'\left(\frac{1}{8}\right) + \frac{1}{4} \left(1 - 2 \sqrt{2}\right)
{\bf C}
-\frac{1}{16} \left( 1 + \sqrt{2}\right) \pi^2 \\ 
{\sf Im}\Li_2(e_{12}) &=& \frac{\sqrt{3}}{24} \psi'\left(\frac{1}{3}\right) + \frac{2}{3} {\bf C} - 
\frac{\sqrt{3}}{36} \pi^2~.
\end{eqnarray}
%--------------------------------------------------------------------------------
The new basis elements spanning the dilogarithms of the $l$th roots of unity for {\sf l $\leq$ 20} are~:
%--------------------------------------------------------------------------------
\begin{eqnarray}
{\sf l = 1,2} & & - \\
{\sf l = 3}   & & {\sf Im}(\Li_2(e_3)) \\
{\sf l = 4}   & & {\bf C} \\
{\sf l = 5}   & &{\sf Im}(\Li_2(e_{5})),  {\sf Im}(\Li_2(e_{5}^2))\\
{\sf l = 6}   & & -  \\
{\sf l = 7}   & & \left. {\sf Im}(\Li_2(e_7^k))\right|_{k=1}^3\\
{\sf l = 8}   & & {\sf Im}(\Li_2(e_8)) \\
{\sf l = 9}   & &  {\sf Im}(\Li_2(e_{9})),
{\sf Im}(\Li_2(e_{9}^2)) \\
{\sf l = 10}  & & -  \\
{\sf l = 11}  & & \left.
{\sf Im}(\Li_2(e_{11}^k))\right|_{k=1}^5\\  
{\sf l = 12}  & & {\sf Im}(\Li_2(e_{12}))\\
{\sf l = 13}  & & \left.
{\sf Im}(\Li_2(e_{13}))\right|_{k=1}^6 \\  
{\sf l = 14}   & & - \\
{\sf l = 15}   & &  
{\sf Im}(\Li_2(e_{15}))
\\
{\sf l = 16}   & & {\sf Im}(\Li_2(e_{16})),  {\sf Im}(\Li_2(e_{16}^3))\\
{\sf l = 17}  & & \left.
{\sf Im}(\Li_2(e_{17}))\right|_{k=1}^8\\  
{\sf l = 18}   & & - \\
{\sf l = 19}  & & \left.
{\sf Im}(\Li_2(e_{19}))\right|_{k=1}^9  \\
{\sf l = 20}   & & {\sf Im}(\Li_2(e_{20}))~.
\end{eqnarray}
%--------------------------------------------------------------------------------

Let us now turn to all convergent sums $\sigma_{1,1}(x,y)$, $x,y \in {\cal C}_n$. These sums belong to
cyclotomy {\sf l} if $x = e_l^{k_1}, y = e_l^{k_2}$ and $k_1, k_2 \in \mathbb{N}_+, k_1 < 
l, k_2 < l$.

We consider first the case $x \neq 
1, y = 1$,
%--------------------------------------------------------------------------------
\begin{eqnarray}
\sigma_{1,1}(x,1) &=& \Li_2(x) + \frac{1}{2} \Li_1^2(x) \nonumber\\
                  &=& \frac{1}{2} {\sf Re}(\Li_1^2(x)) + \left(r_2 - \frac{1}{2} r_1^2\right) \pi^2
                      + i \left[r_1 \pi {\sf Re}(\Li_1(x)) + {\sf Im}(\Li_2(x))\right]~,   
\end{eqnarray}
%--------------------------------------------------------------------------------
with
%--------------------------------------------------------------------------------
\begin{eqnarray}
r_1 &=& {\sf Im}(-\ln(1-x)) \\
r_2 &=& {\sf Re}(\Li_2(x))~.
\end{eqnarray}
%--------------------------------------------------------------------------------
Including the basis elements of {\sf w = 1} up to {\sf l}, no new basis element  is obtained.
Furthermore, 
%--------------------------------------------------------------------------------
\begin{eqnarray}
\label{eq:sigD}
\sigma_{1,1}(e_2,x) &=& -\frac{1}{2} \pi^2 +\frac{1}{2} \ln^2(2) + \Li_2\left(\frac{1-x}{2}\right)\\
\sigma_{1,1}(x,e_2) &=& -\Li_1(x) \ln(2) + \frac{1}{2} \Li_2(x^2) - \Li_2(x) + \frac{1}{2} \left[\pi^2 +
\ln^2(2) \right] - \Li_2\left(\frac{1-x}{2}\right)
\end{eqnarray}
%--------------------------------------------------------------------------------
hold. $\Li_2\left((1+x)/{2}\right)$ may yield a new basis element.
In some cases besides $x$ also $-x$ is element of the the cycle of the roots of unity which have to be 
considered. Here, however, $\Li_2\left((1+x)/{2}\right)$ is given by
%--------------------------------------------------------------------------------
\begin{eqnarray}
\Li_2\left(\frac{1+x}{2}\right) = - \Li_2\left(\frac{1-x}{2}\right) + \frac{\pi^2}{6} - \Li_1(x) \Li_1(-x) - \ln^2(2)
- \ln(2) \left[\Li_1(x) + \Li_1(-x)\right]~.
\nonumber\\
\end{eqnarray}
%--------------------------------------------------------------------------------
Also the elements $x = e_n^k$ and $e_1 - x \equiv 1 - x$ occur in the cycles, for which
%--------------------------------------------------------------------------------
\begin{eqnarray}
\label{eq:LI2c}
\Li_2(1-e_n^k) = - \Li_2(e_n^k) + 2 \pi i \frac{k}{n} \Li_1(e_n^k) + \frac{\pi^2}{6}
\end{eqnarray}
%--------------------------------------------------------------------------------
holds. 

For {\sf l = 2} one obtains
%--------------------------------------------------------------------------------
\begin{eqnarray}
\sigma_{1,1}(e_2,1) = -\frac{1}{2} \pi^2 + \frac{1}{2} \ln^2(2)~.
\end{eqnarray}
%--------------------------------------------------------------------------------
Because of (\ref{eq:LI2a}) two basis elements contribute. If a corresponding special number 
has occured already at {\sf w = 1} at the same value of {\sf l}, it is not counted as new. 
$\pi$ occurs at {\sf w = 2, l = 1}, unlike for {\sf w = 1} where it contributes first at  
{\sf l = 3}. No new basis element  occurs at {\sf l=2}. For {\sf w = 3,4} 
the new basis elements occur for $\Li_2(e_n^k)$ only. 
We apply the relations, cf. e.g. \cite{LEWIN1},
%--------------------------------------------------------------------------------
\begin{eqnarray}
\Li_2\left(\frac{1}{1+x}\right) &=& \Li_2(-x) + \ln(x) \ln(1+x) - \frac{1}{2} \ln^2(1+x) + \zeta_2 \\ 
\Li_2\left(\frac{x}{1+x}\right) &=& - \Li_2(-x) - \frac{1}{2} \ln^2(1+x)~.
\end{eqnarray}
%--------------------------------------------------------------------------------
At {\sf l = 5} the above relations lead to corresponding reductions and the two functions
%--------------------------------------------------------------------------------
\begin{eqnarray}
\Li_2\left(-\frac{e_5^3(1-e_5)}{1-e_5^3}\right),~~~~~~~~~~~~~\Li_2\left(-\frac{e_5^2}{1+e_5} \right)
\end{eqnarray}
%--------------------------------------------------------------------------------
remain. For the first function the identity
%--------------------------------------------------------------------------------
\begin{eqnarray}
\Li_2\left(-\frac{e_5^3(1-e_5)}{1-e_5^3}\right) = \Li_2\left(\frac{e_5^4}{1+e_5^4}\right)
= - \Li_2(-e_5^4) - \frac{1}{2} \ln^2(1+e_5^4)
\end{eqnarray}
%--------------------------------------------------------------------------------
holds, through which the corresponding sum $\sigma_{1,1}(e_5^3,e_5)$ can be expressed by
%--------------------------------------------------------------------------------
\begin{eqnarray}
\sigma_{1,1}(e_5^3,e_5) = \frac{1}{2} \Li_2^*(e_5^2)                         
                          - \Li_2^*(e_5) 
                          - \frac{1}{2} \left(\Li_1^2(e_5)\right)^* 
                          + \Li_1(e_5)^*\Li_1(e_5^2)^*~.  
\end{eqnarray}
%--------------------------------------------------------------------------------
The second dilogarithm can be transformed in the following way~:
%--------------------------------------------------------------------------------
\begin{eqnarray}
\Li_2\left(-\frac{e_5^2}{1+e_5} \right) &=& 
- \Li_2\left(\frac{1+e_5+e_5^2}{1+e_5}\right)-\ln\left(-\frac{e_5^2}{1+e_5}\right) 
\ln\left(\frac{1+e_5+e_5^2}{1+e_5}\right)+ 
\zeta_2~.
\end{eqnarray}
%--------------------------------------------------------------------------------
Furthermore,
%--------------------------------------------------------------------------------
\begin{eqnarray}
\Li_2\left(\frac{1+e_5+e_5^2}{1+e_5}\right) = \Li_2(-e_5^3)
\end{eqnarray}
%--------------------------------------------------------------------------------
holds, through which
%--------------------------------------------------------------------------------
\begin{eqnarray}
\Li_2\left(-\frac{e_5^2}{1+e_5} \right) &=& 
- \frac{1}{2} \Li_2(e_5)
+ \Li_2^*(e_5^2) + \frac{19}{150} \pi^2 - \frac{i \pi}{5}\left[\Li_1(e_5^2) - \Li_1(e_5)\right]
\end{eqnarray}
%--------------------------------------------------------------------------------
is obtained.

The representations at {\sf w = 6} were given in \cite{DB1}, with $\Li_4(1/2)$ 
the new basis element. The Clausen function ${\rm Cl}_2(\pi/3)$ used there is given by
%--------------------------------------------------------------------------------
\begin{eqnarray}
{\rm Cl}_n(x) &=& \left\{\begin{array}{l} 
\frac{i}{2}\left[\Li_n(\exp(-ix)) - \Li_n(\exp(ix)) \right],~~~n~\text{even}
\\
\frac{1}{2}\left[\Li_n(\exp(-ix)) + \Li_n(\exp(ix)) \right],~~~n~\text{odd}
\end{array} \right.
\\
{\rm Cl}_2\left(\frac{\pi}{3}\right) &=& {\rm Ls}_{2}^{(0)}\left(\frac{\pi}{3}\right) =
\frac{3}{2} {\sf Im}(\Li_2(e_3)) 
\end{eqnarray}
%--------------------------------------------------------------------------------
with
%--------------------------------------------------------------------------------
\begin{eqnarray}
{\rm Ls}_j^{(k)}(\theta) = - \int_0^\theta dt t^k \ln^{j-k-1}\left|2 \sin\left(\frac{t}{2}\right)\right|~,
\end{eqnarray}
%--------------------------------------------------------------------------------
cf.~\cite{KK}. In Table~7 we summarize the number of basis elements found at {\sf w = 2} using the relations in 
Section~4.3, the distribution relations (\ref{eq:DISTR}, \ref{eq:DISTR1}), and the relations for 
dilogarithms given above. We also list the number of basis elements for the class of dilogarithms at roots
of unity, and in both cases the number of new elements beyond those being obtained at {\sf w = 1} at the 
same value of {\sf l}.

%-----------------------------------------------------------------------------------------
\begin{table}[H]\centering
%\begin{tabular}{l*{6}{c}r}
{\small
\begin{tabular}{|c|r|r|r|r|r|r|r|r|r|r|r|r|r|r|r|r|r|r|r|r|}
\hline
\multicolumn{1}{|c}{\sf l}             &
\multicolumn{1}{|c}{1}                 &
\multicolumn{1}{|c}{2}                 &
\multicolumn{1}{|c}{3}                 &
\multicolumn{1}{|c}{4}                 &
\multicolumn{1}{|c}{5}                 &
\multicolumn{1}{|c}{6}                 &
\multicolumn{1}{|c}{7}                 &
\multicolumn{1}{|c}{8}                 &
\multicolumn{1}{|c}{9}                 &
\multicolumn{1}{|c}{10}                &
\multicolumn{1}{|c}{11}                &
\multicolumn{1}{|c}{12}                &
\multicolumn{1}{|c}{13}                &
\multicolumn{1}{|c}{14}                &

\multicolumn{1}{|c}{15}                &
\multicolumn{1}{|c}{16}                &
\multicolumn{1}{|c}{17}                &
\multicolumn{1}{|c}{18}                &
\multicolumn{1}{|c}{19}                &
\multicolumn{1}{|c|}{20}               \\
\hline
%{\sf w = 2} 
%&  &   &   &   &   &   &   &   &   &   &   &   &   &   &   &   &   &   &    & \\
{\sf $\Li_2$~~basis} 
& 1 & 1 & 2 & 2 & 3 & 2 & 4 & 3 & 4 & 3 & 6 & 3 & 7 & 4 & 5 & 5 & 9 & 4 & 10 & 5   
\\
{\sf $\Li_2$~~new} 
& 1 & 0 & 1 & 1 & 2 & 0 & 3 & 1 & 2 & 0 & 5 & 0 & 6 & 0 & 1 & 2 & 8 & 0 & 9  & 1   
\\
{\sf basis} 
& 1 & 2 & 3 & 3 & 5  & 5 & 8  & 7  & 10  & 10  & 16  & 12  & 21  & 17  & 21  & 21  & 33  & 23  & 40   & 29   
\\
{\sf Ref.~\cite{RAC}} 
& 1 & 1 & 1 & 1 & 2 & 2 & 4 & 4 & 7 & 6 & 10&   &   &   &   &   &   &   &    &  
\\
{\sf  new} 
&  &   &   &   &   &   &   &   &   &   &   &   &   &   &   &   &   &   &    & \\
{\sf  elements} 
& 1 & 0 & 1 & 1 & 2 & 1 & 4 & 3 & 5  & 4  & 10 & 5  & 14 & 8  & 12  & 12  & 24 & 11 & 30  & 16   
\\
\hline
\end{tabular}
}
\label{TT:XXb}
\caption[]{\sf The number of the basis elements spanning the dilogarithms resp. {\sf w = 2} cyclotomic 
harmonic sums at $l$th roots of unity  up to {\sf  20}.}.
\end{table}

At {\sf w = 2} the respective new basis elements are~:
%--------------------------------------------------------------------------------
\begin{eqnarray}
{\sf l = 1}   & & \pi \\
{\sf l = 2}   & & - \\
{\sf l = 3}   & & {\sf Im}(\Li_2(e_3)) \\
{\sf l = 4}   & & {\bf C} \\
{\sf l = 5}   & & {\sf Im}(\Li_2(e_{5})),  {\sf Im}(\Li_2(e_{5}^2)) \\
{\sf l = 6}   & & \Li_4\left(\frac{1}{2}\right)  \\
{\sf l = 7}   & & \left. {\sf Im}(\Li_2(e_7^k))\right|_{k=1}^3,
\sigma_{1,1}(e_7,e_7^2) \\
{\sf l = 8}   & & {\sf Im}(\Li_2(e_8)),
\sigma_{1,1}(e_8,e_4),
\sigma_{1,1}(e_8,e_8^3)\\
{\sf l = 9}   & &  {\sf Im}(\Li_2(e_{9})),
{\sf Im}(\Li_2(e_{9}^2)), 
\sigma_{1,1}(e_9,e_9^2),
\sigma_{1,1}(e_9,e_3),
\sigma_{1,1}(e_9^2,e_3)\\
{\sf l = 10}  & & 
\sigma_{1,1}(e_5,e_2),
\sigma_{1,1}(e_5^2,e_2),
\sigma_{1,1}(e_{10},e_5),
\sigma_{1,1}(e_{10},e_{10}^3) \\
{\sf l = 11}  & &  \left.
{\sf Im}(\Li_2(e_{11}^k))\right|_{k=1}^5, 
\left. \sigma_{1,1}(e_{11}, e_{11}^k)\right|_{k=2}^4,
\left. \sigma_{1,1}(e_{11}^2, e_{11}^k)\right|_{k=3}^4,~\text{etc.} 
\end{eqnarray}
%--------------------------------------------------------------------------------
\noindent
We mention that
counting relations for majorants of the motivic numbers, which are claimed to be related 
to the bases of the sums $\sigma_{k_1,...,k_m}(x_1, ..., x_m), x_j \in {\cal C}_n$, were
given in \cite{DG}. The dimension of the respective $\mathbb{Q}$-vector space is majorized by the expansion 
coefficients $D_n({\sf w})$ of
%--------------------------------------------------------------------------------
\begin{eqnarray}
G_{{\sf w}}(x) &=& \sum_{n=0}^\infty D_n({\sf w}) x^n 
\end{eqnarray}
%--------------------------------------------------------------------------------
with
%--------------------------------------------------------------------------------
\begin{eqnarray}
G_{\sf 1}(x)      &=& \frac{1}{1 - x^2 - x^3} \nonumber\\
G_{\sf 2}(x)      &=& \frac{1}{1 - x - x^2} \nonumber\\
G_{\sf k}(x)      &=& \frac{1}{1 - \frac{1}{2}\left[\varphi(k) + \nu\right] x - (\nu-1) 
x^2},~~ k \geq 3~. 
\end{eqnarray}
%--------------------------------------------------------------------------------
Here, $\varphi(k)$ denotes Euler's totient function \cite{TOTIENT} and $\nu$ is the number of prime
factors of $k$. 

%%%%%%%%%%%%%%%%%%%%%%%%%%%%%%%%%%%%%%%%%%%%%%%%%%%%%%%%%%%%%%%%%%%%%%%%
\section{Conclusions}
\label{sec:7}
%%%%%%%%%%%%%%%%%%%%%%%%%%%%%%%%%%%%%%%%%%%%%%%%%%%%%%%%%%%%%%%%%%%%%%%%

\vspace{1mm}
\noindent
In evaluating massive 3-loop integrals with local operator insertions
nested sums occur, containing denominators which belong to the class of
harmonic sums generated by cyclotomic polynomials. To simplify the
computations, the relations between these quantities have to be know and
used in computer algebra codes such as {\sf Sigma} \cite{SIGMA} and
{\sf HarmonicSums} \cite{HASUM} to allow for an efficient reduction in
the corresponding summation problem. In integrals of similar structure
we expect even more general terms (\ref{eq:NEWT}) to occur.

The usual harmonic sums \cite{HS1,HS2} and polylogarithms \cite{VR} were
thus generalized to cover the newly occurring structures. We started with
the harmonic polylogarithms extending the usual alphabet of denominators
ranging to $\Phi_2(x)$ to general cyclotomic polynomials $\Phi_n(x), n \geq 3$.
The cyclotomic harmonic polylogarithms form a shuffle algebra. They have
support $x \in [0,1]$ and one may define a Mellin transform, usually of argument
$k N$, with $k,N \in \mathbb{N}_+$ which span the finite cyclotomic
harmonic sums (\ref{eq:MSU}) together with special values as the cyclotomic harmonic sums
in the limit $N \rightarrow \infty$.

The cyclotomic harmonic sums are meromorphic functions with poles at the non-positive
integers. They obey recurrence relations in terms of sums of lower depth and one may
derive analytical asymptotic representations. In this way the cyclotomic harmonic
sums are analytically continued from integer values of the sum index $N$ to $N \in \mathbb{C}$.
The cyclotomic harmonic sums form a quasi-shuffle algebra, cf.~\cite{HOFF}.
After the analytic continuation they obey differentiation relations, accounting for their values at
$N \rightarrow \infty$. Furthermore, three multiple argument relations apply.
Using these relations one may reduce the number of cyclotomic harmonic sums vastly growing
with the weight to smaller bases. We study the case of the sums following from iteration of the
denominators (\ref{eqSel}) up to weight {\sf w = 5}. Corresponding counting relations for
the number of basis elements are obtained.

The values of the cyclotomic harmonic sums for $N \rightarrow \infty$, resp. the values of the
cyclotomic harmonic polylogarithms at $x=1$, are of special interest. They are linearly related
to the infinite nested harmonic sums with roots of unity as numerator weight factors. Already in
case of the single infinite cyclotomic harmonic sums a large number of new constants contribute
beyond those know in the case of multiple zeta values and Euler-Zagier values \cite{MZV}.
The quasi-shuffle and multiple argument relations of the cyclotomic harmonic sums and the shuffle
relations of the cyclotomic harmonic polylogarithms allow to derive basis representations induced by 
these relations. We studied
in this respect the case of the infinite cyclotomic harmonic sums based on the iteration of the summands (\ref{eqSel})
to weight {\sf w = 6} and the sums of weight {\sf w = 1,2} for cyclotomy {$\sf l \leq 20$}.
Using computer algebra methods the explicit representation of all infinite cyclotomic harmonic sums
were derived as well. For wide classes of relations explicit counting relations of the basis elements
were given. The corresponding representations were derived with the {\sf Mathematica}-based
computer algebra system {\sf HarmonicSums} \cite{HASUM}.

The present investigations can be readily extended to cyclotomic harmonic sums and polylogarithms of 
higher weight and
cyclotomy, both in the case of finite values of $N$ and for $N \rightarrow \infty$, using the present methods.
The requested computational time and storage resources grow accordingly. This applies in particular to the
derivation of the explicit representations of all sums over the corresponding bases.

We finally considered also the generalized harmonic sums of weight {\sf w = 1,2} at $l$th roots of unity
for $1 \leq l \leq 20$. They obey dilogarithmic reprentations. Besides the shuffle and stuffle relations,
they obey distribution relations and the known relations for dilogarithms. We used these relations to derive
corresponding basis representations. Compared to the case of the infinite cyclotomic harmonic sums these
sums obey more symmetries. Thus at a given weight and cyclotomy $l$ they are represented by a lower number 
of basis elements. We compared to results in literature.

\vspace*{4mm}
\noindent
{\bf Acknowledgment.}~
{We thank D. Broadhurst for providing us the relations (\ref{eq:DB1}) and (\ref{eq:DB2}). For discussions we would
like to thank D. Broadhurst, F. Brown and D. Kreimer, and N.J.A. Sloan and M. Kauers for remarks on
special issues. We would like to thank H. Kawamura for Ref.~\cite{BERNOULLI1}. This work
has been supported in part by DFG Sonderforschungsbereich Transregio 9, Computergest\"utzte Theoretische Teilchenphysik, by the Austrian Science Fund (FWF) grant P20162-N18, and by the
EU Network {\sf LHCPHENOnet}
PITN-GA-2010-264564.}

\newpage
\appendix
%%%%%%%%%%%%%%%%%%%%%%%%%%%%%%%%%%%%%%%%%%%%%%%%%%%%%%%%%%%%%%%%%%%%%%%%
\section{Appendix}
\label{sec:APP1}
%%%%%%%%%%%%%%%%%%%%%%%%%%%%%%%%%%%%%%%%%%%%%%%%%%%%%%%%%%%%%%%%%%%%%%%%

\vspace{1mm}
\noindent
In the following we summarize some technical aspects needed to represent expressions
used in the present paper.

First we summarize some aspects on cyclotomic polynomials, \cite{LANG}.
We give the decompositions of the polynomials
%----------------------------------------------------------------------------------------------
\begin{eqnarray}
x^l + 1,~~~l \in \mathbb{N} \backslash \{0\}
\end{eqnarray}
%----------------------------------------------------------------------------------------------
in terms of cyclotomic polynomials for $l \leq 20$~:
%----------------------------------------------------------------------------------------------
\begin{eqnarray}
\label{eq:A1}
x^3 + 1 &=& \Phi_2(x) \Phi_6(x) \\
x^5 + 1 &=& \Phi_2(x) \Phi_{10}(x) \\
x^6 + 1 &=& \Phi_4(x) \Phi_{12}(x) \\
x^7 + 1 &=& \Phi_2(x) \Phi_{14}(x) \\
\label{eq:A1a}
x^9 + 1 &=& \Phi_2(x) \Phi_{6}(x) \Phi_{18}(x) \\
x^{10} + 1 &=& \Phi_4(x) \Phi_{20}(x) \\
x^{11} + 1 &=& \Phi_2(x) \Phi_{22}(x) \\
x^{12} + 1 &=& \Phi_8(x) \Phi_{24}(x) \\
x^{13} + 1 &=& \Phi_2(x) \Phi_{26}(x) \\
x^{14} + 1 &=& \Phi_4(x) \Phi_{28}(x) \\
\label{eq:A1b}
x^{15} + 1 &=& \Phi_2(x) \Phi_6(x) \Phi_{10}(x) \Phi_{30}(x)\\
x^{17} + 1 &=& \Phi_2(x) \Phi_{34}(x) \\
\label{eq:A1c}
x^{18} + 1 &=& \Phi_4(x) \Phi_{12}(x) \Phi_{36}(x) \\
x^{19} + 1 &=& \Phi_2(x) \Phi_{39}(x) \\
\label{eq:A2}
x^{20} + 1 &=& \Phi_8(x) \Phi_{40}(x)~.
\end{eqnarray}
%----------------------------------------------------------------------------------------------
For odd values of $n$,
%----------------------------------------------------------------------------------------------
\begin{eqnarray}
\Phi_{2n}(x) = \Phi_n(-x)
\end{eqnarray}
%----------------------------------------------------------------------------------------------
holds. The decomposition
%----------------------------------------------------------------------------------------------
\begin{eqnarray}
x^{2k+1} - 1 = (x-1) \prod_i \Phi_i(x)
\end{eqnarray}
%----------------------------------------------------------------------------------------------
results thus into
%----------------------------------------------------------------------------------------------
\begin{eqnarray}
x^{2k+1} + 1 = \Phi_2(x) \prod_i \Phi_{2i}(x)~.
\end{eqnarray}
%----------------------------------------------------------------------------------------------
From
%----------------------------------------------------------------------------------------------
\begin{eqnarray}
\Phi_{2n}(x)~|~(x^{2n} - 1)~~~~~~\text{and}~~~~~~\Phi_{2n}(x)~\not|~(x^{n} - 1)
\end{eqnarray}
%----------------------------------------------------------------------------------------------
it follows
%----------------------------------------------------------------------------------------------
\begin{eqnarray}
\Phi_{2n}(x)~|~(x^{n} + 1)~.
\end{eqnarray}
%----------------------------------------------------------------------------------------------
If $p$ is a prime and $p|n$ then~\cite{NAG}
%----------------------------------------------------------------------------------------------
\begin{eqnarray}
\Phi_{pn}(x) = \Phi_n(x^p)~.
\end{eqnarray}
%----------------------------------------------------------------------------------------------
For $n = 2^k, k \in \mathbb{N}_+$ it follows that all $\Phi_{2^k}(x)$ are cyclotomic
polynomials. In (\ref{eq:A1a},\ref{eq:A1b},\ref{eq:A1c}) more factors than just $\Phi_{2n}(x)$ occur.
They originate due to power-rescaling, i.e.,
%----------------------------------------------------------------------------------------------
\begin{eqnarray}
\frac{x^{15}+1}{x^3+1} &=& \frac{y^{5}+1}{y+1} = \Phi_{10}(y) \\
\frac{x^{15}+1}{x^5+1} &=& \frac{y^{3}+1}{y+1} = \Phi_{6}(y)~.
\end{eqnarray}
%----------------------------------------------------------------------------------------------
Therefore, all the factors of $(x^5+1)$ and $(x^3+1)$ have to emerge in the decomposition, and similarly
for other $N$ with more non-trivial divisors. For $N = 2^k \cdot n$ where the integer $n > 1$ is odd we get
%----------------------------------------------------------------------------------------------
\begin{eqnarray}
\frac{x^{2^k \cdot n}+1}{x^{2^k}+1} &=& \frac{y^{n}+1}{y+1} = \Phi_{2n}(y)~.
\end{eqnarray}
%----------------------------------------------------------------------------------------------
The argument remains valid if $n$ is a product of  odd primes.
Therefore the only cyclotomic polynomials of the structure $x^a+1$ are those with $a = 2^k, k \in \mathbb{N}_+$.

For a proper definition of the cyclotomic harmonic polynomials which appear in sum representations like
(\ref{eq:SISU})
%----------------------------------------------------------------------------------------------
\begin{eqnarray}
\frac{1}{x^l \pm 1},~~~l \in \mathbb{N}_+
\end{eqnarray}
%----------------------------------------------------------------------------------------------
we perform partial fractioning. In the following we provide the corresponding expressions in terms of the
words $f_k^l(x)$ forming the cyclotomic harmonic polylogarithms up to $l = 6$~:
%----------------------------------------------------------------------------------------------
\begin{eqnarray}
\label{eq:A3}
%--------------------------------
(x - 1)^{-1} &=& f_1^0(x) \\
%--------------------------------
(x + 1)^{-1} &=& f_2^0(x) \\
%--------------------------------
(x^2 - 1)^{-1} &=& \frac{1}{2} \left[f_1^0(x) - f_2^0(x)\right] \\
%--------------------------------
(x^2 + 1)^{-1} &=& f_4^0(x)\\
%--------------------------------
(x^3 - 1)^{-1} &=& \frac{1}{3} \left[f_1^0(x) - 2 f_3^1(x) - f_3^0(x) \right] \\
%--------------------------------
(x^3 + 1)^{-1} &=& \frac{1}{3} \left[f_2^0(x) - f_6^1(x) + 2 f_6^0(x) \right] \\
%--------------------------------
(x^4 - 1)^{-1} &=& \frac{1}{4} \left[f_1^0(x) - f_2^0(x) - 2 f_4^0(x) \right] \\
%--------------------------------
(x^4 + 1)^{-1} &=& f_8^0(x) \\
%--------------------------------
(x^5 - 1)^{-1} &=& \frac{1}{5} \left[f_1^0(x)
- \frac{4}{5} f_5^0(x)
- \frac{3}{5} f_5^1(x)
- \frac{2}{5} f_5^2(x)
- \frac{1}{5} f_5^3(x)
\right] \\
%--------------------------------
(x^5 + 1)^{-1} &=& \frac{1}{5} \left[f_2^0(x)
+ \frac{4}{5} f_5^0(x)
- \frac{3}{5} f_5^1(x)
+ \frac{2}{5} f_5^2(x)
- \frac{1}{5} f_5^3(x)
\right] \\
%--------------------------------
(x^6 - 1)^{-1} &=& \frac{1}{6} \left[
  f_1^0(x)
- f_2^0(x)
- 2 f_3^0(x)
-   f_3^1(x)
- 2 f_6^0(x)
+   f_6^1(x)
\right] \\
%--------------------------------
\label{eq:A4}
(x^6 + 1)^{-1} &=& \frac{1}{3} \left[
  f_4^0(x)
+ 2 f_{12}^0(x)
-   f_{12}^2(x)
\right] ~~ \text{etc.}
\end{eqnarray}
%----------------------------------------------------------------------------------------------
%%%%%%%%%%%%%%%%%%%%%%%%%%%%%%%%%%%%%%%%%%%%%%%%%%%%%%%%%%%%%%%%%%%%%%%%
\section{Appendix}
\label{sec:APP2}
%%%%%%%%%%%%%%%%%%%%%%%%%%%%%%%%%%%%%%%%%%%%%%%%%%%%%%%%%%%%%%%%%%%%%%%%

\vspace{1mm}
\noindent
{\sf \underline{Proof of Eq.~(\ref{halfint}).}}\\
%\begin{proof}

\vspace{1mm}
\noindent
We proceed by induction on $m.$ Let $m=1:$
%----------------------------------------------------------------------------------------------
\begin{eqnarray*}
\X{\{a_1,b_1,c_1\}}{2n}+\X{\{a_1,b_1,-c_1\}}{2n}&=&\sum_{i=1}^{2 n}\frac{1}{(a_1 i
+b_1)^{c_1}}+\sum_{i=1}^{2
n}\frac{(-1)^i}{(a_1 i +b_1)^{c_1}}\\
		&=&\sum_{i=1}^{n}\left( \frac{1}{(a_1 2i +b_1)^{c_1}}+\frac{1}{(a_1 (2i -1) +b_1)^{c_1}}\right)\\
		&&+\sum_{i=1}^{n}\left( \frac{1}{(a_1 2i +b_1)^{c_1}}-\frac{1}{(a_1 (2i -1) +b_1)^{c_1}}\right)\\
		&=&2 \sum_{i=1}^{n}\frac{1}{(2 a_1 i +b_1)^{c_1}}=2 \X{\{2 a_1,b_1,c_1\}}{n}.\\
\end{eqnarray*}
%----------------------------------------------------------------------------------------------
In the following we use the abbreviation:
%----------------------------------------------------------------------------------------------
$$
A(n):=\sum{\X{\{a_m,b_m,\pm c_m\},\ldots,\{a_1,b_1,\pm c_1\}}{n}}.
$$
%----------------------------------------------------------------------------------------------
Now we assume that (\ref{halfint}) holds for $m:$
%----------------------------------------------------------------------------------------------
\begin{eqnarray*}
&&\sum{\X{\{a_{m+1},b_{m+1},\pm c_{m+1}\},\ldots,\{a_1,b_1,\pm c_1\}}{2n}}\\
	 &&\hspace{1cm}=\sum_{i=1}^{2n}\frac{1}{(a_{m+1}i+b_{m+1})^{c_{m+1}}}\sum{\X{\{a_m,b_m,\pm
c_m\},\ldots,\{a_1,b_1,\pm c_1\}}{i}}\\
	&&\hspace{1cm} \ +\sum_{i=1}^{2n}\frac{(-1)^i}{(a_{m+1}i+b_{m+1})^{c_{m+1}}}\sum{\X{\{a_m,b_m,\pm
c_m\},\ldots,\{a_1,b_1,\pm c_1\}}{i}}\\
	 &&\hspace{1cm}=\sum_{i=1}^{2n}\frac{A(i)}{(a_{m+1}i+b_{m+1})^{c_{m+1}}}+\sum_{i=1}^{2n}\frac{(-1)^i A(i)}{(a_{m+1}i+b_{m+1})^{c_{m+1}}}\\
	&&\hspace{1cm}=\sum_{i=1}^{n}\left(\frac{A(2i)}{(2 a_{m+1}i+b_{m+1})^{c_{m+1}}}+\frac{A(2 i-1)}{((2 i-1) a_{m+1}+b_{m+1})^{c_{m+1}}}\right)\\
	&&\hspace{1cm} \ +\sum_{i=1}^{n}\left(\frac{(-1)^{2i}A(2i)}{(2 a_{m+1}i+b_{m+1})^{c_{m+1}}}+\frac{(-1)^{2i-1}A(2 i-1)}{( (2 i-1) a_{m+1}+b_{m+1})^{c_{m+1}}}\right)\\
	&&\hspace{1cm}= 2 \sum_{i=1}^{n}\frac{A(2i)}{(2 a_{m+1}i+b_{m+1})^{c_{m+1}}}\\
	&&\hspace{1cm}= 2 \sum_{i=1}^{n}\frac{1}{(2 a_{m+1}i+b_{m+1})^{c_{m+1}}} 2^m \X{\{2 a_m,b_m,c_m\},\ldots,\{2
a_1,b_1,c_1\}}{i}\\
	&&\hspace{1cm}= 2^{m+1} \X{\{2 a_{m+1},b_{m+1},c_{m+1}\},\ldots,\{2 a_1,b_1,
c_1\}}{n}~~~~\Box
\end{eqnarray*}
%----------------------------------------------------------------------------------------------
%\end{proof}

\newpage
\vspace{1mm}
\noindent
{\sf \underline{Proof of Eq.~(\ref{halfint2}).}}\\
%\begin{proof}

\vspace{1mm}
\noindent
%\begin{proof}
We proceed by induction on $m.$ Let $m=1:$
%----------------------------------------------------------------------------------------------
\begin{eqnarray*}
\X{\{a_1,b_1,c_1\}}{2n}-\X{\{a_1,b_1,-c_1\}}{2n}&=&\sum_{i=1}^{2 n}\frac{1}{(a_1 i +b_1)^{c_1}}+\sum_{i=1}^{2
n}\frac{(-1)^i}{(a_1 i +b_1)^{c_1}}\\
		&=&\sum_{i=1}^{n}\left( \frac{1}{(a_1 2i +b_1)^{c_1}}+\frac{1}{(a_1 (2i -1) +b_1)^{c_1}}\right)\\
		&&-\sum_{i=1}^{n}\left( \frac{1}{(a_1 2i +b_1)^{c_1}}-\frac{1}{(a_1 (2i -1) +b_1)^{c_1}}\right)\\
		&=&2 \sum_{i=1}^{n}\frac{1}{((2i -1) a_1+b_1)^{c_1}}=2 \X{\{2 a_1,b_1-a_1,c_1\}}{n}.\\
\end{eqnarray*}
%----------------------------------------------------------------------------------------------
In the following we use the abbreviation:
%----------------------------------------------------------------------------------------------
$$
A(n):=\sum{d_m \cdots d_1 \X{\{a_m,b_m,d_m c_m\},\ldots,\{a_1,b_1, d_1 c_1\}}{n}}.
$$
%----------------------------------------------------------------------------------------------
Note that
%----------------------------------------------------------------------------------------------
%>JA
\begin{eqnarray*}
A(2 n-1)&=&A(2n)-\sum{d_m \cdots d_1 \frac{d_m^{2n} \X{\{a_{m-1},b_{m-1},d_{m-1} c_{m-1}\},\ldots,\{a_1,b_1, d_1
c_1\}}{2n} } {(a_m 2n+b_m)^{c_m}}}\\
	&=&A(2n)-\sum{d_{m-1} \cdots d_1 \frac{\X{\{a_{m-1},b_{m-1},d_{m-1} c_{m-1}\},\ldots,\{a_1,b_1, d_1 c_1\}}{2n} }
{(a_m 2n+b_m)^{c_m}}}\\
	&&+\sum{d_{m-1} \cdots d_1 \frac{\X{\{a_{m-1},b_{m-1},d_{m-1} c_{m-1}\},\ldots,\{a_1,b_1, d_1 c_1\}}{2n} } {(a_m
2n+b_m)^{c_m}}}=A(2n).
\end{eqnarray*}
%<JA
%----------------------------------------------------------------------------------------------
Now we assume that (\ref{halfint2}) holds for $m:$
%----------------------------------------------------------------------------------------------
\begin{eqnarray*}
&&\sum{d_{m+1} \cdots d_1 \X{\{a_{m+1},b_{m+1},d_{m+1} c_{m+1}\},\ldots,\{a_1,b_1,d_1 c_1\}}{2n}}\\
	&&\hspace{1cm}=\sum_{i=1}^{2n}\frac{1}{(a_{m+1}i+b_{m+1})^{c_{m+1}}}\sum{d_m
\cdots d_1\X{\{a_m,b_m,d_m c_m\},\ldots,\{a_1,b_1,d_1 c_1\}}{i}}\\
	&&\hspace{1cm} \
-\sum_{i=1}^{2n}\frac{(-1)^i}{(a_{m+1}i+b_{m+1})^{c_{m+1}}}\sum{d_m \cdots d_1\X{\{a_m,b_m,d_m c_m\},\ldots,\{a_1,b_1,d_1 c_1\}}{i}}\\
	 &&\hspace{1cm}=\sum_{i=1}^{2n}\frac{A(i)}{(a_{m+1}i+b_{m+1})^{c_{m+1}}}-\sum_{i=1}^{2n}\frac{(-1)^i A(i)}{(a_{m+1}i+b_{m+1})^{c_{m+1}}}\\
	&&\hspace{1cm}=\sum_{i=1}^{n}\left(\frac{A(2i)}{(2 a_{m+1}i+b_{m+1})^{c_{m+1}}}+\frac{A(2 i-1)}{((2 i-1) a_{m+1}+b_{m+1})^{c_{m+1}}}\right)\\
	&&\hspace{1cm} \ -\sum_{i=1}^{n}\left(\frac{(-1)^{2i}A(2i)}{(2 a_{m+1}i+b_{m+1})^{c_{m+1}}}+\frac{(-1)^{2i-1}A(2 i-1)}{( (2 i-1) a_{m+1}+b_{m+1})^{c_{m+1}}}\right)\\
	&&\hspace{1cm}= 2 \sum_{i=1}^{n}\frac{A(2i-1)}{((2 i -1) a_{m+1}+b_{m+1})^{c_{m+1}}}= 2 \sum_{i=1}^{n}\frac{A(2i)}{((2 i -1) a_{m+1}+b_{m+1})^{c_{m+1}}}\\
	&&\hspace{1cm}= 2 \sum_{i=1}^{n}\frac{1}{(2 a_{m+1}i+b_{m+1}- a_{m+1})^{c_{m+1}}}
2^m \X{\{2 a_m,b_m-a_m, c_m\},\ldots,\{2 a_1,b_1-a_1, c_1\}}{n}\\
\end{eqnarray*}
\begin{eqnarray*}
	&&\hspace{1cm}= 2^{m+1} \X{\{2 a_{m+1},b_{m+1}-a_{m+1},c_{m+1}\},\ldots,\{2
a_1,b_1-a_1,c_1\}}{n} \Box
\end{eqnarray*}
%----------------------------------------------------------------------------------------------
%\end{proof}

\vspace{7mm}
\noindent
{\sf \underline{Proof of Eq.~(\ref{eq:Cf00}--\ref{eq:Cf41}).}}\\
%\begin{proof}

\vspace{4mm}
\noindent
We start with $C_{f_0^0,\vec{m}}(x)$. We consider integrals of the form
%----------------------------------------------------------------------------------------------
\begin{eqnarray*}
\int_0^x{\frac{1}{y}\sum_{i=1}^{\infty}\frac{\sigma^iy^{2 i+c_j}}{(2 i+c_j)^a}\X{\ve
n}i}dy&=&\sum_{i=1}^{\infty}\frac{\sigma^i}{(2 i+c_j)^a}\X{\ve n}i\int_0^x{y^{2
i+c_j-1}}dy\\
	&=&\sum_{i=1}^{\infty}\frac{\sigma^i}{(2 i+c_j)^a}\X{\ve n}i\frac{x^{2 i+c_j}}{2
i+c_j}dy\\
	&=&\sum_{i=1}^{\infty}\frac{\sigma^i x^{2 i+c_j}}{(2 i+c_j)^{a+1}}\X{\ve n}i~.
\end{eqnarray*}
%----------------------------------------------------------------------------------------------
Summing over $j$ yields the desired result.

\vspace*{5mm}
\noindent
For $C_{f_4^0,\vec{m}}(x)$ we  consider the integrals
%----------------------------------------------------------------------------------------------
\begin{eqnarray*}
\int_0^x{\frac{1}{1+y^2}\sum_{i=1}^{\infty}\frac{\sigma^iy^{2 i+c_j}}{(2 i+c_j)^a}\X{\ve
n}i}dy&=&
	\int_0^x{\sum_{i=1}^{\infty}{(-1)^iy^{2
i}}\sum_{i=0}^{\infty}\frac{\sigma^{i+1}y^{2 i+c_j+2}}{(2 i+c_j+2)^a}\X{\ve n}{i+1}}dy\\
	&=&\int_0^x{\sum_{i=1}^{\infty}\sum_{k=0}^{i}{(-1)^{i-k}y^{2 i- 2
k}}\frac{\sigma^{k+1}y^{2 k+c_j+2}}{(2 k+c_j+2)^a}\X{\ve n}{k+1}}dy\\
	&=&\int_0^x{\sum_{i=1}^{\infty}{(-1)^{i+1}y^{2 i +c_j
+2}}\sum_{k=0}^{i}\frac{(-\sigma)^{k+1}}{(2 k+c_j+2)^a}\X{\ve n}{k+1}}dy\\
	&=&\sum_{i=1}^{\infty}\frac{(-1)^{i+1}y^{2 i +c_j +3}}{2 i +c_j
+3}\sum_{k=1}^{i+1}\frac{(-\sigma)^{k}}{(2 k+c_j)^a}\X{\ve n}{k}\\
	&=&\sum_{i=1}^{\infty}\frac{(-1)^i x^{2 i+c_j+1}}{(2 i+c_j+1)}\X{\{2,c_j,-\sigma
a\},\ve n}i~.
\end{eqnarray*}
%----------------------------------------------------------------------------------------------
Summing over $j$ yields the desired result. The case $C_{f_4^1,\vec{m}}(x)$ follows analogously.

\vspace*{5mm}
\noindent
For $C_{f_2^0,\vec{m}}(x)$ we consider the integrals of the form
%----------------------------------------------------------------------------------------------
\begin{eqnarray*}
\int_0^x{\frac{1}{1+y}\sum_{i=1}^{\infty}\frac{\sigma^iy^{2 i+c_j}}{(2 i+c_j)^a}\X{\ve
n}i}dy&=&
	\int_0^x{\frac{1-y}{1-y^2}\sum_{i=1}^{\infty}\frac{\sigma^iy^{2 i+c_j}}{(2
i+c_j)^a}\X{\ve n}i}dy\\
	&=&\int_0^x{\frac{1}{1-y^2}\sum_{i=1}^{\infty}\frac{\sigma^iy^{2 i+c_j}}{(2
i+c_j)^a}\X{\ve n}i}dy\\
	&&-\int_0^x{\frac{y}{1-y^2}\sum_{i=1}^{\infty}\frac{\sigma^iy^{2 i+c_j}}{(2 i+c_j)^a}
\X{\ve n}i}dy\\
	&=&\int_0^x{\sum_{i=0}^{\infty}{y^{2 i}}
\sum_{i=0}^{\infty}\frac{\sigma^{i+1}y^{2 i+c_j+2}}{(2 i+c_j+2)^a}\X{\ve n}{i+1}}dy\\
\end{eqnarray*}
\begin{eqnarray*}
	&&-\int_0^x{\sum_{i=0}^{\infty}{y^{2 i+1}}
\sum_{i=0}^{\infty}\frac{\sigma^{i+1}y^{2 i+c_j+2}}{(2 i+c_j+2)^a}\X{\ve n}{i+1}}dy\\
	&=&\int_0^x{\sum_{i=0}^{\infty} \sum_{k=0}^{i} {y^{2 i - 2 k}}
\frac{\sigma^{k+1}y^{2 k+c_j+2}}{(2 k+c_j+2)^a}\X{\ve n}{k+1}}dy\\
	&&-\int_0^x{\sum_{i=0}^{\infty} \sum_{k=0}^{i} {y^{2 i - 2 k +1}}
\frac{\sigma^{k+1}y^{2 k+c_j+2}}{(2 k+c_j+2)^a}\X{\ve n}{k+1}}dy\\
	&=&\int_0^x\sum_{i=0}^{\infty} {y^{2 i +c_j +1}} \X{\{2,c_j,\sigma a\},\ve
n}{i+1} dy\\
	&&-\int_0^x\sum_{i=0}^{\infty} {y^{2 i +c_j +3}} \X{\{2,c_j,\sigma a\},\ve
n}{i+1} dy\\
	&=&\sum_{i=1}^{\infty}\frac{x^{2 i+c_j+1}}{(2 i+c_j+1)}\X{\{2,c_j,\sigma a\},\ve
n}i\\
	&&-\sum_{i=1}^{\infty}\frac{x^{2 i+c_j+2}}{(2 i+c_j+2)}\X{\{2,c_j,\sigma a\},\ve
n}i~.
\end{eqnarray*}
%----------------------------------------------------------------------------------------------
Summing over $j$ yields the desired result. The case $C_{f_1^0,,\vec{m}}(x)$ follows analogously.

\newpage
%%%%%%%%%%%%%%%%%%%%%%%%%%%%%%%%%%%%%%%%%%%%%%%%%%%%%%%%%%%%%%%%%%%%%%%%%

%%%%%%%%%%%%%%%%%%%%%%%%%%%%%%%%%%%%%%%%%%%%%%%%%%%%%%%%%%%%%%%%%%%%%%%%%%%%%%%%%%

%%%%%%%%%%%%%%%%%%%%%%%%%%%%%%%%%%%%%%%%%%%%%%%%%%%%%%%%%%%%%%%%%%%%%%%%%%%%%%%%%%
\end{document}